\documentclass[11pt,twoside]{article}
\usepackage{graphicx,amssymb,epsfig,epsf}
\usepackage{times}
\usepackage{a4wide}
\usepackage{cite}
\usepackage{amsmath}
\usepackage{dsfont}
\usepackage{floatflt}
\usepackage{cancel}
\usepackage{axodraw}
\usepackage{slashed}
\usepackage[utf8]{inputenc}

\allowdisplaybreaks[1]

\usepackage{fancyhdr}
\advance \headheight by 3.0truept       
\newlength{\nseparation}
\newenvironment{nfigure}[1]
        {\begin{figure}[#1]\vspace{-3\nseparation}\hrule\vspace{\nseparation}\par}
        {\vspace{\nseparation}\par \hrule \end{figure}}
\newenvironment{ntable}[1]
        {\begin{table}[#1]\hrule\vspace{\nseparation}\par}
        {\vspace{\nseparation}\par \hrule \end{table}}
\newenvironment{nfloatingfigure}[2][]
        {\begin{floatingfigure}[#1]{#2}\hrule\vspace{\nseparation}\par}
        {\vspace{\nseparation}\par \hrule 
         \vspace{\nseparation} \end{floatingfigure}}

\usepackage{axodraw,pstricks,color,caption,subfigure}

\def\be {\begin{equation}}
\def\ee {\end{equation}}
\def\bea {\begin{eqnarray}}
\def\eea {\end{eqnarray}}

\def\issue(#1,#2,#3){{\bf #1}, #2 (#3)}
\def\opcit(#1){ {\em op. cit.}, #1}

\def\APP(#1,#2,#3){Acta Phys.\ Polon.\ \issue(#1,#2,#3)}
\def\ARNPS(#1,#2,#3){Ann.\ Rev.\ Nucl.\ Part.\ Sci.\ \issue(#1,#2,#3)}
\def\CPC(#1,#2,#3){Comp.\ Phys.\ Comm.\ \issue(#1,#2,#3)}
\def\CIP(#1,#2,#3){Comput.\ Phys.\ \issue(#1,#2,#3)}
\def\EPJC(#1,#2,#3){Eur.\ Phys.\ J.\ C\ \issue(#1,#2,#3)}
\def\EPJD(#1,#2,#3){Eur.\ Phys.\ J. Direct\ C\ \issue(#1,#2,#3)}
\def\IEEETNS(#1,#2,#3){IEEE Trans.\ Nucl.\ Sci.\ \issue(#1,#2,#3)}
\def\IJMP(#1,#2,#3){Int.\ J.\ Mod.\ Phys. \issue(#1,#2,#3)}
\def\JHEP(#1,#2,#3){J.\ High Energy Physics \issue(#1,#2,#3)}
\def\JPG(#1,#2,#3){J.\ Phys.\ G \issue(#1,#2,#3)}
\def\MPL(#1,#2,#3){Mod.\ Phys.\ Lett.\ \issue(#1,#2,#3)}
\def\NP(#1,#2,#3){Nucl.\ Phys.\ \issue(#1,#2,#3)}
\def\NIM(#1,#2,#3){Nucl.\ Instrum.\ Meth.\ \issue(#1,#2,#3)}
\def\PL(#1,#2,#3){Phys.\ Lett.\ \issue(#1,#2,#3)}
\def\PRD(#1,#2,#3){Phys.\ Rev.\ D \issue(#1,#2,#3)}
\def\PRL(#1,#2,#3){Phys.\ Rev.\ Lett.\ \issue(#1,#2,#3)}
\def\SJNP(#1,#2,#3){Sov.\ J. Nucl.\ Phys.\ \issue(#1,#2,#3)}
\def\ZPC(#1,#2,#3){Zeit.\ Phys.\ C \issue(#1,#2,#3)}
\def\EPL(#1,#2,#3){Europhys.\ Lett.\ \issue(#1,#2,#3)}

\newcommand{\bb}{\ensuremath{B\!-\!\Bbar{}\,}}
\newcommand{\bbm}{\bb\ mixing}
\newcommand{\bbs}{\ensuremath{B_s\!-\!\Bbar{}_s\,}}
\newcommand{\bbms}{\bbs\ mixing}

\newcommand{\eq}[1]{Eq.~(\ref{#1})}
\newcommand{\eqsand}[2]{Eqs.~(\ref{#1}) and (\ref{#2})}
\newcommand{\eqsto}[2]{Eqs.~(\ref{#1}--\ref{#2})}

\newcommand{\Bbar}{\,\overline{\!B}}

\newcommand{\nn}{\nonumber \\}
\newcommand{\ov}[1]{\overline{#1}}
\newcommand{\lt}{\left}
\newcommand{\rt}{\right}

\newcommand{\mususy}{\ensuremath { \mu_{\scriptscriptstyle \text{SUSY} }}}
\newcommand{\msusy}{\ensuremath{M_{\rm SUSY}}}

\newcommand{\epsfc}{\ensuremath { \epsilon_{\scriptscriptstyle \text{FC} }}}

\newcommand{\gev}{\,\mbox{GeV}}

\newcommand{\fig}[1]{Fig.~\ref{#1}}
\newcommand{\ds}{\displaystyle}

\begin{document}
\thispagestyle{plain}
\parbox[t]{0.4\textwidth}{TTP09-26 \\
SFB/CPP-09-68}
\hfill July 2009 \\

\begin{center}
\boldmath
{\Large \bf Resummation of 
    $\tan\beta$-enhanced supersymmetric loop corrections\\[3mm] beyond the
    decoupling limit}\\
\unboldmath
\vspace*{1cm}
\renewcommand{\thefootnote}{\fnsymbol{footnote}}
Lars Hofer, Ulrich Nierste and Dominik Scherer\\
\vspace{10pt}
{\small
    {\em Institut f\"ur Theoretische Teilchenphysik\\ 
         Karlsruhe Institute of Technology, 
         Universit\"at Karlsruhe\\ 
         D-76128 Karlsruhe, Germany}}
 
\normalsize
\end{center}

\begin{abstract}
  We study the Minimal Supersymmetric Standard Model with Minimal
  Flavour Violation for the case of a large parameter $\tan\beta$ and
  arbitrary values of the supersymmetric mass parameters. We derive
  several resummation formulae for $\tan\beta$-enhanced loop
  corrections, which were previously only known in the limit of
  supersymmetric masses far above the electroweak scale. Studying first
  the renormalisation-scheme dependence of the resummation formula for
  the bottom Yukawa coupling, we clarify the use of the sbottom mixing
  angle in the supersymmetric loop factor $\Delta_b$. As a new feature,
  we find $\tan\beta$-enhanced loop-induced flavour-changing neutral
  current (FCNC) couplings of gluinos and neutralinos which in turn give
  rise to new effects in the renormalisation of the
  Cabibbo-Kobayashi-Maskawa matrix and in FCNC processes of $B$ mesons.
  For the chromomagnetic Wilson coefficient $C_8$, these gluino-squark
  loops can be of the same size as the known chargino-squark
  contribution. We discuss the phenomenological consequences for the
  mixing-induced CP asymmetry in $B_d\to\phi K_S$. We further quote
  formulae for $B_s\to \mu^+\mu^-$ and \bbms\ valid beyond the
  decoupling limit and find a new contribution affecting the phase of
  the \bbms\ amplitude. Our resummed $\tan\beta$-enhanced effects are
  cast into Feynman rules permitting an easy implementation in automatic
  calculations.
\end{abstract}

\tableofcontents

\section{Introduction}
The Minimal Supersymmetric Standard Model (MSSM) contains two Higgs doublets
$H_u$ and $H_d$, whose Yukawa couplings to quarks are given by 
\begin{equation}
   \mathcal{L}_y\; =\; - y_u^{ij}\, \bar{u}^i_R Q_j^T \epsilon H_u 
                    \; + \; 
                         y_d^{ij}\, \bar{d}^i_R Q_j^T \epsilon H_d 
                \; +\; \mbox{h.c.} 
   \label{eq:YukLag}
\end{equation}
Here $Q_j$, $u^i_R$ and $d^i_R$ are the usual left-handed doublet and
right-handed singlet quark fields, 
$\epsilon$ is the antisymmetric $2\times 2$ matrix with
$\epsilon_{12}=-\epsilon_{21}=1$, and $y_u$ and $y_d$ are Yukawa matrices with
generation indices $i,j=1,2,3$.  The holomorphy of the superpotential forbids
couplings of $H_u$ to $d_R$ and of $H_d$ to $u_R$, so that the Yukawa
Lagrangian of \eq{eq:YukLag} is that of a two-Higgs-doublet model (2HDM) of
type II.  The neutral components of the Higgs doublets acquire vacuum
expectation values (vevs) $v_u$ and $v_d$ with $v=\sqrt{v_u^2+v_d^2}\approx
174\,\gev$ leading to quark mass matrices $M_u=y_u v_u$ and $M_d=y_d
v_d$. Unitary rotations of the quark
fields in flavour space diagonalise these matrices, the resulting basis of
mass eigenstates is no more a weak basis (with manifest SU(2) symmetry) and
the familiar Cabibbo-Kobayashi-Maskawa (CKM) matrix appears in the couplings
of the W boson to the quark fields. As long as only the tree-level couplings
of $ \mathcal{L}_y$ are considered the Yukawa couplings are diagonal in
flavour space, $y_q^{ij}=y_{q_j} \delta_{ij}$ (no sum over $j$). At this point
no flavour-changing couplings of neutral Higgs bosons occur and the diagonal
Yukawa couplings are easily expressed in terms of quark masses $m_{q_j}$ and
$\tan\beta\equiv v_u/v_d$: $y_{d_j}=m_{d_j}/v_d = m_{d_j}/(v
\cos\beta)$ and $y_{u_j}=m_{u_j}/v_u = m_{u_j}/(v \sin\beta)$. If $\tan\beta $
is large, the Higgs couplings to down-type fermions can be enhanced to a level
which is detectable in present-day B physics experiments. In particular, for
$\tan\beta ={\cal O} (50)$ the bottom Yukawa coupling $y_b=y_{d_3}$ can be of
order 1.  A theoretical motivation of such large values of $\tan\beta$ is
given by bottom-top Yukawa unification, which occurs in SO(10) GUT models with
minimal Yukawa sector.  Phenomenologically, the anomalous magnetic moment of
the muon invites large values of $\tan\beta $ \cite{Bennett:2006fi}, but the
current situation is inconclusive in the light of recent experimental data on
the hadroproduction cross section measured by BaBar \cite{davier:08}.

Once soft supersymmetry-breaking terms are considered, the pattern
described above changes dramatically: As pointed out first by Banks,
one-loop diagrams induce an effective coupling of $H_u$ to $d^j_R$
\cite{Banks:1987iu}. Hall, Rattazzi and Sarid then discovered the
relevance of this loop contribution for large-$\tan\beta$ phenomenology
\cite{Hall:1993gn,Hempfling:1993kv,Carena:1994bv}.  If $M_{\rm SUSY}$,
the mass scale of the supersymmetry-breaking terms, is much larger than
the masses and vevs of the Higgs sector, we can integrate out the SUSY
particles. The resulting effective Lagrangian is that of a general 2HDM,
different from the type-II 2HDM which we encounter at tree-level. In the
Super-CKM basis for the quark and squark fields, in which
$y_d^{ij}=y_{d_i}\delta^{ij}$, the Yukawa couplings of down-type quarks
are given by the effective Lagrangian
\begin{equation}
   \mathcal{L}^{\rm eff}_{y,d} \; =\;  
     y_{d_i}\, \bar{d}^i_R Q_i^T  \epsilon H_d \; -\; 
  \widetilde{y}_d^{ij}\, \bar{d}^i_R Q_j^T H_u^* 
  \; +\; \mbox{h.c.}  \label{eq:EffLag}
\end{equation}
In this paper we restrict ourselves to the case that the soft
SUSY-breaking terms are flavour-diagonal in the Super-CKM basis.  As a
consequence, all gluino-squark-quark and neutralino-squark-quark
couplings in the MSSM Lagrangian are flavour-diagonal. Further the
chargino-squark-quark couplings come with the same CKM elements as the
corresponding couplings of W bosons or charged Higgs bosons to
(s)quarks. This scenario of naive Minimal Flavour Violation (naive
MFV) occurs if e.g.\ supersymmetry is broken at a low scale by a
flavour-blind mechanism leading to flavour-universal squark mass
matrices.  (A symmetry-based and RG-invariant definition of MFV has
been proposed in \cite{D'Ambrosio:2002ex}.  For a recent analysis see
Ref.~\cite{smith}.)  In our version of naive MFV, however, we slightly
go beyond flavour universality, as we allow the SUSY-breaking terms of
the third generation to be different from those of the first two
generations.  In this way we also include the cases of the
widely-studied CMSSM (see
e.g. Refs.~\cite{Heinemeyer:2008fb,Buchmueller:2007zk} for recent
studies) and mSUGRA
\cite{Nilles:1982ik,Chamseddine:1982jx,Barbieri:1982eh,Ohta:1982wn,
Hall:1983iz,Soni:1983rm} models, in which renormalisation-group (RG)
effects involving the large top and bottom Yukawa couplings destroy
the universal boundary condition imposed at the GUT scale. In such
models with high-scale flavour universality the RG also induces
flavour-violating gluino and neutralino couplings at the electroweak
scale, but their impact on FCNC transitions like \bbm\ and $b\to
s\gamma$ is small \cite{Baer:1997jq,Dudley:2008dp} and the naive MFV
pattern essentially stays intact. On the other hand, the universality
of the flavour-diagonal SUSY-breaking terms is badly broken at low
energies, e.g.\ the trilinear term of the third generation $A_t$
substantially differs from $A_u \simeq A_c$.  We emphasize that no
variant of the MFV assumption forbids flavour-diagonal CP-violating
phases \cite{Ellis:2007kb}. Such phases appear in $A_t$, the higgsino mass parameter
$\mu$, and the gaugino mass terms $M_i$, $i=1,2,3$, which we
consequently always treat as complex quantities throughout our
analysis.

\begin{nfigure}{t}
~\\

  \begin{picture}(200,90) (-200,-40)
    \SetWidth{0.5}
    \SetColor{Black}
    \ArrowArc(-110,0)(30,180,0)
    \DashArrowArcn(-110,0)(30,180,90){5}
    \DashArrowArcn(-110,0)(30,90,0){5}
    \ArrowLine(-170,0)(-140,0)
    \ArrowLine(-80,0)(-50,0)
    \DashLine(-110,30)(-110,60){5}
    \ArrowLine(115,0)(160,0)
    \ArrowLine(60,0)(105,0)
    \DashLine(110,5)(110,60){5}
    \Oval(110,0)(5,5)(0)
    \Line(106.5,-3.5)(113.5,3.5)
    \Line(106.5,3.5)(113.5,-3.5)
    \Text(65,-15)[lb]{\Black{$d_L^i$}}
    \Text(145,-15)[lb]{\Black{$d_R^i$}}
    \Text(113,50)[lb]{\Black{$H_u$}}
    \Text(-165,-15)[lb]{\Black{$d_L^i$}}
    \Text(-65,-15)[lb]{\Black{$d_R^i$}}
    \Text(-107,50)[lb]{\Black{$H_u$}}
    \Text(-112,-45)[lb]{\Black{$\tilde{g}$}}
    \Text(-148,20)[lb]{\Black{$\tilde{d}_L^i$}}
    \Text(-87,20)[lb]{\Black{$\tilde{d}_R^i$}}
    \Text(98,-18)[lb]{\Black{$-i\widetilde{y}_{d_i}$}}
    \LongArrow(-15,10)(15,10)
  \end{picture}

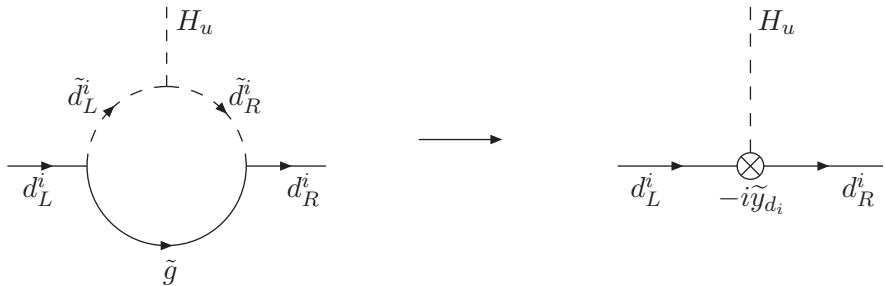
\captionof{figure}{Effective coupling of the down-type quarks to $H_u$}
\label{fig:EffCoup}
\end{nfigure}
The dominant contribution to the effective coupling $\widetilde{y}_d^{ij}$
stems from a gluino-squark loop and is depicted in \fig{fig:EffCoup}.
In naive MFV, the corresponding contribution to  $\widetilde{y}_d^{ij}$ is 
$\widetilde{y}_{d_i}^{\tilde{g}} \delta_{ij}$ with   
\begin{eqnarray}
   \widetilde{y}_{d_i}^{\tilde{g}} &=& y_{d_i} \cdot 
      \epsilon_i^{\tilde{g}}(\mu,m_{\tilde{d}_L^i},m_{\tilde{d}_R^i}),
   \hspace{1cm} \nn
\textrm{and}\qquad\qquad   
\epsilon_i^{\tilde{g}} (\mu,m_{\tilde{d}_L^i},m_{\tilde{d}_R^i}) &=& 
     -\frac{2\alpha_s}{3\pi}m_{\tilde{g}}\mu^*\,
     C_0(m_{\tilde{g}},m_{\tilde{d}^i_L},m_{\tilde{d}^i_R})
   \label{eq:NonHoloCoup} .
\end{eqnarray}
Here $m^2_{\tilde{d}^i_L}$ and $m^2_{\tilde{d}^i_R}$ are the mass terms for
the left-handed and right-handed down-squarks of the $i$-th generation,
respectively, $m_{\tilde g}$ is the gluino mass and the loop integral
$C_0$ is defined in Appendix A.  Accounting for similar contributions
from loops with charginos (still neglecting flavour mixing) or neutralinos 
we write
$\epsilon_i=\epsilon_i^{\tilde{g}}+ \epsilon^{\widetilde{\chi}^{\pm}}_i
+\epsilon^{\widetilde{\chi}^0}_i $.
Both terms in $\mathcal{L}^{\rm
  eff}_{y,d}$ of \eq{eq:EffLag} contribute to the masses of down-type quarks.
The ratio of the two contributions is
\begin{equation}
   \Delta_i\equiv
    \frac{\widetilde{y}_{d_i} v_u}{y_{d_i}v_d} = 
    \epsilon_i \cdot\tan\beta.
   \label{eq:RatioMasses}
\end{equation}
A large value of $\tan\beta$ can compensate for the loop factor
$\epsilon_i$ rendering $\Delta_i={\cal O}(1)$. The relation
between the Yukawa coupling $y_{d_i}$ and the physical quark mass $m_{d_i}$
is therefore modified substantially:
\begin{equation}
   y_{d_i}=\frac{m_{d_i}}{v_d(1+\Delta_i)}.
   \label{eq:YukMassRel}
\end{equation}
Several papers have studied the impact of $\Delta_i$ on Yukawa unification 
\cite{Hall:1993gn,Carena:1994bv}, neutral \cite{Carena:1998gk} and charged 
Higgs \cite{cgnw} phenomenology. 

Later Hamzaoui, Pospelov and Toharia have discovered that
$\widetilde{y}_d^{ij}$ has a profound impact on flavour physics: The
down-quark mass matrix $M_d$ computed from $\mathcal{L}^{\rm eff}_{y,d}$ will
be non-diagonal and conversely a non-diagonal Yukawa coupling $y_d^{ij}$
appears in the basis of mass eigenstates \cite{Hamzaoui:1998nu}.  The
resulting FCNC couplings of the non-standard neutral Higgs bosons $H^0$ and
$A^0$ are loop-suppressed but involve two powers of $\tan \beta$. Thus the new
FCNC couplings may compete in size with the flavour-diagonal tree-level
coupling which involves a single power of $\tan\beta$ and is of order 1 in the
case of the bottom quark.  Importantly, these effects are already highly
relevant in naive MFV, where only chargino-loops contribute to the off-diagonal
entries of $\widetilde{y}_d^{ij}$, which moreover involve the same small CKM
elements as the SM contribution. In our effective theory, the general 2HDM
with $\mathcal{L}^{\rm eff}_{y,d}$ in \eq{eq:EffLag}, FCNC processes proceed
through tree diagrams with $H^0$ or $A^0$ exchange.  Most spectacular effects
occur in $B_{d,s}\to \ell^+\ell^-$ decays, where a priori orders-of-magnitude
effects were possible even in the MSSM with naive MFV \cite{Babu:1999hn}. The dominant
Higgs-mediated contribution to ${\cal B} (B_{d,s}\to \ell^+\ell^-)$ is
proportional to six powers of $\tan \beta$ and ${\cal B} (B_{d,s}\to
\ell^+\ell^-)$ is more sensitive to the large-$\tan\beta$ region of the MSSM
than any other decay rate or cross section. A correlated analysis of ${\cal B}
(B_{d,s}\to \ell^+\ell^-)$ with the muon anomalous magnetic moment has been
performed in Ref.~\cite{ddn}. The presence of $\widetilde{y}_d^{ij}$ in
$\mathcal{L}^{\rm eff}_{y,d }$ further leads to a modification of the relation
between $y_d^{ij}$ and the CKM elements by $\tan \beta$-enhanced loop
corrections. This feature was studied in Ref.~\cite{Blazek:1995nv} in MFV well
before the discovery of the Higgs-mediated FCNC effects.\footnote{Recently,
  this finite CKM renormalisation has been extended to the case of non-minimal
  flavour violation \cite{cn}.}  As a consequence, the couplings of the
charged Higgs boson to down-type fermions are modified, with phenomenological
impact on $B^+\to \tau^+ \nu$ \cite{ip} and $B^+\to D \tau^+ \nu$
\cite{ntw,km}.

\bbm\ plays a special role: The superficially leading contribution from
diagrams with right-handed $b$-quark fields vanishes \cite{Hamzaoui:1998nu},
because the two diagrams with $H^0$ and $A^0$ exchange cancel each other.
Buras et al.\ have discovered that, despite of a suppression factor of
$m_s/m_b$, the analogous diagrams with one right-handed $s$-quark field can
sizably diminish \bbms\ \cite{Buras:2001mb}.  This effect is highly correlated
with ${\cal B} (B_s\to \ell^+\ell^-)$ and today's upper bound on ${\cal B}
(B_s\to \mu^+\mu^-)$ from the Tevatron experiments \cite{cdf,dzero} severely
limits the size of the Higgs-mediated contribution in \bbms\ 
\cite{Buras:2002wq}. In subsequent papers further contributions such as the
charged-Higgs box diagram to \bbm\ \cite{Isidori:2001fv} and contributions to
$\widetilde{y}_d^{ij}$ involving the electroweak gauge couplings were
considered \cite{bcrs,gjnt}. A complete list of all one-loop contributions to
$\widetilde{y}_d^{ij}$ for the case of universal SUSY-breaking terms taking
into account all possible CP phases can be found in Ref.~\cite{gjnt}.  The
absence of the superficially dominant contribution renders \bbm\ vulnerable to
other subleading corrections proportional to other small expansion parameters
such as $\cot\beta$, $v/M_{\rm SUSY}$ or the loop factor $1/(16\pi^2)$. Any of
these corrections could potentially spoil the cancellation and endanger the
correlation found in \cite{Buras:2001mb}. The recent symmetry-based analysis
of Ref.~\cite{gjnt} has revealed that all these subleading corrections are
small and the correlation found in Ref.~\cite{Buras:2002wq} essentially stays
intact.  An important ingredient of this study are contributions to \bbm\ 
stemming from loop corrections to the Higgs potential. At this point the
appropriate definition of $\tan\beta$, which is ill-defined in a general 2HDM,
had to be clarified.  Loop corrections to \bbm\ from the Higgs potential were
also calculated in Ref.~\cite{brs}. In view of the findings of
Refs.~\cite{gjnt,brs} we neglect all radiative contributions to Higgs
self-couplings and work with the tree-level Higgs potential of the MSSM. The
latter is CP-conserving; we can work with the usual Higgs mass eigenstates
with definite CP quantum numbers (i.e.\ the CP-odd $A^0$ and the CP-even
$h^0,H^0$) and all CP-violation discussed in this paper enters through the
(loop-corrected) Yukawa sector.  

The last three paragraphs have addressed Higgs couplings to right-handed
down-type quarks which involve a factor of $\tan \beta$ at tree-level.  A
different type of $\tan\beta$-enhanced corrections occurs in Higgs couplings
of the right-handed top-quark field, which are suppressed by a factor of
$\cot\beta$ at tree level. A prominent example is the $\ov t_R s_L H^+$
coupling entering the charged-Higgs loop in $b\to s\gamma$. Supersymmetric
vertex corrections lift the $\cot\beta$ suppression and the one-loop
correction competes with the tree-level coupling
\cite{Carena:2000uj,Degrassi:2000qf}. In the decoupling limit also these
effects can be easily described by an effective Lagrangian $\mathcal{L}^{\rm
  eff}_{y,u}$, which in addition to the first term in \eq{eq:YukLag} contains
an effective loop-induced coupling $\widetilde{y}_u^{ij}$ involving $H_u^*$.

The appearance of the $\tan\beta$-enhanced supersymmetric loop
correction $\Delta_i$ in the denominator of $y_i$ in
\eq{eq:YukMassRel} signals the resummation of this correction to all
orders in perturbation theory. As a drawback, the
effective-field-theory method is only valid for $M_{\rm SUSY} \gg v,
M_{A^0}, M_{H^0}, M_{H^+}$. This is unsatisfactory, since in
supersymmetry one naturally expects $M_{\rm SUSY} \sim v$, especially
if the electroweak symmetry is broken radiatively. One needs an
unnatural fine-tuning in the Higgs potential to achieve $M_{\rm SUSY}
\gg M_{A^0}, M_{H^0}, M_{H^+}$ \cite{brs}. After all the
widely-studied scenarios with neutralino LSP involve several
supersymmetric particles with masses around and below $v$. Of course,
several authors have discovered $\tan\beta$-enhanced loop corrections
within diagrammatical one-loop calculations in the MSSM
\cite{gjs,Huang}. Yet only four papers have studied
$\tan\beta$-enhanced corrections with their subsequent resummation
beyond the $M_{\rm SUSY}\gg v$ limit: In Ref.~\cite{cgnw} the
$\tan\beta$-enhanced diagrams contributing to $\Delta_i$ have been
identified to all orders and have been explicitly resummed. The result
of Ref.~\cite{cgnw} mimics the form of \eq{eq:YukMassRel}, but
$\Delta_i$ involves squark mass eigenstates and its validity does not
assume any hierarchy between $v$ and $M_{\rm SUSY}$. In
Ref.~\cite{mmns} the method of Ref.~\cite{cgnw} has been applied to
the lepton sector in an analysis of the muon anomalous magnetic
moment. The authors of Ref.~\cite{bcrs} have calculated Higgs-mediated
FCNC processes to one-loop order for arbitrary $M_{\rm SUSY}$, but
relied on the effective-field-theory formalism for the all-order
resummation. In Ref.~\cite{Ellis:2007kb} the $\tan\beta$-enhanced
corrections to the Yukawa sector have been incorporated in an
effective-potential approach, with a proper consideration of all CP
phases of the MSSM. The results of Refs.~\cite{bcrs,Ellis:2007kb}
permit the resummation of the flavour-changing $\tan\beta$-enhanced
corrections through an iterative procedure, which converges if the
magnitude of these resummed corrections are numerically smaller than
1. We present analytical resummation formulae in this paper
corresponding to the limits to which the iterative method converges.

\begin{nfigure}{t}
~\\

  \begin{picture}(200,90) (-210,-40)
    \SetWidth{0.5}
    \SetColor{Black}
    \ArrowArc(-145,0)(25,180,0)
    \DashArrowArcn(-145,0)(25,180,90){5}
    \DashArrowArcn(-145,0)(25,90,0){5}
    \ArrowLine(-200,0)(-170,0)
    \ArrowLine(-120,0)(-90,0)
    \DashLine(-145,25)(-145,55){5}
    \Text(-195,-15)[lb]{\Black{$d_L^i$}}
    \Text(-105,-15)[lb]{\Black{$d_R^i$}}
    \Text(-142,50)[lb]{\Black{$H_u$}}
    \Text(-147,-40)[lb]{\Black{$\tilde{g}$}}
    \Text(-180,17)[lb]{\Black{$\tilde{d}_L^i$}}
    \Text(-125,17)[lb]{\Black{$\tilde{d}_R^i$}}
    \Text(-78,-3)[lb]{\Black{\bf{\Large{+}}}}
    \ArrowArc(-5,0)(25,180,0)
     \DashCArc(-5,0)(25,0,180){5}
    \ArrowLine(-60,0)(-30,0)
    \ArrowLine(20,0)(50,0)
    \DashLine(-24,19)(-46,41){5}
    \DashLine(14,19)(36,41){5}
    \DashLine(-5,25)(-5,55){5}
    \Text(-55,-15)[lb]{\Black{$d_L^i$}}
    \Text(35,-15)[lb]{\Black{$d_R^i$}}
    \Text(-2,50)[lb]{\Black{$H_u$}}
    \Text(-60,35)[lb]{\Black{$H_u$}}
    \Text(37,35)[lb]{\Black{$H_u$}}
    \Text(-7,-40)[lb]{\Black{$\tilde{g}$}}
    \Text(-42,7)[lb]{\Black{$\tilde{d}_L^i$}}
    \Text(20,7)[lb]{\Black{$\tilde{d}_R^i$}}
    \Text(-24,26)[lb]{\Black{$\tilde{d}_R^i$}}
    \Text(3,26)[lb]{\Black{$\tilde{d}_L^i$}}
    \Text(62,-3)[lb]{\Black{\bf{\Large{+}}}}
    \ArrowArc(135,0)(25,180,0)
    \DashCArc(135,0)(25,0,180){5}
    \ArrowLine(80,0)(110,0)
    \ArrowLine(160,0)(190,0)
    \DashLine(113,12.5)(87,27.5){5}
    \DashLine(122.5,22)(107.5,48){5}
    \DashLine(135,25)(135,55){5}
    \DashLine(157,12.5)(183,27.5){5}
    \DashLine(147.5,22)(162.5,48){5}
    \Text(85,-15)[lb]{\Black{$d_L^i$}}
    \Text(175,-15)[lb]{\Black{$d_R^i$}}
    \Text(138,50)[lb]{\Black{$H_u$}}
    \Text(110,45)[lb]{\Black{$H_u$}}    
    \Text(163,45)[lb]{\Black{$H_u$}}
    \Text(78,30)[lb]{\Black{$H_u$}}
    \Text(183,30)[lb]{\Black{$H_u$}}    
    \Text(133,-40)[lb]{\Black{$\tilde{g}$}}
    \Text(202,-3)[lb]{\Black{\bf{\Large{+}}}}
    \Text(220,-3)[lb]{\Black{\bf{\Large{...}}}}
  \end{picture}

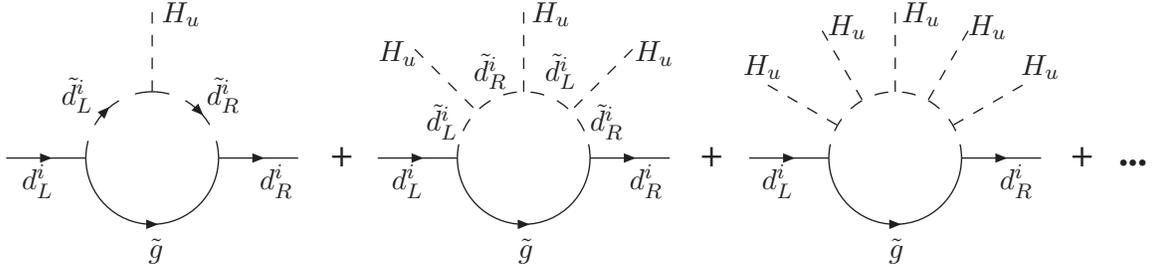
\captionof{figure}{Series of `hedgehog diagrams' contributing to $m_{d_i}$}
\label{fig:Hedgehog}
\end{nfigure}
It is illustrative to consider the extension of the effective-field-theory
formalism to subleading powers in $v^2/M_{\rm SUSY}^2$: To this end we must
add higher-dimensional couplings to $\mathcal{L}^{\rm eff}_{y}$ involving more
$H_u$ fields.  The gluino contributions to these new effective couplings are
shown in \fig{fig:Hedgehog}. Interestingly, in this simple case one can sum
the contributions of these `hedgehog diagrams' to $m_{d_i}$ to all orders in
$v^2/M_{\rm SUSY}^2$: The result has again the form of \eq{eq:YukMassRel} with
$\epsilon_i^{\tilde{g}}$ of \eq{eq:RatioMasses} replaced as
\begin{equation}
   \epsilon_i^{\tilde{g}}(\mu,m_{\tilde{d}_L^i},m_{\tilde{d}_R^i}) 
   \to \epsilon_i^{\tilde{g}}(\mu,m_{\tilde{d}_1^i},m_{\tilde{d}_2^i}), 
\label{eq:resum}
\end{equation}
where $m_{\tilde{d}_{1,2}^i}$ denote the physical squark masses, i.e.\ the
eigenvalues of the squark mass matrix. Using this expression in
\eqsand{eq:RatioMasses}{eq:YukMassRel} reproduces the result of the
diagrammatic resummation of Ref.~\cite{cgnw}. The information encoded
in the diagrams of \fig{fig:Hedgehog} is also contained in the
one-loop effective functional of Ref.~\cite{Ellis:2007kb}. 

In this paper we derive formulae for the resummation of
$\tan\beta$-enhanced corrections which are valid for any value of
$M_{\rm SUSY}$. As in any analysis of radiative corrections this
requires the full control over the renormalisation scheme of the
parameters in the MSSM Lagrangian. This can be achieved with the
diagrammatic method of Ref.~\cite{cgnw}, but is very difficult to
achieve with the effective-field-theory formalism, even if one succeeds
to resum the series in $v^2/M_{\rm SUSY}^2$ as in \eq{eq:resum}. The
origin of this difficulty is readily understood: While resummation
formulae derived from $\mathcal{L}^{\rm eff}_{y}$ correspond to a
decoupling scheme for the MSSM parameters, any two of such schemes may
differ by terms of order $v/M_{\rm SUSY}$ and the corresponding
resummation formulae look different.  The plan of the paper is as
follows: In Sect.~\ref{sect:2} we first recall the diagrammatic
resummation method and then address the open issues of the case without
flavour mixing.  In particular we clarify the renormalisation scheme of
the sbottom mixing angle and derive analytic expressions for
$\Delta_b\equiv\Delta_3$ for three different schemes.  In
Sect.~\ref{flavour_mixing} we resum the $\tan\beta$-enhanced loop
effects in FCNC processes. Sect.~\ref{fcnc} is devoted to an analysis
of $\tan\beta$-enhanced corrections to FCNC processes in B physics.
Sect~\ref{numerics} contains a numerical study of the Wilson
coefficients $C_7$ and $C_8$ and an analysis of novel effects in $B\to
\phi K_S$. Finally we conclude.\\

\boldmath
\section{Diagrammatic resummation: the flavour-conserving case}\label{sect:2} 
\unboldmath%
We use the conventions of the SUSY Les Houches
Accord (SLHA) \cite{Skands:2003cj} for the MSSM parameters. Several of these
parameters carry complex phases, but only certain phase differences are
physical, CP-violating quantities. We choose a phase convention in which the
gluino mass parameter $M_3$ is real and positive, so that $M_3=m_{\tilde g}$.
The phases entering the left-right mixing of squarks are unspecified by the
SLHA and are defined in Appendix~\ref{conventions}, where also our conventions
for the loop integrals can be found. We always work in the Super-CKM basis, in
which the Yukawa matrices are diagonal in flavour space. For definiteness we
consider the quark sector only and in our discussion of flavour-diagonal
effects we usually quote the results for the $b$ quark. The expressions
generalise to the case of the $\tau$ lepton in a straightforward way, by 
dropping the gluino contributions, replacing squarks by sleptons and changing
the hypercharges in the couplings appropriately.

\subsection{The method}\label{flavour_cons}
There are two potential sources of $\tan\beta$-enhanced corrections, 
\begin{itemize}
\item[i)]  the (renormalised) MSSM Lagrangian ${\cal L}$ and 
\item[ii)] the transition matrix element ${\cal M}$ from which the process 
           of interest is calculated. 
\end{itemize}
We first identify the enhanced corrections at one-loop order and turn to
higher orders (and the resummation) afterwards.  To address point i) we
decompose ${\cal L}$ in the usual way as ${\cal L}={\cal L}_{\rm ren}+{\cal
  L}_{\rm ct}$, where ${\cal L}_{\rm ren}$ is obtained from ${\cal L}$ by
replacing bare quantities by renormalised ones and ${\cal L}_{\rm ct}$
contains the counterterms. Loop effects only reside in ${\cal L}_{\rm ct}$ and
the quark mass counterterm $\delta m_b$ is a source of $\tan\beta$-enhanced
corrections.
We write $m_b$ for the renormalised mass, so that the bare mass reads 
$m_b^{(0)}=m_b + \delta m_b$. The mass term in ${\cal L}$ is 
\begin{equation} 
\begin{split}\label{eq:lm}  
  {\cal L}_m =&\; -\; m_b^{(0)} \ov b_R b_L \; -\;  m_b^{(0)*} \ov b_L b_R
             \; = \; - \;
         m_b\, \ov b b \;  - \; 
    \delta m_b \, \ov b_R b_L \; - \; \delta m_b^* \, \ov b_L b_R .
\end{split}
\end{equation}
Here we have taken into account that $\delta m_b$ must be complex to render
$m_b$ real if the loops canceled by $\delta m_b$ involve complex parameters.
We further decompose the self-energy $\Sigma_b(p)$ as
\begin{equation} 
\begin{split}  
  \Sigma_b(p) = \slashed{p} \left[ \Sigma_b^{LL}(p^2)P_L + 
                                 \Sigma_b^{RR}(p^2)P_R \right]& 
   \; + \; \Sigma_b^{RL}(p^2)P_L \, +\,  \Sigma_b^{LR}(p^2)P_R \\
  & \qquad \qquad \textrm{with}\hspace{0.5cm}
  \Sigma_b^{LR}(p^2) =\left(\Sigma_b^{RL}(p^2)\right)^*,
\end{split}
\end{equation} 
where $P_{L,R}=(1\mp\gamma_5)/2$ and $p$ is the external momentum.  If the
mass is renormalised on-shell, i.e.\ if $m_b$ coincides with the pole of the
propagator, the counterterm reads
\begin{equation}
\delta m_b = - \frac{m_b}{2} 
    \lt[ \Sigma^{LL}_b(m_b^2) + \Sigma^{RR}_b(m_b^2) \rt] 
             - \Sigma^{RL}_b(m_b^2) .
\label{mbos}
\end{equation}
The second term $\Sigma^{RL}_b(m_b^2)$ contains pieces proportional to 
$y_b v \sin\beta$ and is therefore $\tan\beta$-enhanced compared to the 
tree-level term $m_b=y_b v \cos \beta$. 
\begin{nfigure}{t}
  \begin{picture}(200,80) (-210,-35)
    \SetWidth{0.5}
    \SetColor{Black}
    \ArrowArc(-130,0)(20,180,0)
    \DashArrowArcn(-130,0)(20,180,0){5}
    \ArrowLine(-180,0)(-150,0)
    \ArrowLine(-110,0)(-80,0)
    \Text(-170,-15)[lb]{\Black{$b_L$}}
    \Text(-100,-15)[lb]{\Black{$b_R$}}
    \Text(-132,-35)[lb]{\Black{$\tilde{g}$}}
    \Text(-139,22)[lb]{\Black{$\tilde{b}_{1,2}$}}

    \ArrowLine(-50,0)(-20,0)
    \ArrowArc(0,0)(20,180,0)
    \DashArrowArcn(0,0)(20,180,0){5}
    \ArrowLine(20,0)(50,0)
    \Text(-40,-15)[lb]{\Black{$b_L$}}
    \Text(30,-15)[lb]{\Black{$b_R$}}
    \Text(-5,20)[lb]{\Black{$\tilde{t}_{1,2}$}}
    \Text(-5,-38)[lb]{\Black{$\widetilde{\chi}_{1,2}^-$}}

    \ArrowLine(80,0)(110,0)
    \ArrowArc(130,0)(20,180,0)
    \DashArrowArcn(130,0)(20,180,0){5}
    \ArrowLine(150,0)(180,0)
    \Text(90,-15)[lb]{\Black{$b_L$}}
    \Text(160,-15)[lb]{\Black{$b_R$}}
    \Text(125,20)[lb]{\Black{$\tilde{b}_{1,2}$}}
    \Text(122,-38)[lb]{\Black{$\widetilde{\chi}_{1..4}^0$}}
  \end{picture}

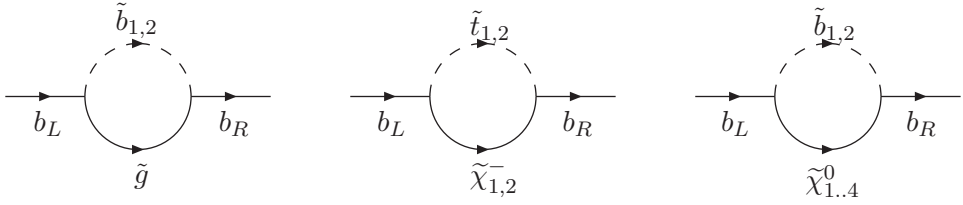
\captionof{figure}{$\tan\beta$-enhanced self-energy diagrams with 
(from left to right) gluinos, charginos and neutralinos.}
\label{fig:SelfDia}
\end{nfigure}
These contributions are depicted in \fig{fig:SelfDia} and read:
\begin{eqnarray}
      \Sigma^{RL}_{b}&=& m_b \Delta_b  
\qquad\qquad\qquad \mbox{with}\qquad 
      \Delta_b \; =\;  \Delta_b^{\tilde{g}} + 
         \Delta_b^{\widetilde{\chi}^{\pm}} + 
         \Delta_b^{\widetilde{\chi}^0} 
\label{eq:Self0}
\end{eqnarray}
and
\begin{eqnarray}
\Delta_{b}^{\tilde{g}} &=& 
     \frac{\alpha_s}{3\pi} \frac{m_{\tilde{g}}}{m_b} 
     \sin2{\tilde{\theta}}_b \, e^{-i{\tilde{\phi}}_b} \cdot 
     \left[ B_0(m_{\tilde{g}},m_{\tilde{b}_1}) - 
            B_0(m_{\tilde{g}},m_{\tilde{b}_2}) \right], 
     \label{eq:Self1}\\
\Delta_{b}^{\widetilde{\chi}^{\pm}} &=& 
  -\, \frac{g^2}{16\pi^2}\frac{1}{\cos\beta} \sum\limits_{m=1}^2 \left\{ 
        \frac{m_{\widetilde{\chi}^{\pm}_m}}{2\sqrt{2}M_W} \, \frac{y_t}{g}
        \, \widetilde{U}^{*}_{m2} \widetilde{V}^{*}_{m2} \, 
        \sin2\tilde{\theta}_t e^{i\tilde{\phi}_t} 
    \right.\nonumber\\
&& \qquad\qquad\qquad  \qquad\qquad  
   \cdot \left[ B_0(m_{\widetilde{\chi}^{\pm}_m},m_{\tilde{t}_1}) -
          B_0(m_{\widetilde{\chi}^{\pm}_m},m_{\tilde{t}_2}) \right] 
    \nn 
    && \left. 
  - \, \frac{m_{\widetilde{\chi}^{\pm}_m}}{\sqrt{2}M_W} \, 
    \widetilde{U}^{*}_{m2} \widetilde{V}^{*}_{m1}
    \left[ \cos^2\tilde{\theta}_t\, 
           B_0(m_{\widetilde{\chi}^{\pm}_m},m_{\tilde{t}_1})
       +   \sin^2\tilde{\theta}_t\, 
           B_0(m_{\widetilde{\chi}^{\pm}_m},m_{\tilde{t}_2}) 
           \right] \right\}, \label{eq:Self2}\\
\Delta_{b}^{\widetilde{\chi}^0} &=& 
   \frac{g^2}{16\pi^2}\frac{1}{\cos\beta}\sum\limits_{m=1}^4 
   \frac{m_{\widetilde{\chi}^{0}_m}}{2M_W}\widetilde{N}_{m2}^{*}
   \widetilde{N}_{m3}^{*} \nn
&& \qquad \qquad\qquad  \qquad\qquad 
  \cdot    \left[ \cos^2\tilde{\theta}_b\,
           B_0(m_{\widetilde{\chi}^0_m},m_{\tilde{b}_1}) + 
          \sin^2\tilde{\theta}_b\, 
           B_0(m_{\widetilde{\chi}^0_m},m_{\tilde{b}_2}) 
    \right] \label{eq:Self3}.
\end{eqnarray}
In (\ref{eq:Self3}) we have neglected some numerically small
contributions: First, a term proportional to $g'^2$ stemming from the
bino component of the neutralinos is omitted. Second, a numerically
small term proportional to $g^2$ (which moreover is suppressed by
$(v/\msusy)^2$ for large $\msusy$ and is therefore absent in the
effective Lagrangian of \eq{eq:EffLag}) is neglected.  Clearly, we have
also discarded terms suppressed by $m_b^2/M_{\rm SUSY}^2$; in particular
$\Sigma^{RL}_{b}$ is evaluated for $p^2=0$.  Whereas in the effective-theory approach the $\tan\beta$-enhancement was easily recognisable by the coupling to $H_u$, in the diagrammatic treatment it is hidden in the
elements of the mixing matrices. Using the analytic expressions for
these matrices listed in Appendix \ref{conventions}, i.e. identities
like \eq{eq:MixingAngle} and \eqsto{eq:ChMixMat1}{eq:ChMixMat2}, we can
derive formulae for the gluino- and chargino-contributions in which the
$\tan\beta$-enhancement becomes explicit. Writing
\begin{eqnarray}
   \Delta_{b}^{K}&=& \epsilon_b^{K}\tan\beta\hspace{0.5cm}
  \qquad \mbox{for }\; K=\tilde{g}, 
   \widetilde{\chi}^{\pm},  \widetilde{\chi}^{0} 
\qquad\qquad \mbox{and}\qquad 
 \epsilon_b \;=\; \epsilon_b^{\tilde g} +   
                  \epsilon_b^{\widetilde{\chi}^{\pm} } + 
                  \epsilon_b^{\widetilde{\chi}^0 } 
\label{eq:dbk}
\end{eqnarray}
we find 
\begin{eqnarray}
\epsilon^{\tilde{g}}_b&=&-\frac{2\alpha_s}{3\pi} m_{\tilde{g}}\mu^* C_0(m_{\tilde{g}},m_{\tilde{b}_1},m_{\tilde{b}_2}),\label{eq:Self4}\\
   \epsilon^{\widetilde{\chi}^{\pm}}_b &=&
   -\frac{y_t^2}{16\pi^2}A_t^*\mu^*\left(
   D_2-|M_2|^2 D_0\right)+\frac{g^2}{16\pi^2}\mu^*M_2^*
   \left( D_2-m_{\tilde{t}_R}^2 D_0\right)\label{eq:Self5},
\end{eqnarray}
where $D_{0,2} = D_{0,2}(m_{\widetilde{\chi}^{\pm}_1},
m_{\widetilde{\chi}^{\pm}_2},m_{\tilde{t}_1},m_{\tilde{t}_2})$. (The $\tan
\beta$-enhancement of $\Delta_{b}^{\widetilde{\chi}^0}$ is already manifest
in \eq{eq:Self3} through the factor $1/\cos\beta \simeq \tan\beta$.) 
Formulae analogous to \eqsto{eq:Self1}{eq:Self5} are also valid for the
corresponding self-energies of the d- and the s-quark with the stop and
sbottom masses appropriately replaced by the corresponding squark masses of
the first or second generation. \eqsto{eq:Self1}{eq:Self5} generalise
the well-known expressions of Ref.~\cite{Pierce:1996zz} to the case of 
complex MSSM parameters.

Different renormalisation schemes correspond to different choices of ${\cal
  L}_{\rm ct}$, hence the analytic form of the $\tan\beta$-enhanced
corrections depends on the chosen scheme.  If we want to use a numerical value
for $m_b$ determined from low-energy data, we must apply an on-shell
subtraction to the supersymmetric loops as in \eq{mbos} (which is the
appropriate ``decoupling scheme'').  To leading order in $\tan\beta$ this
means
\begin{equation}
   \delta m_b \, =\, -\Sigma^{RL}_b \, =\,  - m_b \epsilon_b \tan\beta.
   \label{eq:MassCT}
\end{equation}
At this point we recall that the loops constituting $\epsilon_b$ are finite,
just as all other $\tan\beta$-enhanced loops appearing in this paper.
Therefore all counterterms and all bare quantities discussed are finite as
well.  We write the bare Yukawa couplings as $y_b^{(0)}= y_b + \delta y_b $,
where $y_b$ is the renormalised coupling and $ \delta y_b $ is the
counterterm.  The choice of $ \delta m_b$ fixes $ \delta y_b $ through
\begin{equation}
  \delta y_b=\frac{\delta m_b}{v_d}=-y_b\epsilon_b\tan\beta .
  \label{eq:YukCT}
\end{equation}
The supersymmetric loop effects encoded in $\epsilon_b$ enter physical
observables only through $\delta y_b$. Choosing e.g.\ a minimal subtraction
for $\delta m_b$ would remove the $\tan\beta$-enhanced term from \eq{eq:YukCT}
and there would be nothing to resum. However, in this scheme the input value
for $m_b$ is obtained from the measured bottom mass by adding $ m_b \epsilon_b
\tan\beta$. Thus the inferred value of $y_b=m_b/v_d$ will implicitly contain
the $\tan\beta$-enhanced corrections, so that physical observables are
scheme-independent \cite{cgnw}. In a practical application one must also
address the renormalisation from ordinary QCD corrections. Whenever we refer 
to the $\ov{\rm MS}$ mass $m_b$ we imply that the $\ov{\rm MS}$ prescription 
is applied to the quark-gluon loop only, while we always subtract the 
supersymmetric loops on-shell. 

Now, are there other sources of $\tan\beta$-enhanced one-loop
corrections in ${\cal L}_{\rm ct}$? There are renormalisation schemes
proposed in the literature in which the counterterm to $\tan\beta$ is
proportional to $\tan^2\beta$, so that \eq{eq:YukCT} would receive an
additional contribution. This feature is obviously absent for the
commonly used definition of $\tan\beta$ in the $\ov{\rm DR}$ scheme.
Finally the one-loop renormalisation also involves wave-function
counterterms.  Those of the quark fields are not $\tan\beta$-enhanced
and the wave-function counterterms of the Higgs fields drop out if the
Higgses solely occur in internal lines of the diagrams.  (These
counterterms nevertheless play a role in schemes in which the
counterterm $\delta \tan\beta$ is derived from wave-function
counterterms and counterterms to the vevs.  This subtlety is absent for
the $\ov{\rm DR}$-defined $\delta \tan\beta$.) The issue of $\tan\beta$
renormalisation is thoroughly analysed in
Refs.~\cite{Gamberini:1989jw,Yamada:2001ck,fs} and was recently studied
for quark flavour physics in the context of the effective-field-theory
method \cite{brs,gjnt}. In our diagrammatic approach, where the issue is
somewhat simpler, the topic of $\tan\beta$ renormalisation is briefly
discussed in Ref.~\cite{mmns} in an application to the muon anomalous
magnetic moment. In conclusion, the only source of $\tan\beta$-enhanced
corrections in ${\cal L}_{\rm ct}$ is $ \delta y_b$ of \eq{eq:YukCT}
unless an inappropriate definition of $\tan\beta$ is adopted.

Next we turn to the second point mentioned at the beginning of this
section.  In order to identify $\tan\beta$-enhanced corrections to a
given transition matrix element ${\cal M}$ we must distinguish two
cases: In the first case the leading-order contribution to ${\cal M}$
has no suppression factor of $\cot\beta$ in any coupling.  Examples for
such unsuppressed couplings are those of $A^0$ and $H^0$ to down-type
quarks, the $H^+$ coupling to right-handed down-type quarks or any gauge
coupling. In this situation $\cal M$ can only have a
$\tan\beta$-enhanced correction if the loop integral involves at least
one inverse power of $m_b$, which combines with $y_b \propto m_b \tan
\beta$ to a factor of $\tan\beta$.  The presence of such inverse powers
of $m_b$ is related to the infrared behaviour of ${\cal M}$ for $m_b\to
0$. This behaviour can be studied by matching ${\cal M}$ onto an effective
matrix element ${\cal M}_{\rm eff}$ which is obtained from ${\cal M}$ by
contracting all lines of particles heavier than $m_b$ to a point
\cite{cgnw}. This analysis should not be confused with the
effective-field-theory method described in the Introduction: Here only
$M_{\rm SUSY},v, M_{A^0}, M_{H^0}, M_{H^+} \gg m_b$ is assumed, with no
assumption on the hierarchy between $M_{\rm SUSY}$ and $v$. The result
of Ref.~\cite{cgnw} is that no such $\tan\beta$-enhanced correction from
genuine multi-loop diagrams occur in the first case. The second case
deals with matrix elements ${\cal M}$ with an explicit
$\cot\beta$-suppressed coupling (such as the $h^0$ coupling to down-type
quarks or the $H^+$ coupling to left-handed down-type quarks) in the
leading order. Here the situation is different, but trivial: An explicit
one-loop vertex correction lifts the suppression and this
$\tan\beta$-enhanced correction does not replicate itself in higher
orders \cite{Carena:2000uj,Degrassi:2000qf}.
  
We now discuss higher orders of the perturbative expansion and the
resummation: While no genuine multi-loop diagrams give enhanced
corrections, there are one-loop diagrams involving lower-order
counterterms $\delta y_b$.  We make the $y_b$-dependence of the
self-energy explicit by writing $\Sigma^{RL}_{b} (y_b)$.  The Yukawa
coupling $y_b$ enters $\Sigma^{RL}_{b} (y_b)$ either directly via the
quark-squark-higgsino-vertex or indirectly via the sbottom mixing angle.
Now, let us consider such self-energy diagrams in which one or more of
the couplings $y_b$ are replaced by the counterterm $\delta y_b$.  The
mass counterterm $\delta m_b$ reads
\begin{equation}
   \delta m_b = v_d\delta y_b=-\Sigma_{b}^{RL}(y_b+\delta y_b).
   \label{eq:ResEq}
\end{equation}
to all orders of the perturbative expansion and to leading order in
$\tan\beta$. Let us denote the $n$-th order contribution to $\delta y_b$
by $\delta y_b^{(n)}$. We can solve \eq{eq:ResEq} recursively, by
  expressing $\delta y_b^{(n)}$ in terms of $\delta y_b^{(n-1)}$. 
Effectively $\delta y_b^{(n)}$ is simply computed from the one-loop 
diagrams contributing to $\Sigma_{b}^{RL}$ including all 
possible substitutions of $y_b$ by $\delta y_b^{(k)}$, $k=1,\ldots n-1$.  
Adapting \eq{eq:Self0} and \eqsto{eq:dbk}{eq:Self5} to account for 
the desired higher-order terms we write 
\begin{eqnarray}
   \Sigma_{b}^{RL} &=& m_b^{(0)} \Delta_b \; = \;
                     y_b^{(0)} v \epsilon_b \sin\beta 
   \label{eq:sblr} .
\end{eqnarray}
Whenever $ \Sigma_{b}^{RL}$ is linear in $ y_b^{(0)}$, that is if
$\epsilon_b$ does not depend on $ y_b^{(0)} $, one can easily determine
$\delta y_b$ to all orders: Noting that $y_b=m_b/v_d$ the one-loop
result of \eq{eq:YukCT} is replaced by 
\begin{equation}
\delta y_b = - \frac{m_b}{v_d} 
    \lt[ \epsilon_b\tan\beta - (\epsilon_b\tan\beta )^2 +
    (\epsilon_b\tan\beta )^3 - \ldots \rt] 
           = - \frac{m_b}{v_d} 
    \frac{\epsilon_b\tan\beta}{1+\epsilon_b\tan\beta} .
\label{eq:naivesum}
\end{equation}   
If we discard the neutralino contribution and take $ \epsilon_b^{\tilde
  g}$ and $\epsilon_b^{\widetilde{\chi}^{\pm} }$ from
\eqsand{eq:Self4}{eq:Self5}, we indeed find $\epsilon_b$ independent of
$y_b$. There is a shortcut to \eq{eq:naivesum}: Adding $m_b=y_b v_d$ to
both sides of \eq{eq:ResEq} gives
\begin{equation}
v_d y_b^{(0)} = m_b -  y_b^{(0)} v_d \epsilon_b \tan\beta  
\label{eq:shortcut}
\end{equation}
which is easily solved for $y_b^{(0)}$ resulting in the resummation
formula of Ref.~\cite{cgnw}: 
\begin{equation}
  y_b^{(0)} = \frac{m_b}{v_d(1+\epsilon_b \tan\beta)}.\label{eq:ResForm0}
\end{equation}

The linearity of $\epsilon_b^{\tilde
  g}+\epsilon_b^{\widetilde{\chi}^{\pm}}$ in $y_b$ beyond the decoupling
limit appears to contradict the discussion in the Introduction,
since the hedgehog diagrams of \fig{fig:Hedgehog} contain any odd power
of $y_b$. However, these additional factors of $y_b$ are implicitly
contained in the sbottom mass eigenstates $m_{\tilde{b}_{1,2}}$. From
this observation it becomes clear that for the correct resummation of
the $\tan\beta$-enhanced corrections one must clearly state the
renormalisation scheme for the supersymmetric parameters.
\eq{eq:ResForm0} implies an on-shell scheme for the sbottom masses
meaning here that $m_{\tilde{b}_{1,2}}$ are used as inputs.
By contrast, many supersymmetric analyses use the diagonal elements of
the mass matrix, $m_{\tilde{b}_{L,R}}$ and the $\mu$ parameter (entering
the off-diagonal elements) as inputs. In this scheme $y_b$ enters the
problem explicitly via the mass matrix and \eq{eq:ResForm0} is not
correct. Similarly, \eq{eq:ResForm0} must also be modified if the
sbottom mixing angle $\tilde{\theta}_b$ and the mixing phase
$\tilde{\phi}_b$ are used as input parameters. These parameters are the
natural choice for applications to collider physics, especially once the
bottom squarks are discovered and their properties are to be studied. It
is therefore of utmost importance to control the definition of
$\tilde{\theta}_b$, in particular if constraints from low-energy data
shall be combined with collider physics. We analyse this point in
Sect.~\ref{sect:smr}. 

In summary, whenever ${\cal M}$ does not suffer from $\cot \beta$-suppression in the leading order, all $\tan\beta$-enhanced corrections stem from $\delta y_b$. The dominant contributions from gluino and chargino loops can be resummed to all orders at the Lagrangian level, if an adequate scheme for the sbottom mass parameters is adopted. We stress that the resummed terms are local, so that one can reproduce the resummed effects from an effective Lagrangian. The effective $\ov b_L b_R H^{0}$, $\ov b_L b_R A^{0}$ and $\ov t_L b_R H^{+}$ couplings are simply obtained by replacing the tree-level Yukawa coupling with $ y_b^{(0)}$ in \eq{eq:ResForm0}. That is, the description of these
couplings by an effective Lagrangian \emph{does not}\ require any assumption on the size of $M_{\rm SUSY}$: E.g.\ the use of \eq{eq:ResForm0} also correctly resums the $\tan\beta$-enhanced corrections in high-energy collider processes, even if the momenta of the particles involved are of the order of $M_{\rm SUSY}$.  Further the results of Ref.~\cite{cgnw} also extend to other couplings in the MSSM Lagrangian which are governed by $y_b$: Also in the higgsino
couplings of the charginos and neutralinos the use of \eq{eq:ResForm0} correctly resums the enhanced corrections, irrespective of the sizes of the momenta and masses involved.  The Feynman rules for these effective couplings are listed in Appendix~\ref{feynman}.  However, the situation is different for a $\cot \beta$-suppressed process: Here the enhanced one-loop correction depends on the kinematics of the studied process. For example, the coupling of the Standard-Model-like Higgs boson $h^{0}$ to fermions involves $\tan\beta$-enhanced momentum-dependent one-loop form factors.

\subsection{Sbottom mixing and resummation}\label{sect:smr}
As an introductory remark, we note that the resummation issue is simple 
if one interchanges the roles of $y_b$ and $m_b$: Choosing $\delta y_b$
as input will fix $\delta m_b$ through \eq{eq:ResEq}, there are no
enhanced corrections beyond one-loop order and any non-linear dependence
of $\Sigma_b^{RL}$ on $y_b$ does not pose a problem. This avenue has been
pursued in Sect.~2 of Ref.~\cite{cgnw}.  Yet in any phenomenological
application we face the fact that we have precise data on $m_b$ and not
on $y_b$, so that we are stuck with the task to invert \eq{eq:shortcut}.
We discuss this for three well-motivated schemes for the sbottom mass
matrix here:

\begin{itemize}
\item[(i)] \textbf{Input:} $m_{\tilde{b}_1}^2$, $m_{\tilde{b}_2}^2$;
  $\mu$, $\tan\beta$\\
  If we express the sbottom mixing angle $\tilde{\theta}_b$ and phase
  $\tilde{\phi}_b$ in \eq{eq:Self1} through our input parameters, using
  relation (\ref{eq:MixingAngle}), the bottom mass in
  $\Delta_b^{\tilde{g}}$ cancels and we find the gluino and chargino
  contributions to $\Sigma^{RL}_{b}$ to be linear in $y_b$. This is the
  case used to illustrate the resummation in \eq{eq:naivesum}.  If we
  assume the neutralino contributions to be linear in $y_b$, too, we
  arrive at
\begin{equation}
    y_b^{(0)} =\frac{m_b}{v_d(1+\Delta_b)}.\label{eq:ResForm1}
\end{equation}
The chargino contribution
$\Sigma_{b}^{RL,\widetilde{\chi}^{\pm}}=m_b^{(0)}
\Delta_b^{\widetilde{\chi}^{\pm}}$ is always linear in $y_b$, it is not
influenced by our choice of input parameters since no bottom squarks are
involved. The neutralino contribution
$\Sigma^{RL,\widetilde{\chi}^0}=m_b^{(0)} \Delta_b^{\widetilde{\chi}^0}$
in (\ref{eq:Self3}) can be rewritten as
\begin{eqnarray}
  \Sigma_{b}^{RL,\widetilde{\chi}^0}&=&
      \frac{y_b g}{16\pi^2}\sum\limits_{m=1}^4 
  \frac{m_{\widetilde{\chi}^{0}_m}}{\sqrt{2}}\widetilde{N}_{m2}^{*}
  \widetilde{N}_{m3}^{*} \cdot 
   B_0(m_{\widetilde{\chi}^0_m},m_{\tilde{b}_1})\\
  &&-\frac{y_b g}{16\pi^2}\sum\limits_{m=1}^4 
  \frac{m_{\widetilde{\chi}^{0}_m}}{\sqrt{2}}\widetilde{N}_{m2}^{*}
  \widetilde{N}_{m3}^{*}\sin^2\tilde{\theta}_b\,
   \left(B_0(m_{\widetilde{\chi}^0_m},m_{\tilde{b}_1}) - 
         B_0(m_{\widetilde{\chi}^0_m},m_{\tilde{b}_2})
    \right),\nonumber
\end{eqnarray}
where the first line is linear in $y_b$, but the second line is found to
contain terms of third order and higher in $y_b$ after insertion of
(\ref{eq:MixingAngle}). In the decoupling limit $\msusy\gg v$, these
higher order terms, which are proportional to
$\sin^2\tilde{\theta}_b\propto v^2/\msusy^2$, vanish and the neutralino
contribution is correctly included into (\ref{eq:ResForm1}). For
$\msusy\sim v$ on the other hand, the higher-order terms spoil the
proper resummation because equation (\ref{eq:ResEq}) cannot be solved
analytically anymore. As $\Delta_b^{\widetilde{\chi}^0}$ is small
anyway, formula (\ref{eq:ResForm1}), though not entirely correct in this
case, still holds to a very good approximation.\bigskip
  
\item[(ii)] \textbf{Input:} $m_{\tilde{b}_1}^2$, $m_{\tilde{b}_2}^2$; 
  $\tilde{\theta}_b$, $\tilde{\phi}_b$\\
  Assuming that some day it will be possible to measure
  $\tilde{\theta}_b$ and $\tilde{\phi}_b$, we could take these
  quantities as our input instead of $\mu$ and $\tan\beta$. In
  \eqsand{eq:Self1}{eq:Self3} $\Delta^{\tilde{g}}_b$ and
  $\Delta^{\widetilde{\chi}^0}_b$ are directly given as a function of
  $\tilde{\theta}_b$ and $\tilde{\phi}_b$. Obviously,
  $\Sigma^{RL,\tilde{g}}_b = m_b^{(0)} \Delta_b^{\tilde{g}}$ does not
  exhibit any explicit $y_b$-dependence in this case, so that no
  reinsertion of $\delta y_b$ into $\Sigma^{RL,\tilde{g}}_{b}$ is
  possible (it is absorbed into the physical mixing angle). The
  neutralino contribution $\Sigma^{RL,\widetilde{\chi}^0}_b$ on the
  other hand is linear in $y_b$ if we choose $\tilde{\theta}_b$ as
  input and it can be properly resummed now, in contrast to case (i).
  The modified relation between $y_b^{(0)}$ and $m_b$ then reads
  \begin{equation}
    y_b^{(0)}= y_b+\delta y_b = 
      \frac{m_b}{v_d}
   \frac{1-\Delta_b^{\tilde{g}}}{1+\Delta_b^{\widetilde{\chi}^{\pm}}
      +\Delta_b^{\widetilde{\chi}^0}}.\label{eq:ResForm2}\bigskip
  \end{equation}
Note that this scheme does not involve an explicit  $\tan\beta$-enhanced 
counterterm to $\tilde{\theta}_b$. The implicit resummation encoded in 
a ``measured'' value of  $\tilde{\theta}_b$ must, however, be taken into
account in a proper analysis of the MSSM parameter space: In the  
large-$\tan\beta$ limit \eqsand{eq:xb}{eq:MixingAngle} imply a
correlation between $ y_b^{(0)}$, $\mu$ and our input parameters:
\begin{align}\label{eq:MixingAngle2}
e^{i\tilde{\phi}_{b}}\, \sin 2\tilde{\theta}_{b}  =&   
   - \frac{2 y_b^{(0)*} \mu v_u}{m_{\tilde b_1}^2 - m_{\tilde b_2}^2} 
\end{align}
That is, in scheme (ii) $\mu$ inherits the large correction from $
y_b^{(0)}$ because the product $y_b^{(0)*} \mu$ is fixed.  Since
$\mu$ enters the chargino and neutralino mass matrices
$\mathcal{M}_{\widetilde{\chi}^{\pm,0}}$, one should solve
\eq{eq:MixingAngle2} for $\mu$, use the value in
$\widetilde{\chi}^{\pm,0}$ and repeat the steps iteratively until
\eqsand{eq:ResForm2}{eq:MixingAngle2} are sufficiently (i.e.\ up to
the neglected $\cot\beta$-suppressed correction proportional to $A_b$)
compatible. As a corollary we remark that a measurement of $m_{\tilde
b_{1,2}}$, $\tilde{\theta}_{b}$ and $\mu$ (which can be inferred from
chargino or neutralino masses) completely fixes $|y_b^{(0)}|$ through
\eq{eq:MixingAngle2} if $\tan\beta$ is large. Once $|y_b^{(0)}|$ is
known the coupling strengths of $A^{0}$ and $H^{0}$ to bottom quarks
are fixed.  $|y_b^{(0)}|$ enters the production cross sections of
these particles and cannot be studied in $A^{0}$,$H^{0}$ decays to $b$
quarks at the LHC because of the large $b\ov b$ background from QCD
processes.

\item[(iii)] \textbf{Input:} $m_{\tilde{b}_{L}}^2$,
  $m_{\tilde{b}_{R}}^2$; $\mu$, $\tan\beta$\\
  As the masses and mixing angles of the SUSY particles are not measured
  yet, this set is the most prominent one because its elements directly
  appear in the Lagrangian.  In terms of these input parameters, the
  mixing angle can be expressed with the help of 
  \begin{equation}
    e^{i\tilde{\phi}_{b}}\, \tan 2\tilde{\theta}_{b}  =   
   - \frac{2 y_b^{(0)*} \mu v_u}{m_{\tilde b_L}^2 - m_{\tilde b_R}^2}
                   \label{eq:TanMix}
  \end{equation}
  Since $\Delta_b^{\tilde{g}}$ is proportional to
  $\sin2\tilde{\theta}_b= \tan2\tilde{\theta}_b/(\sqrt{1+\tan^2
    2\tilde{\theta}_b})$ and in addition the squark masses appearing in
  the loop functions have to be replaced by $m_{\tilde{b}_{L}}^2$ and
  $m_{\tilde{b}_{R}}^2$ via (\ref{eq:SquarkMasses}), the
  $y_b$-dependence of $\Delta_b^{\tilde{g}}$ gets so complicated that
  (\ref{eq:ResEq}) cannot be solved analytically anymore. This problem
  can be avoided in the following way: In a first approximation, we
  determine $m_{\tilde{b}_{1,2}}^2$ from (\ref{eq:SquarkMasses}) using
  the tree level value for $y_b$. Now we can calculate $\Delta_b$ as a
  function of the parameter set (i). In a next step, the resulting
  modified Yukawa coupling (\ref{eq:ResForm1}) can be reinserted into
  (\ref{eq:SquarkMasses}) to get corrected values for
  $m_{\tilde{b}_{1,2}}^2$. This procedure has to be repeated until the
  value of $\Delta_b$ converges. The resummed Yukawa coupling is then
  given by (\ref{eq:ResForm1}). Alternatively, we could calculate
  $\Delta_b^{\tilde{g}}$ and $\Delta_b^{\widetilde{\chi}^0}$ iteratively
  as a function of the input parameters (ii), determining
  $\sin2\tilde{\theta}_b$ from \eq{eq:TanMix}. In that case, 
  \eq{eq:ResForm2} provides the resummed Yukawa coupling.
\end{itemize}
\eq{eq:ResForm1} has the same form as the widely-used relation between
$y_b^{(0)}$ and $m_b$ valid in the decoupling limit and quoted in
\eq{eq:YukMassRel}. Therefore we will take parameter set (i) as the
physical input from now on.

\section{Flavour mixing at large \boldmath $\tan\beta$\unboldmath 
        }\label{flavour_mixing}
In the effective-field-theory approach the resummation of $\tan\beta$-enhanced
effects in flavour-changing transitions is achieved in the same way as in the
flavour-conserving case: One calculates loop-induced couplings of $H_u$ to
quarks, now taking flavour mixing into account.  After the Higgs doublets
acquire their vevs the down-quark mass matrix is diagonalised.  In the basis
of quark mass eigenstates we face flavour-non-diagonal Yukawa couplings, as
expected in a general 2HDM
\cite{Hamzaoui:1998nu,Babu:1999hn,Buras:2001mb,Isidori:2001fv}.  This method
is correct for $\msusy\gg v,M_{A^0,H^0,H^\pm}$.  In this chapter we extend the
resummation of $\tan\beta$-enhanced effects to the case of any hierarchy
between $\msusy$ and $v$ to cover the natural situation $\msusy\sim
M_{A^0,H^0,H^\pm} \sim v$. First, our results allow us to assess the accuracy
of the decoupling limit used in the literature. Second, we access a new field
and calculate the $\tan\beta$-enhanced loop corrections to genuine
supersymmetric couplings: For instance, the gluino-quark-squark coupling,
which is flavour-diagonal at tree-level, receives enhanced FCNC loop
corrections just as the neutral Higgs bosons $A^0$ and $H^0$ do. These
effective FCNC couplings of supersymmetric particles cannot be studied with
the effective-field-theory approach, because these particles are treated as
heavy and are integrated out.

Our diagrammatic treatment of $\tan\beta$-enhanced loop corrections can easily
be generalised to the flavour off-diagonal case. In the naive MFV framework,
$\tan\beta$-enhanced flavour transitions  only arise from self-energies of
down-type quarks involving chargino-squark exchange (see 
\fig{char_sq_loop}). 
\begin{nfigure}{t}
\centering
\includegraphics[height=2.6cm]{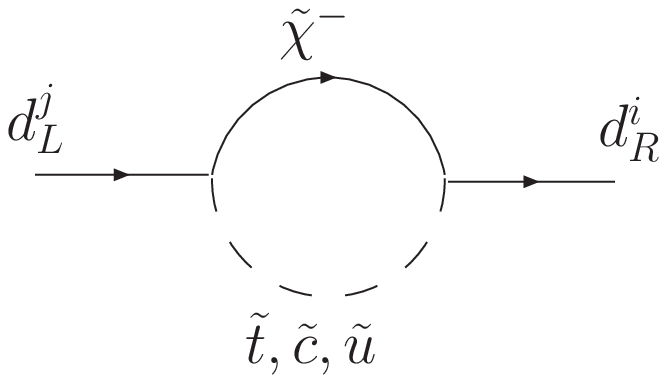}
\caption{$\tan\beta$-enhanced flavour-changing self-energy}
\label{char_sq_loop}
\end{nfigure}
In the case of $d$-$s$-transitions, the stop contribution is suppressed by
$V_{ts}^*V_{td}$. Since we neglect the small Yukawa couplings of up and charm 
and take degenerate masses for $\tilde u$ and $\tilde c$ squarks, 
the $\tilde u$ and $\tilde c$ contributions to $d$-$s$-transitions vanish because of a GIM cancellation. For
the flavour-changing self-energies involving a bottom quark we find
\begin{equation}\label{eq:FCSelf}
   \Sigma^{RL}_{ij}(V) =V_{ti}^*V_{tj} 
      \frac{m_i\epsfc\tan\beta}{1+\epsilon_i\tan\beta},
   \qquad\qquad\mbox{for~~~}(i,j)=(3,1),(3,2),(1,3),(2,3).
\end{equation}
Here the unitarity of the CKM matrix and the mass degeneracy of the
$\tilde u$ and $\tilde c$ squarks have been used to factor out the CKM
combination $V_{ti}^*V_{tj}$.  The explicit expression for $\epsfc$ in
terms of the stop mixing-parameters $\tilde{\theta}_t$, $\tilde{\phi}_t$
and the chargino mixing matrices $\widetilde{U}$, $\widetilde{V}$ reads
\begin{align}\nonumber
  \epsfc = & -\frac{1}{16\pi^2}\frac{g}{\sqrt{2}M_W\sin\beta} \sum_{m=1}^2
  m_{\tilde\chi^{\pm}_m} \widetilde{U}_{m2}^*\left[ \frac{y_t}{2}
    \widetilde{V}_{m2}^* \sin 2\tilde{\theta}_t e^{i\tilde{\phi}_t} \left(
      B_0(m_{\tilde \chi^{\pm}_m}, m_{\tilde t_1}) - B_0(m_{\tilde
        \chi^{\pm}_m}, m_{\tilde t_2}) \right) \right. \\ & \left.  - g
    \widetilde{V}_{m1}^* \left( \cos^2\tilde{\theta}_t B_0(m_{\tilde
        \chi^{\pm}_m}, m_{\tilde t_1}) + \sin^2 \tilde{\theta}_t B_0(m_{\tilde
        \chi^{\pm}_m}, m_{\tilde t_2}) - B_0(m_{\tilde \chi^{\pm}_m},
      m_{\tilde q}) \right) \right],
\end{align}
with $m_{\tilde{q}}$ denoting the common mass of the left-handed first
and second generation squarks.  If one wants to express $\epsfc$ in
terms of the SUSY-breaking parameters instead, one can use the relations
given in Appendix \ref{conventions} to find
\begin{align}
\epsfc = & -\frac{y_t^2}{16\pi^2} A_t^*\mu^*  
    \left( D_2 - |M_2|^2 D_0 \right)
           + \frac{g^2}{16\pi^2} M_2^* \mu^* 
\left( D_2 - m_{\tilde t_{R}}^2 D_0 - C_0\right)\label{epsilonfc}
\end{align}
where $D_{0,2} = D_{0,2}(m_{\widetilde{\chi}^{\pm}_1},
m_{\widetilde{\chi}^{\pm}_2}, m_{\tilde{t}_1},m_{\tilde{t}_2})$ and
$C_0=C_0(m_{\widetilde{\chi}^{\pm}_1}, m_{\widetilde{\chi}^{\pm}_2},
m_{\tilde{q}})$. \eq{epsilonfc} makes clear that $\epsfc$ and thus also the
$\tan\beta$-enhanced flavour-changing self-energies are directly linked to the
SUSY-breaking sector of the Lagrangian. They vanish if $M_2$ and $A_t$ are set
to zero. The part of $\epsfc$ which is proportional to $g^2$ is absent if the
left-right mixing of the top squarks is neglected and in addition universality
for the mass terms of the left-handed squarks is assumed.
We next present two different ways to account for $\epsfc$ in practical
calculations of low-energy flavour observables. The first option,
explained in Sect.~\ref{self_en_approach}, is to consider self-energy
corrections in external quark legs. The second possibility, discussed in
Sect.~\ref{ct_approach}, involves a flavour-non-diagonal wave-function
renormalisation for the quark fields, which enters the Feynman rules of
the couplings of quarks to SUSY particles and Higgs fields.

\begin{nfigure}{t}
  \begin{picture}(200,70) (-220,-35)
    \SetWidth{0.5}
    \SetColor{Black}
    \ArrowLine(-170,0)(-140,0)
    \ArrowLine(-100,0)(-80,0)
    \ArrowLine(-80,0)(-60,0)
    \Vertex(-80,0){2}
    \GOval(-120,0)(20,20)(0){0.882}
    \Line(-60,-20)(-60,20)
    \Line(-60,10)(-50,20)
    \Line(-60,0)(-50,10)
    \Line(-60,-10)(-50,0)
    \Line(-60,-20)(-50,-10)
    \Text(-165,5)[lb]{\Black{$s_L$}}
    \Text(-95,5)[lb]{\Black{$b_R$}}
    \Text(-75,5)[lb]{\Black{$b_L$}}
    \Text(-130,-8)[lb]{\Black{$\Sigma^{RL}_{bs}$}}
    \Text(-120,-40)[lb]{(1)}

    \ArrowLine(20,0)(40,0)
    \ArrowLine(40,0)(60,0)
    \ArrowLine(100,0)(130,0)
    \Vertex(40,0){2}
    \GOval(80,0)(20,20)(0){0.882}
    \Line(130,-20)(130,20)
    \Line(130,10)(140,20)
    \Line(130,0)(140,10)
    \Line(130,-10)(140,0)
    \Line(130,-20)(140,-10)
    \Text(25,5)[lb]{\Black{$b_L$}}
    \Text(45,5)[lb]{\Black{$b_R$}}
    \Text(110,5)[lb]{\Black{$s_L$}}
    \Text(70,-8)[lb]{\Black{${\Sigma^{RL*}_{bs}}$}}
    \Text(75,-40)[lb]{(2)}
  \end{picture}

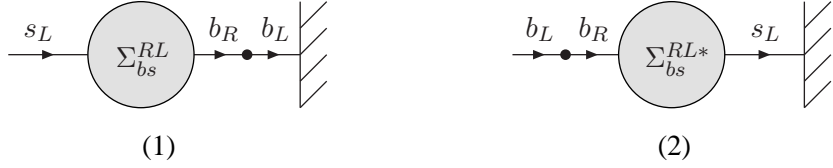
\captionof{figure}{Feynman diagrams with flavour-changing self-energy in  
 an external leg.}
\label{fig:ExtLegSelf}
\end{nfigure}

\subsection{Flavour-changing self-energies in 
  external legs}\label{self_en_approach}%
Consider the generic situation of a self-energy subdiagram in an external
quark leg of some Feynman diagram, as displayed in \fig{fig:ExtLegSelf} for
the case of an external $s$ quark.  In flavour-conserving transitions such
self-energies in external legs are truncated, they instead enter the S-matrix
elements through the LSZ factor (``external wave-function renormalisation'').
However, if the truncation affects a particle with a different mass than the
external particle, the diagram with the external self-energy can be treated in
the same way as a 1PI vertex correction \cite{ln}, provided that the mass
difference is much larger than the self-energy diagram. Despite of the
$\tan\beta$-enhancement, this condition, which reads $m_b-m_s \gg
|\Sigma_{bs}|$ in our case, is certainly fulfilled because the self-energy
$\Sigma_{bs}$ is CKM-suppressed by a factor of $V_{ts} V_{tb}^*$.
Treating external self-energies as 1PI diagrams makes 
the origin of the large effects most obvious. The alternative approach, 
which truncates all self-energies and introduces 
flavour-non-diagonal wave-function renormalisation, is discussed below in 
Sect.~\ref{ct_approach}. Of course, both methods lead to the same results for 
physical amplitudes. 

For definiteness we consider diagrams with external $s$ or $b$ quarks
(\fig{fig:ExtLegSelf}).  The case of $b$-$d$ transitions is obtained by
obvious replacements.  For $m_s=0$ the Feynman amplitudes are given by
\begin{eqnarray}
   \mathcal{M}_1& = &
  \mathcal{M}^{\textrm{rest}}_1\cdot{ \left. 
    \frac{i(\slashed{p}+m_b)}{p^2-m_b^2}\right|}_{\slashed{p}=0} 
       (-i\Sigma^{RL}_{bs}) = 
    -\mathcal{M}^{\textrm{rest}}_1\cdot V_{ts}V_{tb}^*
   \frac{\epsfc\tan\beta}{1+\epsilon_b\tan\beta},\label{eq:ExtLegSelf1}\\
   \mathcal{M}_2& =& 
  \mathcal{M}^{\textrm{rest}}_2\cdot{ \left.
    \frac{i(\slashed{p}+m_s)}{p^2-m_s^2}\right|}_{\slashed{p} = 
   m_b^{\textrm{pole}}} (-i\Sigma^{RL*}_{bs}) = 
  +\mathcal{M}^{\textrm{rest}}_2\cdot V_{ts}^*V_{tb}
   \frac{\epsfc^*\tan\beta}{1+\epsilon_b^*\tan\beta}.\label{eq:ExtLegSelf2}
\end{eqnarray}
Here, $\mathcal{M}_i^{\textrm{rest}}$ stands for the part of the Feynman
amplitude corresponding to the truncated diagram. The expressions
(\ref{eq:ExtLegSelf1}) and (\ref{eq:ExtLegSelf2}) are of order
$\mathcal{O}(\epsfc\tan\beta)$. Thus, if a large value of $\tan\beta$
compensates for the small $\epsfc$, it is possible to get a $b\to s$
transition without paying the price of a loop suppression. 

There is one important physical process for which even diagrams with two self-energies in external lines must be considered: In $b\to s\gamma$ the expansion of the diagrams to lowest order in $m_b/M_{SUSY}$ understood in \eqsand{eq:ExtLegSelf1}{eq:ExtLegSelf2} gives zero. One therefore has to consider contributions of higher order in this ratio. This means that in \eq{mbos} the right-hand side has to be expanded to order $m_b^2/M_{SUSY}^2$ in order to find the appropriate counterterm $\delta m_b$, whereas only the leading term was kept in chapter \ref{sect:2}. We stress that this expansion does not spoil the resummation of the counterterm. Now let us have a look at the $b\to s\gamma$-diagrams in \fig{fig:bsg}. We observe that an insertion of $\delta m_b$ like in the lower-left diagram (denoted by a cross) cancels only partially with a corresponding flavour-conserving self-energy insertion like in the upper-left diagram if we perform an on-shell calculation of the amplitude. The reason is that $\delta m_b$ is determined at $p^2=m_b^2$ while the self-energy is probed at $p^2=0$. The remnant is of order $\mathcal O(m_b^2/M_{SUSY}^2)$, just as the contribution that we find from the vertex correction in the upper-right diagram. For completeness, we mention that some non-$\tan\beta$-enhanced contributions are canceled by insertions of on-shell wave-function counterterms of the bottom quark like the one shown in the lower-right diagram (also denoted by a cross). Summing up all the diagrams yields a gauge-invariant result of the order $(m_b/M_{SUSY})^2\, \epsfc^* \tan^2\beta$  times another loop factor, which is the same order as the leading supersymmetric one-loop contribution to $b\to s\gamma$.

\begin{nfigure}{t}
\centering
\includegraphics[
                 width=0.9\textwidth]{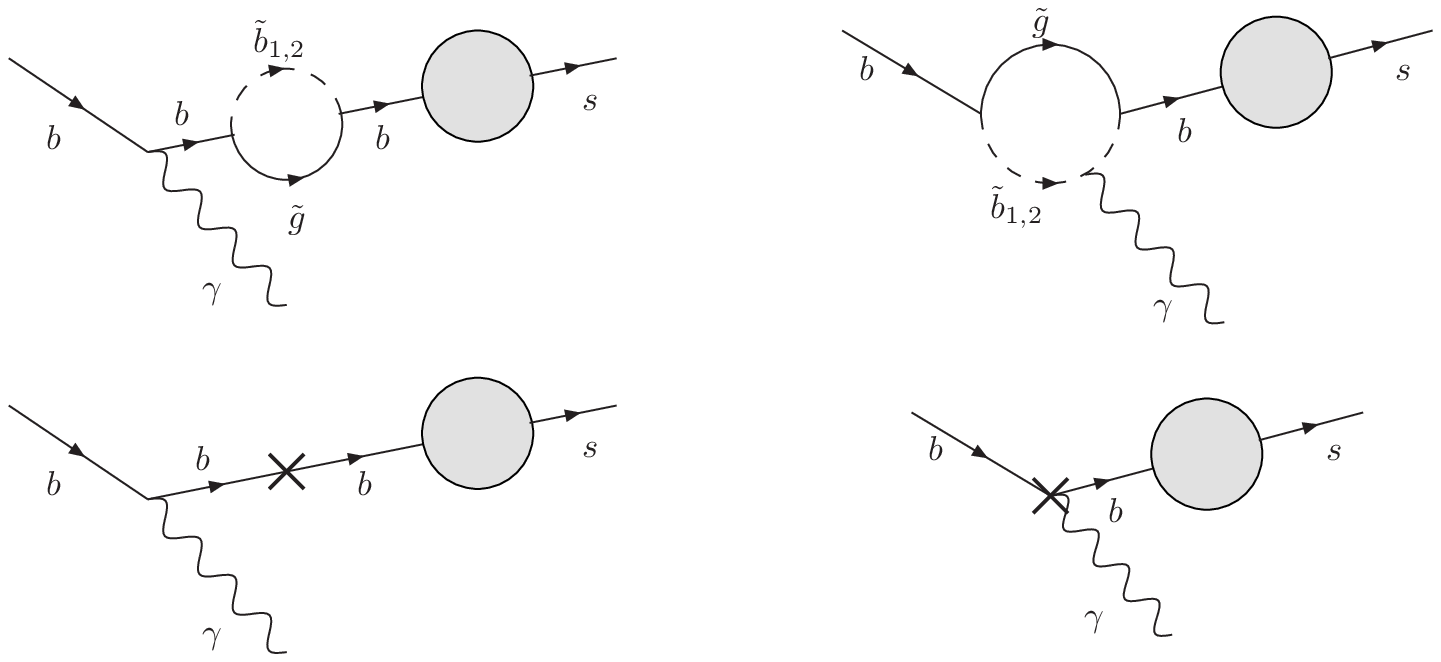}
                 \caption{Some diagrams with self-energies in external lines for the process  $b\to
                   s\gamma$  \label{fig:bsg}}
\end{nfigure}

It is natural to ask whether the above effect, i.e.\ the generation of
$\tan\beta$-enhanced $b\to s$ transitions via self-energy insertions, also
occurs for internal quark lines. It is important to notice that the
$\tan\beta$-enhancement in \eqsand{eq:ExtLegSelf1}{eq:ExtLegSelf2} is
generated by the fact that the quark propagator $-i/m_b$ cancels a factor 
of $m_b$ in $\Sigma^{RL}_{bs}$.
A potential $1/m_b$-dependence of some loop integral would originate
from the low momentum region $p^2\ll M_{SUSY}^2$, but we have
constructed the mass counterterm $\delta m_b$ in Section
\ref{flavour_cons} in such a way that it subtracts the self-energy
insertion in this momentum region. Therefore we only need to worry about
situations similar to $b\to s\gamma$, in which higher orders of
$m_b/M_{SUSY}$ are relevant. However, we are not aware of a meaningful
physical process in which an internal $b$ line is responsible for a
$1/m_b$ singularity in this way and do not consider this possibility
further. 

Before investigating the further consequences of the $\tan\beta$-enhanced
flavour transitions, we want to point out a subtlety of equation
(\ref{eq:ExtLegSelf2}). The $b$-quark mass which enters the propagator via the
equation of motion is the pole mass $m_b^{\textrm{pole}}$. The $b$-quark mass
appearing in $\Sigma_{bs}^{RL}$, on the other hand, is the
$\overline{\rm MS}$-mass $m_b$. However, if QCD-corrections to the diagrams of
\fig{fig:ExtLegSelf} are taken into account, additional contributions
add to the $\overline{\rm MS}$-mass in $\Sigma^{RL}_{bs}$ to give the pole mass
$m_b^{\textrm{pole}}$. Therefore the $b$-quark mass correctly cancels from 
\eq{eq:ExtLegSelf2}. A detailed analysis of this feature can be
found in Appendix~\ref{sec:QCDcorrections}.
\begin{nfloatingfigure}[l]{4cm}
\centering
\vspace{-8pt}
\includegraphics[width=4cm]{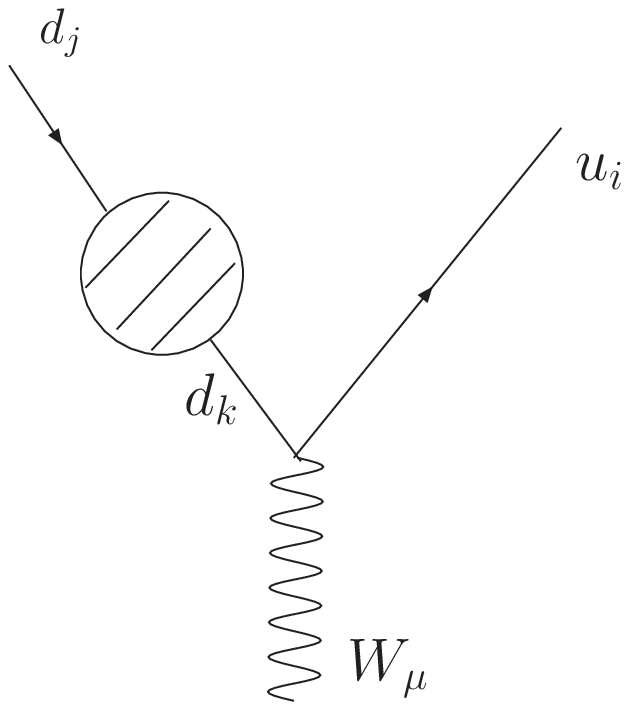}
\caption{Generic enhanced correction to $V_{ij}$}
\label{CKM_generic}
\end{nfloatingfigure}

Now, let us consider the $\tan\beta$-enhanced corrections to the
$u_i$-$d_j$-$W$-vertex (see \fig{CKM_generic}). We apply an on-shell
renormalisation condition to $V_{ij}$ and cancel the contribution from the
self-energy diagram at $p^2=0$ by a counterterm $\delta V_{ij}$. In this way
the renormalised $V$ corresponds to the CKM matrix measured from low energy
data.\footnote{Therefore our $V$ corresponds to $V^{\rm eff}$ of
  Ref~\cite{bcrs}.}  We find
\begin{eqnarray}\label{eq:CKMCounter}
    \delta V_{ij}&=&-V_{ik}\Lambda_{kj},\qquad\qquad\textrm{with}\nn
    \Lambda_{kj} (V)&=&
 \left\{\begin{array}{cl}\ds \frac{m_{d_j}}{m_{d_j}^2-m_{d_k}^2} 
  \Sigma_{kj}^{LR}+\frac{m_{d_k}}{m_{d_j}^2-m_{d_k}^2}\Sigma_{kj}^{RL}
                  &\ds ,\,k\not= j \\ \ds 0 &\ds ,\,k=j
        \end{array}\right. \quad
\end{eqnarray}
Note that $\delta V_{ij}$ never involves less powers of the Wolfenstein
parameter $\lambda$ than $V_{ij}$. The bare CKM matrix $ V^{(0)}$ reads
\begin{equation}
   V^{(0)}=V+\delta V=V(1-\Lambda)\approx V e^{-\Lambda}.
\end{equation}
This shows that the chosen renormalisation condition preserves the 
unitarity of the CKM matrix because the matrix $\Lambda$ is anti-hermitian.

From eq. (\ref{eq:FCSelf}) we find that the corrections $\delta V_{td}$,
$\delta V_{ts}$, $\delta V_{ub}$ and $\delta V_{cb}$ are of order
$\mathcal O(\epsfc\tan\beta)$ and so can be comparable in size to the
corresponding tree-level quantities $V_{ij}$. Hence, the situation is
the same as it was for the flavour-conserving self-energies in Section
\ref{flavour_cons}: Reinsertion of the counterterms $\delta V_{ij}$ into
the diagram of \fig{CKM_generic} leads to contributions which are
formally of higher loop order but also of higher order in $\tan\beta$.
To resum these corrections we generalise \eq{eq:CKMCounter} to all orders in
perturbation theory as
\begin{equation}\label{eq:ResCKM}
 \delta V_{ij}=-(V_{ik}+\delta V_{ik})\cdot\Lambda_{kj}(V+\delta V),
\end{equation}
which is in complete analogy with eq. (\ref{eq:ResEq}) for the flavour
conserving case. Note that the enhanced flavour-conserving corrections
associated with $y_b$ are already properly resummed in \eq{eq:FCSelf} through
the factor of $1/(1+\epsilon_i\tan\beta)$.  We have two possibilities to deal
with \eq{eq:ResCKM}. Firstly, we can expand the RHS order by order, deduce a
recursive relation between the CKM counterterms $\delta V_{ij}^{(n)}$ and
$\delta V_{ij}^{(n-1)}$ and perform the resummation explicitly.  Secondly, we
can add $V_{ij}$ to both sides of \eq{eq:ResCKM} and solve the
resulting matrix equation
\begin{equation}\label{eq:rme}
    V^{(0)}=V-V^{(0)}\cdot\Lambda(V^{(0)})
\end{equation}
for $V^{(0)}$. Inserting $\Lambda_{kj}(V^{(0)})$ from \eq{eq:CKMCounter} 
with $\Sigma_{ij}^{RL}=\Sigma_{ji}^{LR*}$ from \eq{eq:FCSelf} into 
\eq{eq:rme} yields
\begin{equation}\label{eq:vsum}
  V^{(0)}_{ij}\;=\;
    V_{ij} \,-\,\sum_{k\neq j}\, V_{ik}^{(0)} V_{tk}^{(0)*} V_{tj}^{(0)} \, 
               \frac{1}{m_j^2-m_k^2}  \,
           \lt[ \frac{m_j^2\epsfc^*\tan\beta}{1+\epsilon_j^*\tan\beta} 
                  \, +\, 
                 \frac{m_k^2\epsfc\tan\beta}{1+\epsilon_k\tan\beta} \rt].
\end{equation}  
Neglecting small quark mass ratios and ignoring the tiny corrections to the
Cabibbo matrix we obtain the solution
\begin{equation}\label{ckm_renorm}
V^{(0)} =  \begin{pmatrix}V_{ud} & V_{us} & K^* V_{ub} \\ 
 V_{cd} & V_{cs} & K^* V_{cb} \\
 K V_{td} & K V_{ts} & V_{tb} \end{pmatrix},\hspace{1cm}\textrm{with}\hspace{0.5cm}
K = \frac{ 1 + \epsilon_b\tan\beta }{ 1+(\epsilon_b - \epsfc)\tan\beta }.
 \end{equation}
 We recognise that this amounts to a 
 renormalisation of the Wolfenstein parameter $A$,
\begin{equation}
A^{(0)} = \left|\frac{ 1 + \epsilon_b\tan\beta }{ 
                1+(\epsilon_b - \epsfc)\tan\beta }\right| \, A.
\end{equation}
Possible complex phases can be absorbed by the usual rephasing of the
top-quark and bottom-quark fields (with the same phase for the left- and
right-handed fields).    In order to
preserve supersymmetry, one should then perform the same rephasing also for
the stop and sbottom fields.

Comparing \eq{ckm_renorm} to results of calculations in effective-theory
approaches \cite{Babu:1999hn,Blazek:1995nv,bcrs,gjnt}, where the SUSY
particles are integrated out at a scale much higher than the electroweak
scale, we see that the results are identical in the limit $\msusy\gg
v$, as they should be. Yet our result \eq{ckm_renorm} provides an
explicit resummation of the $\tan\beta$-enhanced flavour-changing
effects to all orders in perturbation theory and is also valid in the
case where the SUSY mass-scale is similar to the electroweak scale.

\subsection{Renormalisation of the flavour-changing 
  self-energies}\label{ct_approach}

The second possibility to deal with flavour-changing self-energies is to
absorb them into wave-function counterterms. In this approach, no
external-leg corrections have to be taken into account in the
calculation of transition amplitudes. Instead, the effect of
flavour-changing self-energies now resides in the wave-function
counterterms, which enter the various couplings of the quark fields. In
particular, the wave-function counterterms render couplings which are
flavour-diagonal at tree-level flavour-changing.  Furthermore, this
method permits an easy incorporation of the resummed
$\tan\beta$-enhanced effects into explicit Feynman rules for the MSSM.
These Feynman rules are collected in Appendix~\ref{feynman} and can be
readily implemented into computer programs like FeynArts
\cite{feynarts,feynarts2}.  They include for example flavour-changing gluino couplings, which have previously been found by Degrassi, Gambino and Slavich in Ref.~\cite{Degrassi:2006eh}. We will see that these
counterterm couplings are indeed enhanced by a factor of $\tan\beta$ and
therefore determine them to all orders in the perturbative expansion,
which has not been done in Ref.~\cite{Degrassi:2006eh}. The scope of
Ref.~\cite{Degrassi:2006eh} is the calculation of the supersymmetric
strong corrections to $ b\rightarrow s\gamma$ for all values of
$\tan\beta$, while we are interested in the leading power of $\tan\beta$
only, albeit to all orders in perturbation theory and with the effects
of all gauge couplings and of the large Yukawa couplings $y_t$ and
$y_b$.

We next present the flavour-changing wave-function counterterms and
reproduce the result for the renormalised CKM matrix of the previous
section: The renormalisation of the CKM matrix with the help of
wave-function counterterms has been first studied by Denner and Sack in
Ref.~\cite{Denner:1990yz} for the Standard Model, where an on-shell
scheme has been chosen.  That is to say, the wave-function counterterms
have been defined in a proper way to cancel flavour-changing
self-energies when one of the external quarks is put on the mass shell.
Later Gambino, Grassi and Madricardo \cite{Gambino:1998ec} have argued
that this on-shell prescription can lead to gauge-noninvariant results
and have given a renormalisation prescription for the flavour-changing
two-point functions at zero external momentum $p$. As long as we neglect
the external momenta in the calculation of the SUSY self-energy
diagrams, there is no difference between the two approaches and the
naive on-shell subtraction of flavour-changing self-energies in external
quark legs gives gauge invariant results.  Only chirality-flipping
self-energies $\Sigma^{RL}_{ij}$ in the down sector are
$\tan\beta$-enhanced. Therefore only down-quark fields have to be
renormalised according to
\begin{equation}\label{wf_ct1}
d_{i,L}^{(0)} = \left( \delta_{ij} +  \frac{1}{2}\delta Z^{L}_{ij} \right) 
                d_{j,L}, \qquad \qquad 
d_{i,R}^{(0)} = \left( \delta_{ij} +  \frac{1}{2}\delta Z^{R}_{ij} \right) 
                d_{j,R}
\end{equation}
and their wave-function counterterms are anti-hermitian:
\begin{equation}\label{wf_ct2}
  \delta  Z^L_{ij}=-\delta Z^{L*}_{ji}, \qquad 
  \delta Z^R_{ij}=-\delta Z^{R*}_{ji}.
\end{equation}
The wave-function renormalisation (\ref{wf_ct1}) corresponds to a unitary
transformation of the down-type quark fields in flavour space. We will see in
the following that this implies, in combination with a suitable
renormalisation of the CKM matrix, that couplings of the Standard-Model
particles to one another are unaffected by our renormalisation. In this way,
no flavour violation occurs in the couplings of the photon, of the $Z^0$ boson,
or of the gluon, as required by the decoupling theorem. 

The rotation of the quark fields in \eq{wf_ct1} affects the down-quark 
mass terms of the Lagrangian (cf.\ \eq{eq:lm}) as 
\begin{equation}\label{eq:lmrot}
{\cal L}_m = -\; m_{d_j}^{(0)} \ov d_{j,R}^{(0)} d_{j,L}^{(0)} 
       + \mbox{h.c.} = 
       - \lt[ m_{d_j}^{(0)} \delta_{jk} + 
              \frac12 m_{d_j}^{(0)} \delta Z^{L}_{jk} 
            - \frac12 m_{d_k}^{(0)} \delta Z^{R}_{jk}  \rt]
          \ov d_{j,R} d_{k,L} + \mbox{h.c.} 
\end{equation}
Subtraction of the flavour-changing self-energies at vanishing external
momentum amounts to the condition
\begin{align}\label{wfr_cond1}
\Sigma^{RL}_{ij} + m^{(0)}_{d_i}\frac{\delta Z^L_{ij}}{2} - m^{(0)}_{d_j} 
         \frac{\delta Z^R_{ij}}{2} =0, &
\qquad\qquad i\neq j,
\end{align}
for $\delta Z^{L,R}_{ij}$ with $\Sigma^{RL}_{ij}$ given in (\ref{eq:FCSelf}).
Here the bare masses $ m^{(0)}_{d_i}= m_{d_i}+\delta  m_{d_i}$ 
contain the $\tan\beta$-enhanced corrections associated with the  
mass counterterms $\delta  m_{d_i}$ calculated in section \ref{flavour_cons}.

The explicit expressions for the anti-hermitian
one-loop counterterms in our scheme follow directly from the condition
(\ref{wfr_cond1}) and its complex-conjugate version. We find
\begin{align}\label{eq:deltaZL}
  \frac{\delta Z_{ij}^{L}}{2} &= \frac{m_{d_i}^{(0)*} \Sigma^{RL}_{ij} +
    m_{d_j}^{(0)} \Sigma^{LR}_{ij}} {|m_{d_j}^{(0)}|^2 - |m_{d_i}^{(0)}|^2 }
  \qquad\qquad\mbox{for $i\neq j$.} \\ \label{eq:deltaZR} 
 \frac{ \delta Z_{ij}^{R}}{2} &=
  \frac{m_{d_i}^{(0)} \Sigma^{LR}_{ij} + m_{d_j}^{0*} \Sigma^{RL}_{ij} }{
    |m_{d_j}^{(0)}|^2 - |m_{d_i}^{(0)}|^2 } 
  \qquad\qquad  \; \;\mbox{for $i\neq j$.}
\end{align}
From these formulae it is obvious that the counterterms $\delta
Z^{L,R}_{ij}$ are $\tan\beta$-enhanced. However, the strong hierarchy of
the quark masses implies that $\delta Z_{ij}^{R}$ is always suppressed
by a small ratio of masses whereas $\delta Z_{ij}^{L}$ is not.

We want to stress that in the expression for $\Sigma^{RL}_{ij}$ in
\eq{eq:FCSelf} the momenta of the external quarks are neglected. As a
consequence self-energies in external quark lines are subtracted by the
counterterms $\delta Z^{L,R}_{ij}$ only up to terms suppressed by the small
ratio $m_{d_i}/M_{SUSY}$. Therefore in calculations where higher order terms
in the momentum expansion are relevant one has to take into account the
corresponding one-particle-reducible diagrams explicitly. One example for such
a process is $b\rightarrow s\gamma$.

\begin{nfigure}{t}
\centering
\includegraphics[width=15cm]{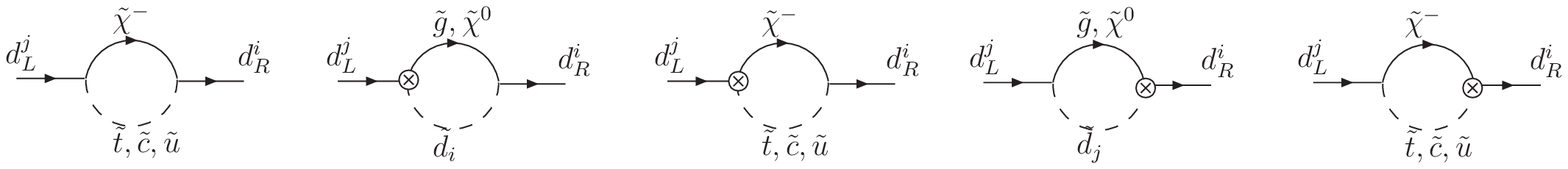}
\caption{ Higher-order $\tan\beta$-enhanced contributions to 
 $\Sigma_{ij}^{RL}$.}
\label{WFR_cond}
\end{nfigure}
Up to now we have considered the flavour-changing self-energies only at the
one-loop level. Are there also higher loop contributions which are
$\tan\beta$-enhanced? In the flavour-conserving case such contributions stem
from insertions of the counterterm $\delta y_b$ into the self-energy diagrams
and are already included in \eq{eq:FCSelf}. To study the new flavour-changing
effects let us now consider self-energy diagrams with wave-function
counterterms $\delta Z^L_{ij}$ and $\delta Z^R_{ij}$ at vertices involving a
gluino, a chargino, or a neutralino. These diagrams 
generate further contributions to $\Sigma^{RL}_{ij}$ (see \fig{WFR_cond}).
The resulting diagrams are $\tan\beta$-enhanced and of the same order in the
Wolfenstein parameter $\lambda$ as the original flavour-changing chargino
diagram. Formula (\ref{eq:FCSelf}) for $\Sigma^{RL}_{ij}$ is then generalised
to all orders in perturbation theory as
\begin{equation}\label{eq:45}
   \Sigma^{RL}_{ij}(\delta Z^L_{ij},\delta Z^R_{ij})= 
    V^{(0)*}_{ti}V^{(0)}_{tj}m_{d_i}^{(0)}\epsfc\tan\beta
     +\frac{\delta Z^L_{ij}}{2}m_{d_i}^{(0)}\epsilon_i\tan\beta - 
   \frac{\delta Z^R_{ij}}{2}m_{d_j}^{(0)}\epsilon_j\tan\beta.
\end{equation}
In writing $V_{ij}^{(0)}$ we have anticipated that the CKM elements will
obtain $\tan\beta$-enhanced counterterms which then also should be included
into the self-energies. Replacing $\Sigma^{RL}_{ij}$ and $\Sigma^{LR}_{ij}$ in
\eqsand{eq:deltaZL}{eq:deltaZR} by $\Sigma^{RL}_{ij}(\delta Z^L_{ij},\delta
Z^R_{ij})$ and $\Sigma^{LR}_{ij}(\delta Z^L_{ij},\delta Z^R_{ij})$ gives us
equations for the determination of the wave-function counterterms which are
valid to all orders in the perturbative expansion. Again, they can be solved
either order-by-order through explicit resummation or simply by solving the
coupled equations for the resummed counterterms $\delta Z^{L,R}_{ij}$
obtained by inserting \eq{eq:45} into \eqsand{eq:deltaZL}{eq:deltaZR}. For
$i=d,s$ we find to leading order in $m_{d_i}/m_b$:
\begin{align}\label{eq:dzlr}
\frac{\delta Z^{L}_{bi} }{2}  = - \frac{\delta Z^{L*}_{ib}}{2}
   &= -\frac{\epsfc \tan\beta}{1+\epsilon_b\tan\beta} 
      V_{tb}^{(0)*} V_{ti}^{(0)} ,\\
\frac{\delta Z^{R}_{bi} }{2} = - \frac{\delta Z^{R*}_{ib}}{2}
   &= -\frac{m_{d_i}}{m_b} 
  \left[   \frac{ \epsfc \tan\beta}{1+\epsilon_b\tan\beta} + 
  \frac{  \epsfc^*\tan\beta } {(1+\epsilon^*_i \tan\beta)} 
   \right] V_{tb}^{(0)*} V_{ti}^{(0)}. \label{eq:dzlr2}
\end{align}
The elements of $\delta Z^{L,R}_{ij}$ which do not involve the third
generation vanish.

Now we can renormalise the CKM matrix with the help of the resummed
left-handed wave-function counterterms, using the prescription of 
Ref.~\cite{Denner:1990yz} and neglecting the up-type counterterms:
\begin{equation}\label{ckm_ct_denner}
\delta V_{ij} = -\sum_{k} V^{(0)}_{ik} \frac{\delta Z^{L}_{kj}}{2}
\end{equation}
On the right-hand side we have again replaced $V_{ik}$ by $V^{(0)}_{ik}$ 
to properly account for the enhanced higher-order effects. 

The resummed CKM counter-terms fixed by this condition exactly cancel
the effect of the field renormalisation of the down-type quarks in their
couplings to the W boson so that only the tree-level coupling survives.
We can now insert \eq{eq:dzlr} into \eq{ckm_ct_denner} and (using
$V^{(0)}_{ij}=V_{ij}+\delta V_{ij}$) solve for $\delta V_{ij}$.  We
obtain the same relation between $V^{(0)}_{ij}$ and
$V_{ij}$ as found in \eq{ckm_renorm} with the method of the previous
section.  We may now express $\delta Z_{bi}^{L,R}$ in terms of the
physical CKM elements: Inserting \eq{ckm_renorm} into
\eqsand{eq:dzlr}{eq:dzlr2} gives 
\begin{align}\label{eq:dzphys}
\frac{\delta Z^{L}_{bi} }{2}  = - \frac{\delta Z^{L*}_{ib}}{2}
   &= - V_{tb}^* V_{ti} \frac{\epsfc \tan\beta}
                             { 1+(\epsilon_b - \epsfc)\tan\beta }  
      ,\\
\frac{\delta Z^{R}_{bi} }{2} = - \frac{\delta Z^{R*}_{ib}}{2}
   &= - V_{tb}^* V_{ti}
 \frac{m_{d_i}}{m_b} 
  \left[   \frac{ \epsfc \tan\beta}{1+\epsilon_b\tan\beta} + 
  \frac{  \epsfc^*\tan\beta } {(1+\epsilon^*_i \tan\beta)} 
   \right] 
      \frac{ 1 + \epsilon_b\tan\beta }{ 
             1+(\epsilon_b - \epsfc)\tan\beta }.
\end{align}
The renormalisation of the CKM matrix beyond the decoupling limit has also
been studied in the second chapter of Ref.~\cite{bcrs}, where an iterative
procedure has been used to incorporate the $\tan\beta$-enhanced higher-order
corrections.  We find that our unitary transformations in
\eqsand{wf_ct1}{wf_ct2} are formally equivalent to this procedure.  Our result
in \eq{eq:dzphys} is the analytic expression for the limit to which the
iterative calculation of Ref.~\cite{bcrs} converges. 


To summarise, in the previous section we found $\tan\beta$-enhanced
$b\to s$ ($b\to d$) transitions from self-energy insertions into
external legs of Feynman diagrams. In the approach used in this section
these self-energy insertions are absorbed into the wave-function
counterterms. 

\boldmath
\subsection{Formulation of Feynman rules for the large-$\tan\beta$ scenario}
\unboldmath
We are now in a position to study the influence of $\tan\beta$-enhanced flavour transitions on MSSM vertices by means of
the counterterms defined above. In particular, we can give Feynman rules
for the large-$\tan\beta$ framework in which the enhanced loop
corrections are included and resummed to all orders.

First of all, as already stated above, we have chosen a renormalisation
scheme such that the standard-model vertices remain unaffected by
enhanced corrections. In the couplings of quarks to the neutral gauge
bosons, the wave-function counterterms drop out by means of their
antihermiticity. The W boson couplings are indeed affected by the field
renormalisation but the renormalised CKM matrix is defined such that the
coupling is given only by a physical matrix element $V_{ij}$. As an
example, the coupling of the W to top- and strange-quark reads
\begin{equation}
-\frac{ig}{\sqrt 2}\gamma_\mu P_L \left( V_{ts} + \delta V_{ts} + V_{tb}\frac{\delta Z^L_{bs}}{2}  \right) = -\frac{ig}{\sqrt 2}\gamma_\mu P_L V_{ts}.
\end{equation}
 
Since we renormalise only the quark fields and not their superpartners,
we cannot expect that the SUSY equivalents of standard-model vertices
follow the same pattern. This is inevitable since the flavour-changing
effects which we want to include in our Feynman rules arise from the
SUSY-breaking sector (see Sect.~\ref{flavour_mixing}). The most
striking example for this property is the misalignment between the
flavour-diagonal quark-gluon vertices and the quark-squark-gluino
couplings which receive flavour-changing contributions. From the unitary
transformations in \eq{wf_ct1} we can read off e.g.\ 
\begin{equation}
\mathcal L \; \supset \;  -i \sqrt{2} g_s T^a  
                          \, \tilde b_L^* \tilde g^a b_L^{(0)}  
    \; = \;  
 -i \sqrt{2} g_s T^a  \, \tilde b_L^* \tilde g^a 
  \left( b_L + \frac{\delta Z^L_{bs}}{2} s_L + 
         \frac{\delta Z^L_{bd}}{2} d_L\right) ,
\end{equation}
which implies the existence of a flavour-violating gluino coupling to a
sbottom and a down- (strange-) quark via the $\tan\beta$-enhanced
counterterm $\delta Z^L_{bd(s)}$. In the approach of section
\ref{self_en_approach}, these corrections would arise via
$\tan\beta$-enhanced flavour-changing self-energies in the external quark line.

In addition to the gluino couplings, also chargino-, neutralino- and
Higgs-couplings to quarks are affected by $\tan\beta$-enhanced
corrections. Moreover, the bare CKM factors in various flavour-changing
squark couplings (not involving quarks) have to be related to their
physical counterparts by means of \eq{ckm_renorm}.  We summarise all
these effects in explicit Feynman rules for the large-$\tan\beta$
scenario in Appendix~\ref{feynman}. These rules are useful for
\begin{itemize}
\item calculations 
  of low-energy processes involving virtual
  SUSY particles and 
\item calculations in collider physics with external SUSY particles.
\end{itemize}
As an example, we give here the result for a flavour-changing gluino
decay. In the approximation $m_b/M_\text{SUSY}\approx 0$, the decay rate
of $\tilde g\rightarrow \tilde b_i \,b$ is at tree-level
\begin{equation}
  \Gamma(\tilde g\rightarrow \tilde b_i \,b) = 
  \frac{\alpha_s}{8\pi}(m_{\tilde g}^2-m_{\tilde b_i}^2)^2.
\end{equation}
For the flavour-violating decay $\tilde g\rightarrow \tilde b_i \,s$, we
find
\begin{equation}
  \frac{\Gamma(\tilde g\rightarrow \tilde b_i \,s)}{\Gamma(\tilde
    g\rightarrow \tilde b_i \,b)} = 
  \left| \frac{\delta Z^L_{bs}}{2} \tilde R^b_{i1}\right|^2 + 
  \left| \frac{\delta Z^R_{bs}}{2} \tilde R^b_{i2}\right|^2 \approx
  \left| \frac{\delta Z^L_{bs}}{2} \tilde R^b_{i1}\right|^2.
\end{equation}
Numerically, this ratio is given by
\begin{equation}
  \left| \frac{\epsfc
      \tan\beta}{1+(\epsilon_b-\epsfc)\tan\beta}\right|^2 \, 
  \left| V_{tb} V_{ts}\right|^2 \, 
  \left| \tilde R^b_{i1}\right|^2   \sim \mathcal O (10^{-4}).
\end{equation}

\section{Phenomenology: FCNC processes}\label{fcnc}

With the knowledge from the previous chapters one can now study the
effects of $\tan\beta$-enhanced SUSY corrections in FCNC processes. It
is well known that even under the MFV assumption, supersymmetric
contributions to FCNC observables in B physics can be sizeable if
$\tan\beta$ is large. The most prominent example is the rare decay $B_s
\to \mu^+\mu^-$, in which the supersymmetric contribution can largely
exceed the Standard-Model rate and can saturate the experimental
bound \cite{Huang,Babu:1999hn,Isidori:2001fv,ddn,bcrs,gjnt}.  In
this section we apply the effective Feynman rules for the
large-$\tan\beta$ scenario listed in Appendix~\ref{feynman} to FCNC
processes.

Most importantly, in this scenario flavour-changing transitions are no
longer mediated exclusively by W bosons, charged Higgs particles and
charginos but also by neutral Higgs particles, gluinos and neutralinos.
For the case of the neutral Higgs bosons, this fact has been realised
first in the framework of the effective 2HDM valid for $\msusy \gg v$
\cite{Hamzaoui:1998nu}.  With our effective Feynman rules, we can on the
one hand calculate the neutral Higgs contributions to FCNC processes for
the case $\msusy\sim\mathcal O(v)$ and on the other hand derive
contributions from other neutral virtual particles, where we will
restrict the discussion to gluinos and neglect the weakly interacting
neutralinos.

Since all the flavour-violating neutral couplings are generated by
$\tan\beta$-enhanced flavour-changing self-energies (or equivalently by
the counterterms $\delta Z^L_{bi}$ and $\delta Z^R_{bi}$ ($i=d,s$) from
Sect.~\ref{ct_approach}), their numerical importance crucially depends
on the parameter $\epsfc \tan\beta$. Since $\delta Z^R_{bi}$ is
suppressed by a small ratio of quark masses, the most important new
contributions are proportional to $\delta Z^L_{bi}$ in \eq{eq:dzphys} 
and thus to the parameter combination
\begin{equation}
    \frac{\epsfc \tan\beta}{1+(\epsilon_b-\epsfc)\tan\beta}.
    \label{eq:ParaCombi}
\end{equation}
It is thus useful to have a first estimate of the size of this
parameter. For this purpose, we neglect the weak contributions to
$\epsilon_b$ and $\epsfc$, focus on the non-decoupling part of
expressions (\ref{eq:Self4}) and (\ref{epsilonfc}) for
$\epsilon^{\tilde{g}}_b$ and $\epsfc$ and set all the SUSY mass
parameters as well as $|\mu|$ and $|A_t|$ equal to a single mass scale
$\msusy$. In this case, the mass dependence drops out and we find
\begin{eqnarray}
  |\epsfc\tan\beta| &=& \frac{y_t(\msusy)^2}{32\pi^2}\tan\beta, \\
  |(\epsilon_b-\epsfc)\tan\beta| &=& 
  |\epsilon_b^{\tilde{g}} \tan\beta | = 
  \frac{\alpha_s(\msusy)}{3\pi} \tan\beta .
\end{eqnarray}
For $\tan\beta=50$ and $\msusy = 500$ GeV, we find typical numerical values of 
\begin{equation}\label{epsfcnum}
|\epsfc\tan\beta| \sim 0.12,\hspace{2cm} |(\epsilon_b-\epsfc)\tan\beta|\sim 0.5 \, .
\end{equation}
Taking $\mu$ real here the 
parameter combination in \eq{eq:ParaCombi} evaluates to
\begin{eqnarray}
  \left|\frac{\epsfc \tan\beta}{1+(\epsilon_b-\epsfc)\tan\beta}\right|&\sim& 0.08,\hspace{0.5cm}
  \textrm{for positive }\mu,\\
  \left|\frac{\epsfc \tan\beta}{1+(\epsilon_b-\epsfc)\tan\beta}\right|&\sim& 0.24,\hspace{0.5cm}
  \textrm{for negative }\mu.   
\end{eqnarray}
Values larger than this for $\epsfc$ and thus for the combination
(\ref{eq:ParaCombi}) occur if $|A_t|$ is significantly larger than the
masses of stops and charginos. If one requires $|A_t|\lesssim
3m_{\tilde{q}}$ (where $m_{\tilde{q}}$ is an average squark mass) to
avoid colour-breaking minima \cite{Casas:1995pd,Casas:1996de},
$\epsfc\tan\beta$ gets constrained to
$|\epsfc\tan\beta|_{\textrm{max}}\sim 0.4$. Experimentally, the size of
$A_t$ is further limited by $\mathcal B(\overline B\rightarrow
X_s\gamma)$ via the $\tan\beta$-enhanced chargino contribution to this
process. However, when the complex phase of $A_t$ is taken into account,
this bound is much weaker \cite{Pokorski:1999hz}. Moreover, this bound
from $\mathcal B(\overline B\rightarrow X_s\gamma)$ may shift when the
gluino contribution, which a priori is expected to be of order
$|\epsfc\tan\beta|$ times the chargino contribution, is taken into
account.

\boldmath
\subsection{The effective $|\Delta B|=1$ Hamiltonian}\label{sec:deltabeq1}
\unboldmath%
Weak $|\Delta B|=|\Delta S|=1$ decays are usually described by an
effective Hamiltonian
\begin{equation}
  \mathcal H_\text{eff} = -\frac{4 G_F}{\sqrt 2}V_{tb}V_{ts}^* 
  \sum_i C_i \mathcal O_i + h.c.
\end{equation}
In the SM the operator basis for radiative and hadronic B decays consists of the four quark operators
\begin{align}
  \mathcal O_1 &= (\bar s_\alpha \gamma_\mu P_L c_\beta)(\bar c_\beta
  \gamma^\mu P_L b_\alpha)
  &\mathcal O_2 &= (\bar s_\alpha \gamma_\mu P_L c_\alpha)(\bar c_\beta \gamma^\mu P_L b_\beta)\\
  \mathcal O_3 &= (\bar s_\alpha \gamma_\mu P_L b_\alpha)\sum_q(\bar
  q_\beta \gamma^\mu P_L q_\beta)
  &\mathcal O_4 &= (\bar s_\alpha \gamma_\mu P_L b_\beta)\sum_q(\bar q_\beta \gamma^\mu P_L q_\alpha)\\
  \mathcal O_5 &= (\bar s_\alpha \gamma_\mu P_L b_\alpha)\sum_q(\bar
  q_\beta \gamma^\mu P_R q_\beta) &\mathcal O_6 &= (\bar s_\alpha
  \gamma_\mu P_L b_\beta)\sum_q(\bar q_\beta \gamma^\mu P_R q_\alpha)
\end{align}
and the magnetic and chromo-magnetic operators
\begin{align}
  \mathcal O_7 &= \frac{e}{16\pi^2}\overline m_b (\bar s
  \sigma^{\mu\nu}P_R b)F_{\mu\nu} &\mathcal O_8 &=
  \frac{g_s}{16\pi^2}\overline m_b (\bar s \sigma^{\mu\nu} T^a P_R b)
  G^a_{\mu\nu}.
\end{align}
In the MSSM with large $\tan\beta$ flavour-changing couplings of the
neutral Higgs bosons to the down-type quarks are generated. For this
reason the operator basis has to be extended to include four quark
operators with scalar, pseudoscalar and tensor structure, namely
\begin{align}
  \mathcal O_{11}^q &= (\bar s_\alpha P_R b_\alpha)(\bar q_\beta P_L
  q_\beta)
  &\mathcal O_{12}^q &= 
          (\bar s_\alpha P_R b_\beta)(\bar q_\beta P_L q_\alpha)\\
  \mathcal O_{13}^q &= (\bar s_\alpha P_R b_\alpha)(\bar q_\beta P_R
  q_\beta)
  &\mathcal O_{14}^q &= 
          (\bar s_\alpha P_R b_\beta)(\bar q_\beta P_R q_\alpha)\\
  \mathcal O_{15}^q &= (\bar s_\alpha \sigma^{\mu\nu}P_R b_\alpha)(\bar
  q_\beta \sigma_{\mu\nu}P_R q_\beta) &\mathcal O_{16}^q &= (\bar
  s_\alpha \sigma^{\mu\nu}P_R b_\beta)(\bar q_\beta \sigma_{\mu\nu}P_R
  q_\alpha).
\end{align}
Note that the operators $\mathcal O_{11}^q \ldots\mathcal O_{16}^q$ are
not linearly independent for $q=b$ or $q=s$. In theses cases $\mathcal
O_{15}^q$ and $\mathcal O_{16}^q$ can be expressed as linear
combinations of the remaining operators using Fierz identities.  We have
checked that these operators have a negligible impact on radiative
decays. The same feature was found for hadronic two-body decays in
Ref.~\cite{Beneke:2009eb}.  The effective Hamiltonian for $|\Delta
B|=|\Delta D|=1$ processes can be found from the $|\Delta B|=|\Delta
S|=1$ one by the replacement $s\to d$.

Let us now have a look at SUSY contributions to the Wilson coefficients
of the operators $\mathcal O_7$ and $\mathcal O_8$: In the SM $\mathcal
O_{7,8}$ involves a chirality flip in the external $b$-quark leg so that
$C_{7,8}$ is proportional to $m_b\propto \cos\beta$.  Therefore SUSY
contributions can be $\tan\beta$-enhanced with respect to the SM
amplitude if the chirality flip stems from a factor of $y_b$ in the
loop.  At the one-loop level the well-known contributions growing with
$\tan\beta$ are loops with charginos and up-type squarks. In this
context often also the diagrams involving a charged Higgs boson and a
top quark are discussed. These contributions are not
$\tan\beta$-enhanced due to the $\cos\beta$-suppression of the
charged-Higgs coupling to the right-handed top. Since this coupling has
vertex-corrections proportional to $\sin\beta$, such diagrams require a
different treatment and are not discussed here. They have been studied
by various authors either in an effective-field-theory approach
\cite{Carena:2000uj,Degrassi:2000qf,Demir:2001yz,Gomez:2006uv} or in an
explicit two-loop calculation \cite{Degrassi:2006eh}. Here we firstly
focus on the chargino contribution. Using our effective Feynman rules we
find
\begin{align}
  C_{7,8,\widetilde\chi^\pm} =&
  \frac{1}{\cos\beta(1+\epsilon_b^*\tan\beta)}\sum_{a=1,2} \left\{
    \frac{\widetilde{U}_{a2}\widetilde{V}_{a1}M_W}{\sqrt{2}
      m_{\widetilde\chi^\pm_a}} \left[ K^* f_{1,2}(x_{\tilde{q}\,
        \widetilde\chi^\pm_a})- c_{\tilde t}^2 \, f_{1,2}
      (x_{{\tilde{t}_1}\,\widetilde\chi^\pm_a}) \rt.\rt. \nn %
      & \lt. \lt.  -s_{\tilde t}^2 \,f_{1,2}
      (x_{{\tilde{t}_2}\,\widetilde\chi^\pm_a}) \right] + s_{\tilde
      t}\,c_{\tilde t} \, e^{-i\phi_{\tilde t}} \frac{\widetilde
      {U}_{a2}\, \widetilde{V}_{a2} \, m_t }{2\sin\beta
      \,m_{\widetilde\chi^\pm_a}} \left[ f_{1,2}
      (x_{\tilde{t}_1\,\widetilde\chi^\pm_a}) - f_{1,2}
      (x_{\tilde{t}_2\,\widetilde\chi^\pm_a})\right] \right \}.
\end{align}
with
\begin{equation}
   s_{\tilde{t}}=\sin\tilde{\theta}_t,\qquad c_{\tilde{t}}=\cos\tilde{\theta}_t,\qquad x_{ij}=m_i/m_j.
\end{equation}
All loop functions are given in appendix \ref{loopfunctions}. Again we have assumed that the squarks of the first two generations are
degenerate in mass and denoted their common mass by $m_{\tilde{q}}$.
Our result differs from the one in \cite{Degrassi:2000qf} only by a
factor of $K^*$ (defined in \eq{ckm_renorm}) in the numerically small up
and charm squark contribution. The stop contribution remains unaffected
because the corrections from the wave function and the CKM counterterm
cancel each other.
 
Besides the well-known chargino and charged-Higgs diagrams, there are
now $\tan\beta$-enhanced gluino-sbottom diagrams contributing to $C_7$
and $C_8$ (\fig{bsgamma_gluino}), which have never been discussed
before in the context of minimal flavour-violation at large
$\tan\beta$. Like the chargino diagrams these contributions vanish for
$\msusy \gg v$, but can be computed with proper resummation of the
enhanced corrections within our framework.

\begin{nfigure}
\centering
\includegraphics[height=3.7cm]{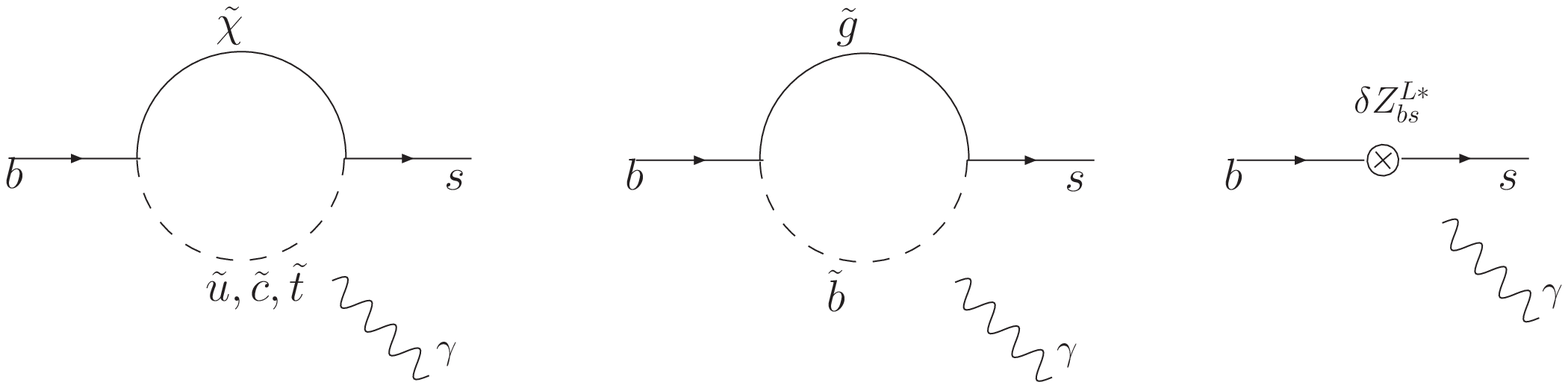}\qquad
\caption{Gluino and chargino diagrams contributing to $C_7$. The photon can couple to any particle except for the gluino. The contributions to $C_8$ are found by replacing the photon by a gluon (which can also couple to the gluino).  }
\label{bsgamma_gluino}
\end{nfigure}

The $\tan\beta$-enhanced parts read
\begin{align}\label{c7g}
  C_{7,\tilde g} =& \frac{\sqrt{2}}{4G_F}\frac{C_F g_s^2
    \mu\tan\beta}{3m_{\tilde{g}}(m_{\tilde{b}_1}^2-m_{\tilde{b}_2}^2)}
  \frac{\epsfc^*\tan\beta}{
    (1+\epsilon_b^*\tan\beta)\left(1+(\epsilon_b^*-\epsfc^*)\tan\beta\right)}
  \left( f_2(x_{\tilde b_1 \tilde g}) - f_2(x_{\tilde b_2 \tilde
      g}\right),\\ \label{c8g} C_{8,\tilde g} =
  &-\frac{\sqrt{2}}{4G_F}\frac{g_s^2\mu\tan\beta}{
    m_{\tilde{g}}(m_{\tilde{b}_1}^2-m_{\tilde{b}_2}^2)}
  \frac{\epsfc^*\tan\beta}{(1+\epsilon_b^*\tan\beta)
    \left(1+(\epsilon_b^*-\epsfc^*)\tan\beta\right)}\nonumber\\
  &\times\left[C_F\left(f_2(x_{\tilde b_1 \tilde g}) - f_2(x_{\tilde b_2
        \tilde g})\right) +C_A\left(f_3(x_{\tilde b_1 \tilde g}) -
      f_3(x_{\tilde b_2 \tilde g})\right)\right].
\end{align}
The arguments of the loop functions are again given by $x_{ab}=m_a^2/m_b^2$, the colour factors are $C_F=4/3$ and $C_A=3$. We remark that the diagram with a gluino and a strange squark in the loop has been neglected because its amplitude is suppressed by the strange-quark mass. To have a rough estimate of the size of $C_{7,8,\tilde{g}}$ compared to $C_{7,8,\widetilde{\chi}^{\pm}}$ we again set all SUSY masses (including $|\mu|$ and $|A_t|$) to the same value $\msusy$. In this case we find
\begin{equation}
  \eta_7=\left|\frac{C_{7,\tilde{g}}}{ 
      C_{7,\widetilde{\chi}^{\pm}}}\right|=\frac{8}{15}\frac{g_s^2}{y_t^2}
  \frac{|\epsfc^*\tan\beta|}{|1+(\epsilon_b^*-\epsfc^*)\tan\beta|},
  \qquad
  \eta_8=\left|\frac{C_{8,\tilde{g}}}{ 
      C_{8,\widetilde{\chi}^{\pm}}}\right|=\frac{10}{3}\frac{g_s^2}{y_t^2}
  \frac{|\epsfc^*\tan\beta|}{|1+(\epsilon_b^*-\epsfc^*)\tan\beta|}.
  \label{eta78}
\end{equation}
Using our estimates for expression (\ref{eq:ParaCombi}) we find
$\eta_7\sim 0.07$ and $\eta_8\sim 0.42$ for positive values of $\mu$ and
$\eta_7\sim 0.2$ and $\eta_8\sim 1.3$ for negative values of $\mu$. It
follows that the impact of the gluino contribution on $C_7$ is small
(especially for positive $\mu$) whereas the contribution to $C_8$ can be
sizeable.  Above we argued that the value of $|\epsfc\tan\beta|$ can be
increased up to $|\epsfc\tan\beta|\sim 0.4$ if we choose large values
for $|A_t|$. Of course, the size of $C_{7,8,\tilde{g}}$ gets larger for
increasing values of $|\epsfc\tan\beta|$. Note, however, that
$C_{7,8,\widetilde{\chi}^\pm}$ is proportional to $A_t$ and thus the
ratio $\eta_{7,8}$, i.e. the relative importance of the gluino
contribution, is essentially unaffected.  On the other hand, the gluino
contribution grows with increasing $|\mu|$ whereas the chargino
contribution decreases because it decouples with the chargino mass.
Therefore for large values of $|\mu|$ the gluino contribution becomes
more important. We will perform a more detailed numerical study of the
new coefficients $C_{7,\tilde g}$ and $C_{8,\tilde g}$ in section
\ref{numerics}.

Replacing in \fig{bsgamma_gluino} the gluino by a neutralino, we find $\tan\beta$-enhanced neutralino 
contributions to the (chromo-)magnetic operators. Their analytic expression reads
\begin{equation}\label{c78n}
  C_{7,\tilde \chi^0} = -\frac{\sqrt 2}{4 G_F} \sum_{i,m} \frac{\epsfc^*\tan\beta}{6 m_{\tilde\chi_m^0} m_b  \left(1+(\epsilon_b^*-\epsfc^*)\tan\beta\right)}  X^{L*}_{im} X^{R}_{im} f_2(x_{\tilde b_i \tilde\chi^0_m}) \quad , \quad
  C_{8,\tilde \chi^0} = 1/e_d \, C_{7,\tilde\chi^0}
\end{equation}
with the neutralino-quark-squark couplings
\begin{equation}
  X^{L}_{im} = \sqrt{2} \widetilde{R}^{b}_{i1}\left( \frac{g}{2} \widetilde{N}^*_{m2} - \frac{g'}{6} \widetilde{N}^*_{m1} \right)  - y_{b}^{(0)} \widetilde{R}^{b}_{i2} \widetilde{N}^*_{m3} \quad , \quad 
  X^{R}_{im} = \frac{ \sqrt{2}}{3}g' \widetilde{R}^{b}_{i2} \widetilde{N}_{m1} + y_{b}^{(0)*} \widetilde{R}^{b}_{i1} \widetilde{N}_{m3}.
\end{equation}
In our convention, $e_d=-1/3$ is the charge of the down-type (s)quarks.
The bare Yukawa coupling $y_{b}^{(0)}$ is determined as explained in
section \ref{sect:smr}. We remark that in the product $X^{L*}_{im}
X^{R}_{im}$, additional factors of $\tan\beta$ from sbottom-mixing and
from $y_{b}^{(0)}$ are hidden, but nevertheless we find the neutralino
contributions to be numerically small compared to their counterparts
from chargino and gluino diagrams.

Another one-loop contribution to $C_{7,8}$, stemming from virtual
neutral Higgs-bosons, has been presented in \cite{D'Ambrosio:2002ex} in
the effective-Lagrangian approach with vanishing SUSY CP-phases. In a
full diagrammatic calculation, we find for these coefficients
\begin{equation}
C_{7,H^0} = -\frac{\epsfc^*\tan\beta}{1+(\epsilon_b^*-\epsfc^*)\tan\beta} \,
  \frac{m_b^2\tan^2\beta}{36|1+\epsilon_b \tan\beta|^2 m_{A^0}^2} \quad ,\quad
C_{8,H^0} =  \frac{C_7^{H^0}}{e_d}.
\end{equation}
In the decoupling limit, setting all SUSY phases to zero, this agrees
with \cite{D'Ambrosio:2002ex} up to the factor $1/e_d$. Compared to the
other contributions from SM and MSSM particles, corrections from
neutral-Higgs diagrams to $C_{7,8}$ are at most in the few-percent
range.

In the following, let us leave the magnetic and chromomagnetic operators
and discuss the remaining parts of the effective Hamiltonian. For the
QCD-penguin operators $\mathcal O_{3-6}$, we find contributions from
gluino and neutralino loops to be small because of destructive
interference of the two occurring internal squark flavours $\tilde b$
and $\tilde s$. This is a remarkable difference to chargino loops, where
this GIM-like cancellation is rather inefficient between the up-type
squarks due to their very different Yukawa couplings.  Furthermore, the
usual power-suppression $m_b^2/\msusy^2$ is present and cannot be
alleviated by a factor of $\tan\beta$ from the loop since no chirality
flip is involved, in contrast to $\mathcal O_{7,8}$.

In the semileptonic decay $\bar B\rightarrow X_s\ell^+\ell^-$, two
semileptonic operators usually denoted by $\mathcal O_9$ and $\mathcal
O_{10}$ come into play. Chargino- and charged Higgs-diagrams
contributing to these operators have been evaluated in
\cite{Bobeth:2004jz} (we refer to this publication for the definition of
$\mathcal O_{9,10}$) and it has been found that the corrections to the
SM coefficients are small. Due to the GIM-like suppression described
above, we find gluino and neutralino corrections to be even smaller.

The charged leptonic B decays $B_{q}^+\rightarrow \ell^+ \nu_\ell$
$(q=d,s)$ are dominated by tree-level diagrams with $W$ boson, but may
receive sizeable contributions from charged-Higgs exchange in the MSSM
\cite{ip}. The charged Higgs boson couples to a right-handed $b$ quark
and (neglecting $y_d$ and $y_s$) the only effect of $\tan\beta$-enhanced
corrections stems from $K$ in \eq{ckm_renorm} and $\epsilon_b\tan\beta$
in the Yukawa coupling in \eq{eq:ResForm0}. The corresponding Feynman
rule is given in \eq{feynsp}.  The same remark applies to the other
charged-Higgs analyser $B\to D\tau \nu_\tau$ \cite{ntw,km}.

Their neutral counterparts $B_{q}^0\rightarrow \ell^+ \ell^-$ are
loop-mediated, with a dramatic impact of a large value of $\tan\beta$.
The phenomenologically most important decay in this class,
$B_s^0\rightarrow \mu^+ \mu^-$, is described by the effective
Hamiltonian
\begin{equation}
  \mathcal H_\text{eff} = -\frac{4G_F}{\sqrt{2}} V_{ts}^* V_{tb} 
  \sum_{i=A,S,P} C_i\mathcal O_i + h.c.
\end{equation}
with the operators
\begin{align}
\mathcal O_A &= (\bar s \gamma_\nu P_L b)(\bar\mu \gamma^\nu\gamma_5 \mu)\\
\mathcal O_S &= \overline{m}_b(\bar s  P_L b)(\bar\mu \mu)\\
\mathcal O_P &= \overline{m}_b(\bar s  P_L b)(\bar\mu\gamma_5 \mu).
\end{align}

At large $\tan\beta$, neutral Higgs exchange is known to be dominant
since it occurs at tree-level in the effective theory at the electroweak
scale \cite{Babu:1999hn}, contributing to $C_S$ and $C_P$.\footnote{The
  $\tan\beta$-enhancement was found in a diagrammatic one-loop
  calculation in Ref.~\cite{Huang}.} Making use of the flavour-changing
neutral Higgs couplings from Appendix~\ref{feynman}, we can generalise
the results in the literature to formulae which
\begin{itemize}
\item resum all $\tan\beta$-enhanced mass- and wave-function
  renormalisation effects
\item contain all possible complex phases from the SUSY breaking
  sector\\ and
\item do not resort to the decoupling limit $\msusy \gg v$.
\end{itemize}
Since the LHCb detector may soon precisely measure the $B_s \rightarrow
\mu^+ \mu^-$ branching fraction, an improved treatment of the SUSY
contribution to this decay is desirable now.  
With $m_{H^0}^2 = m_{A^0}^2$ in the
large-$\tan\beta$ limit, this Higgs-mediated
contribution reads\footnote{If $\tan\beta$ is small, Z-penguin and 
box diagrams become important. These contributions can be found in 
Ref.~\cite{Dedes:2008iw}.} 
\begin{align}
  C_S = - C_P =
  -
  \frac{\epsfc^*\tan\beta}{1+(\epsilon_b^*-\epsfc^*)\tan\beta}\,\frac{m_\mu
    \tan^2\beta}{(1+\epsilon_b^*\tan\beta)(1+\epsilon_\mu\tan\beta)2m_{A^0}^2
  }.
\end{align}
Here $\epsilon_\mu$ is the analogue of $\epsilon_b$ for the muon 
(see e.g.\ \cite{mmns,gjnt}).

\boldmath
\subsection{The effective $|\Delta B|=2$ Hamiltonian}
\unboldmath%
In order to study the effects of $\tan\beta$-enhanced flavour
transitions in $B-\bar B$ oscillations, we write the $\Delta B=2$
effective Hamiltonian as
\begin{equation}
  \mathcal H_\text{eff} = \frac{G_F^2 m_W^2}{16\pi^2} 
        (V_{tb}^* V_{tq})^2 \sum_{i} C_i\mathcal O_i
\end{equation}
with $q=d,s$. The dimension-six operators $\mathcal O_i$ are
\begin{align}
\mathcal O^{VLL} &= (\bar b \gamma_\mu P_L q)(\bar b \gamma^\mu P_L q),\\
\mathcal O^{LR}_1 &= (\bar b \gamma_\mu P_L q)(\bar b \gamma^\mu P_R q),\\
\mathcal O^{LR}_2 &= (\bar b  P_L q)(\bar b P_R q),\\
\mathcal O^{SLL}_1 &= (\bar b  P_L q)(\bar b P_L q),\\
\mathcal O^{SLL}_2 &= (\bar b  \sigma_{\mu\nu} P_L q)(\bar b \sigma^{\mu\nu} P_L q)
\end{align}
and $\mathcal O^{VRR}, \mathcal O^{SRR}_1,\mathcal O^{SRR}_2$ which are
obtained by replacing $P_L$ by $P_R$.

Various contributions to \bbm\ have been obtained
in the effective-theory approach in Refs.~\cite{Hamzaoui:1998nu,
  Buras:2001mb,Isidori:2001fv, Buras:2002wq,bcrs,gjnt}.  We specify to
\bbms, which involves numerically important contributions proportional
to $m_s$ \cite{Buras:2001mb}. The first type of contributions to the Wilson coefficients of these
operators which we want to consider are diagrams with neutral Higgs
exchange analogous to the $B_s\rightarrow \mu^+ \mu^-$ diagram in the
previous subsection.   With our Feynman rules we find
\begin{align}
  C_1^{SLL}=&-\frac{16\pi^2m_b^2\tan^2\beta}{\sqrt{2}G_F
    M_W^2}\cdot\frac{\epsfc^2\tan^2\beta}{(1+\epsilon_b\tan\beta)^2 
   \left(1+(\epsilon_b-\epsfc)\tan\beta
    \right)^2}\cdot \mathcal{F}_{-},\\
  C_2^{LR}=&-\frac{32\pi^2m_b m_s \tan^2\beta}{\sqrt{2}G_F
    M_W^2}\cdot\frac{|\epsfc\tan\beta|^2}
  {\left|1+\epsilon_b\tan\beta\right|^2\,
    \left|1+(\epsilon_b-\epsfc)\tan\beta\right|^2}
  \cdot\mathcal{F}_{+}\nonumber\\&\times \left[1+(1-e^{2i\phi})
    \frac{(\epsilon_b^*- \epsfc^*-\epsilon_s^*)\tan\beta}{
      1+\epsilon_s^*\tan\beta} \right] \label{c2lr}
\end{align}
\begin{equation}
  \textrm{with}\qquad\phi= \arg \left\{\epsfc\tan\beta\, 
     \left(1+(\epsilon_b^*-\epsfc^*)\tan\beta\right)\right\}.
\end{equation}
Up to terms suppressed by $\tan^{-1}\beta$, we obtain here
\begin{equation}
\mathcal{F}_{+} = \frac{2}{m_{A^0}^2} \quad , \quad  \mathcal{F}_{-} = 0.
\label{deffpm}
\end{equation}

The contribution from the operator $\mathcal O^{LR}_2 $ is thus
important despite its suppression by $m_s$ since $\mathcal{F}_{-}$
vanishes at large $\tan\beta$ \cite{Hamzaoui:1998nu}. Our result for
$C^{LR}_2$ involves the new term
\begin{equation}
  r=(1-e^{2i\phi})\frac{(\epsilon_b^*-\epsfc^*-\epsilon_s^*)\tan\beta}{1+\epsilon_s^*\tan\beta}. \label{defr}
\end{equation}
Obviously this correction factor $r$ disappears if all parameters are
real. It also vanishes if we go to the decoupling limit and choose all
squark mass terms to be equal because in this case we have
\begin{equation}
    \epsilon_s\to\epsilon_0,\qquad \epsilon_b\to\epsilon_0+\epsfc. 
\end{equation}
For this reason the $r$-term is absent in \cite{Hamzaoui:1998nu,
  Buras:2001mb,Isidori:2001fv, Buras:2002wq,bcrs,gjnt}. Beyond the
decoupling limit $r$ does not vanish even if we set all SUSY-breaking
mass terms to the same value because the squark masses are split due to
electro-weak symmetry breaking. However, this effect is tiny for $\mu>0$
where the correction factor $1/(1+\epsilon_b\tan\beta)$ to the Yukawa
coupling suppresses the off-diagonal element $X_{\tilde b} = -y_b^{(0)*}
v_u \mu$ in the sbottom mass matrix.  In this case we have $|r|\lesssim
0.01$. For $\mu<0$ the off-diagonal element $X_{\tilde{b}}$ is enhanced
and we have $|r|\lesssim 0.1$. Significantly larger values for $r$ can
be achieved if we allow the squark masses of the third generation to be
different from those of the first two generations.\footnote{It should be
  stressed that this is possible for the right-handed bilinear mass
  terms but not for the left-handed ones: In the super-CKM basis one has
  $\widetilde{m}_{d_L}^2 = V^{(0)\dagger}\widetilde{m}_{u_L}^2 V^{(0)}$
  and the naive MFV hypothesis of diagonal $\widetilde{m}_{d_L}^2$,
  $\widetilde{m}_{u_L}^2 $ matrices therefore implies
  $\widetilde{m}_{u_L,d_L}^2 \propto \mathds{1}$.}  In this case the new
term can be important for mixing-induced CP asymmetries, because
$|C_1^{SLL}|$ is much smaller than $|C_2^{LR}|$ (even after loop
corrections to ${\cal F}_-$ in \eq{deffpm} are included \cite{gjnt}) and
the imaginary part of $C_2^{LR}$ in \eq{c2lr} stems solely from $r$.  A
benchmark measurement of LHCb will be $A_{\rm CP}^{\rm mix}(\Bbar_s\to
(J/\psi \phi)_{CP\pm})$ which equals $\mp 0.04
\pm 0.01$ in the SM. In view of the smallness of this SM prediction the
new contribution involving $\text{Im}\, r$ should be taken into
consideration.  The same remark applies to the even smaller SM
prediction of the CP asymmetry in flavour-specific decays \cite{ln}.

With our large-$\tan\beta$ Feynman rules we have further investigated
the contributions to the $\Delta B=2$ Hamiltonian from box-diagrams with
virtual gluinos and down-type squarks depicted in \fig{gluinobox}.
\begin{nfigure}{t}
\centering
\includegraphics[height=3.5cm]{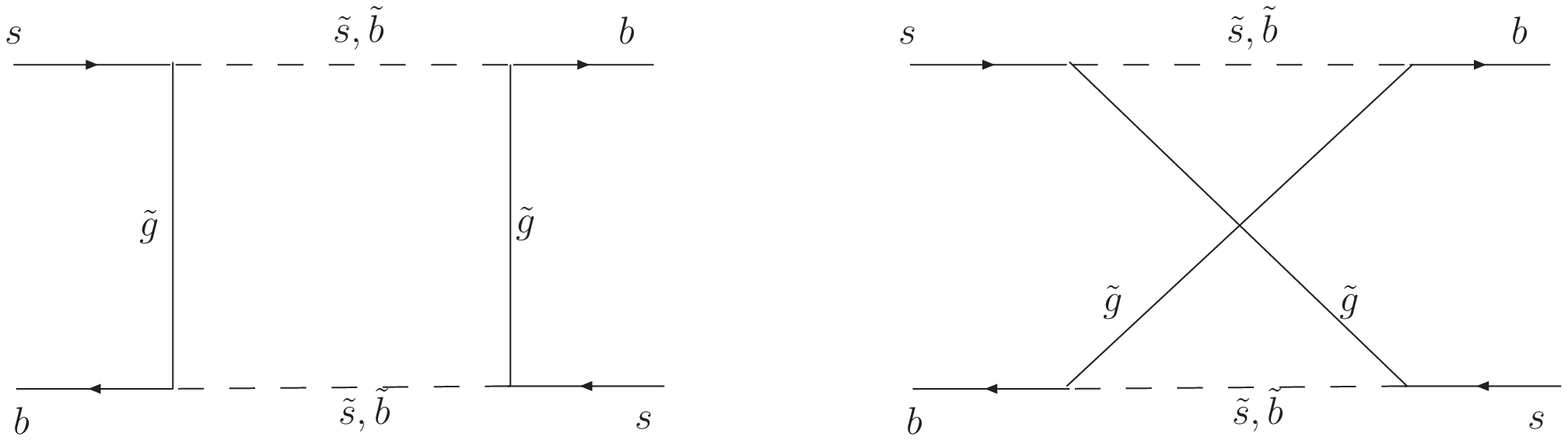}\qquad
\caption{Gluino-box diagrams contributing to the $\Delta B=2$
  Hamiltonian. Two further diagrams are obtained by $90^\circ$ rotations.
 }
\label{gluinobox}
\end{nfigure}
We find that contributions to $C^{LR}_{1,2}$, $C^{VRR}$
and $C^{SRR}_{1,2}$ are always proportional to powers of
$\delta Z^R_{bs}$, thus suppressed by $m_s/m_b$. Contributions to
$C^{VLL}$ and $C^{SLL}_{1,2}$ are proportional to
$(\delta Z^L_{bs})^2$, which is rather small as discussed at the
beginning of Sect.~\ref{fcnc}, and furthermore suffer from destructive
interference between the $\tilde s$ and $\tilde b$ contributions.  These
suppression effects render gluino contributions to the $\Delta B=2$
Hamiltonian numerically negligible compared to other supersymmetric
contributions like e.g.\ those from charginos or neutral Higgs bosons. The same statement holds for the neutralino box diagrams.

\section{\boldmath
         Numerical study of $C_{7,\tilde g}$ and $C_{8,\tilde g}$ 
         and implications for $\bar B^0 \to \phi K_S$ \unboldmath 
         }
\label{numerics}
We have argued in the previous sections that at large
$\tan\beta$ there can be potentially large contributions to the
coefficients of the (chromo-)magnetic $\Delta B=1$ operators $\mathcal
O_7$ and $\mathcal O_8$ from SUSY-QCD. In order to have a clearer
picture of this effect, we now present a numerical study of the
Wilson coefficients $C_7$ and $C_8$ and an application to the 
mixing-induced CP asymmetry $S_{\phi K_S}$.    

As a first step, we have performed a general scan over the MSSM
parameter space and calculated the absolute values and phases of the
various standard-model and supersymmetric contributions to both $C_7$
and $C_8$.
\begin{ntable}{b}
\centering
\begin{tabular}{|c|c|c|}\hline
 & min (GeV) & max (GeV)  \\ \hline
$\tilde m_{Q_L}$, $\tilde m_{u_R}$, $\tilde m_{d_R}$ &  250 & 1000 \\ \hline
$|A_t|, |A_b|$  & 100 & 1000 \\ \hline
$\mu$, $M_1$, $M_2$ & 200 & 1000 \\ \hline
$M_3$ & 300 & 1000\\ \hline
$m_{A^0}$ & 200 & 1000 \\ \hline
\end{tabular}
\caption{Scan ranges used for massive MSSM parameters.}
\label{scanranges}
\end{ntable}
Our ranges for the dimensionful MSSM parameters are given in
Tab.~\ref{scanranges}. We vary the phase of $A_t$ between $0$ and $2\pi$
and $\tan\beta$ between $40$ and $60$. In this section we further take
$\mu$ real and positive.  Only parameter points compatible with the
following constraints have been accepted:
\begin{itemize}
\item All squark masses are larger than $200$ GeV.
 \item The lightest supersymmetric particle (LSP) is charge- and color-neutral.
 \item The experimental $2\sigma$-bound on the lightest Higgs-boson mass
   is respected. 
 \item $\mathcal B(\overline B \rightarrow X_s\gamma)$ is in the 
       experimental $2\sigma$-range.
\end{itemize}
For the last constraint, $\mathcal B(\overline B \rightarrow X_s\gamma)$
has been calculated according to Eq.~(20) of
Ref.~\cite{Kagan:1998ym}.  This results in a severe limitation for large
values of $|A_t|$ since $\mathcal B(\overline B \rightarrow X_s\gamma)$
is dominated by $C_7$, which receives substantial SUSY corrections if
both $|A_t|$ and $\tan\beta$ are large \cite{Ciuchini:1998xy}. In view
of this fact, the question arises how a complex $A_t$ should be treated.
It is often possible to fine-tune its phase in such a way that the sum
of a very large SUSY correction to $C_7$ and the standard model value is
still compatible with the measurements of $\mathcal B(\overline B
\rightarrow X_s\gamma)$. We have decided to consider such a fine-tuning
as unnatural and thus impose another constraint on our scan points.
\begin{itemize} 
\item We reject all points yielding a SUSY correction $\left|
    C_7^\text{\scriptsize{SUSY}}(m_W) \right| >\left| C_7^\text{SM}(m_W)
  \right| \approx 0.22$
\end{itemize}

\begin{nfigure}{t}
\centering
\includegraphics[height=1.5cm]{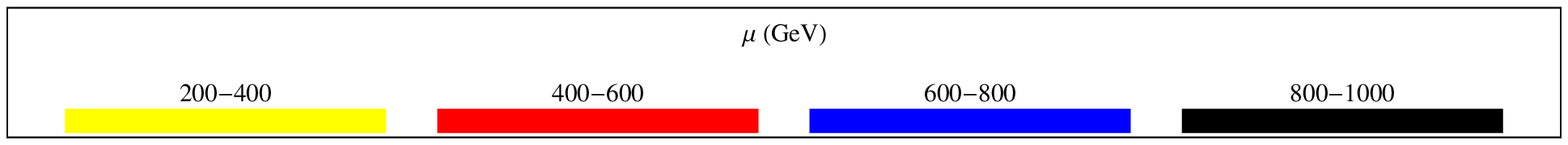}\\
\includegraphics[width=\textwidth]{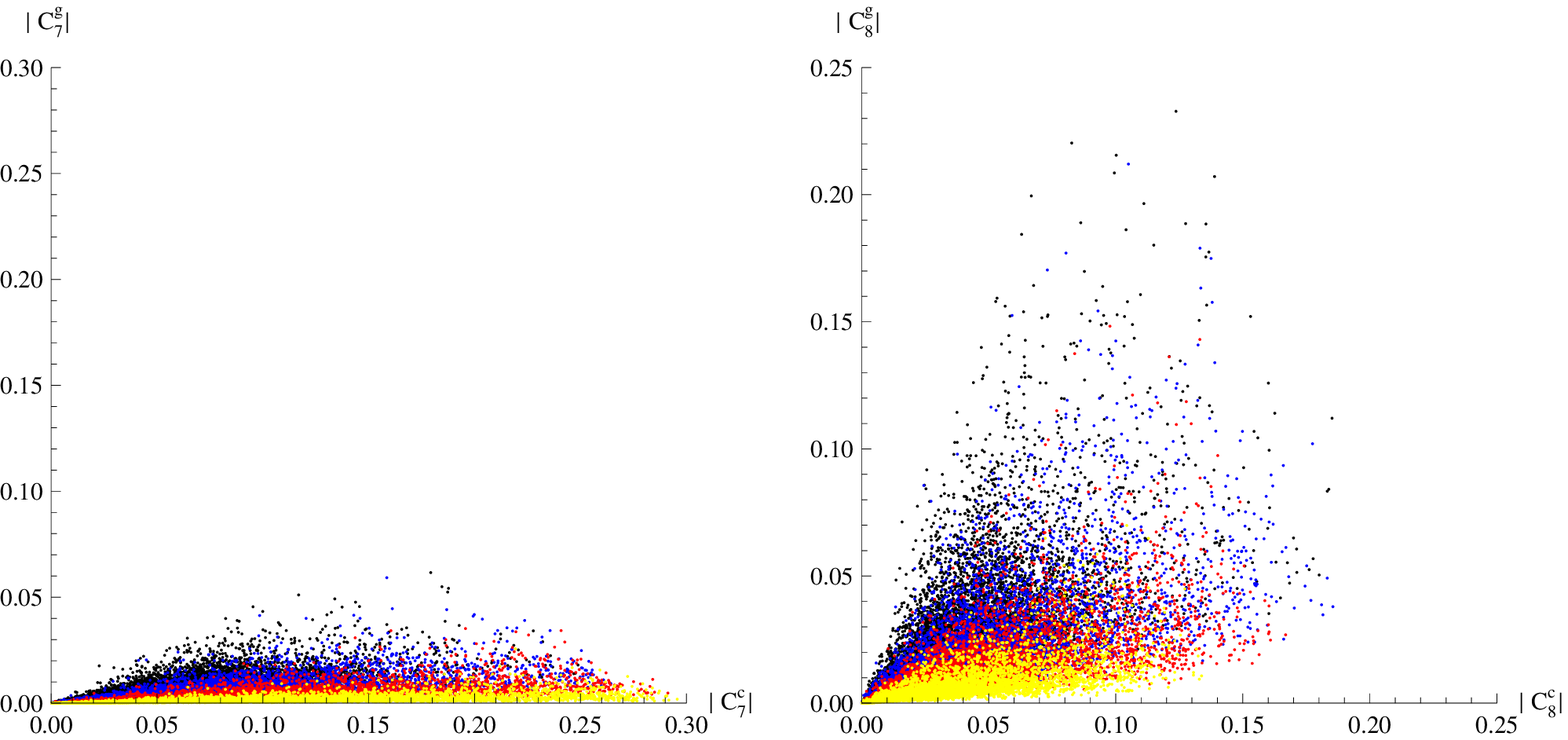}
\caption{Magnitudes of chargino and gluino contributions to $C_7(\mususy)$ and $C_8(\mususy)$ scanned over the MSSM parameter space.}
\label{scanc7c8cg}
\end{nfigure}
\begin{nfigure}{t}
\centering
\includegraphics[width=\textwidth]{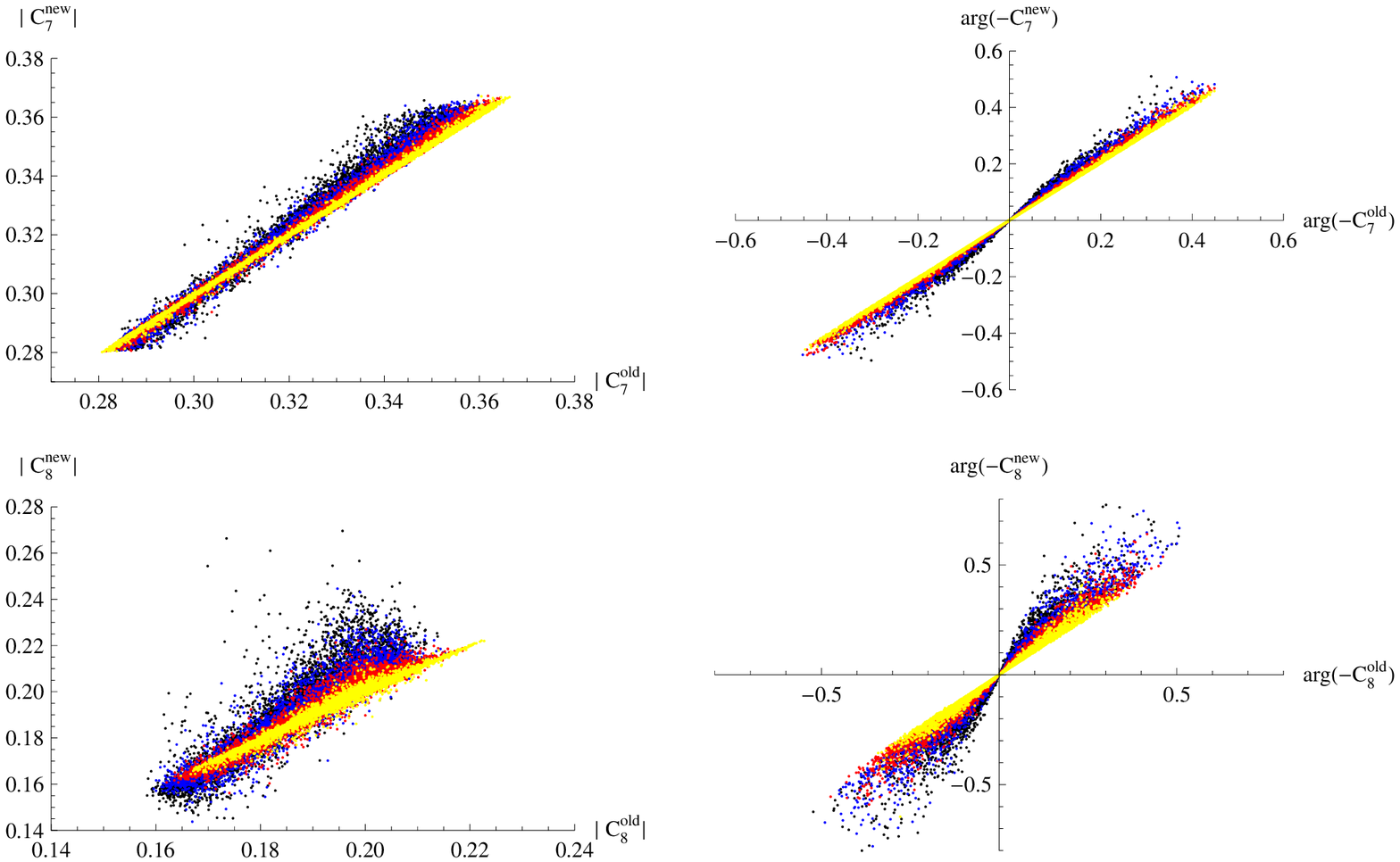}
\caption{Magnitudes and phases of $C_7(m_b)$ and $C_8(m_b)$ scanned over
  the MSSM parameter space. The meaning of the colours is the same as in
  Fig.~\ref{scanc7c8cg}. For further details see text.}
\label{scanc7c8}
\end{nfigure}

The results of the scan are depicted in Figs.~\ref{scanc7c8cg} and
\ref{scanc7c8}.  The plot in Fig.~\ref{scanc7c8cg} is a comparison of
the numerical importance of the well-known chargino contributions
$C_{7,8,\tilde \chi^\pm}(\mususy)$ on the one hand and the new gluino
contribution $C_{7,8,\tilde g}(\mususy)$ on the other hand. We show the
absolute values of these (complex) Wilson coefficients.  The picture
confirms our rough estimate in \eq{eta78}, i.e. it shows that the gluino
contribution to $C_7$ is accidentally suppressed, whereas it is enhanced
for $C_8$ and can yield sizeable corrections, especially for large
values of $|\mu|$. The different colours of the scan points correspond to
different ranges of values for $\mu$ as indicated in the picture legend.

Next, in Fig.~\ref{scanc7c8} we have plotted for each scan point in the
parameter space the absolute values and phases of $C_7(m_b)$ and
$C_8(m_b)$, including the SM and charged-Higgs contributions as well as
the $\tan\beta$-enhanced chargino contributions. The abscissa always
represents our new value, taking into account also the gluino and neutralino contributions from Eqs. (\ref{c7g},\ref{c8g}) and (\ref{c78n}), while the ordinate represents the corresponding ``old'' value, discarding gluino-squark and neutralino-squark diagrams. In this
way, the deviation from the diagonal indicates the relative size of the
new contribution. In the Standard Model both coefficients are negative;
we have plotted here $\arg(-C_{7,8})$ in order to center the phase plots
around the origin.

We can see that the gluino-squark contributions do not cause strong
modifications of $C_7(m_b)$, however they can have a strong impact on
$C_8(m_b)$ for large values of $\mu$. This confirms again the result of
our estimate in section \ref{sec:deltabeq1}. The reason for the
dependence of $C_8(m_b)$ on $\mu$ is the experimental constraint from
$\mathcal B(\overline B \rightarrow X_s\gamma)$. The value of $\mu$
determines the mass of the higgsino component of the charginos. If $|\mu|$
is small, the higgsino is light and gives a potentially large
contribution to $C_7(m_b)$ which is only compatible with data on
$\mathcal B(\overline B \rightarrow X_s\gamma)$ if $|A_t|$ is rather small
and the stops are rather heavy. As discussed above, this reduces in turn
the value of $\epsfc$, to which the gluino contributions to the magnetic
operators are proportional. Conversely, if $|\mu|$ is large, the higgsino
is heavy and larger values of $|A_t|$ and $\epsfc$ are possible.
\begin{ntable}{b}
\centering
\begin{tabular}{|c|c||c|c|}\hline
$\tilde m_{Q_L}$, $\tilde m_{u_R}$, $\tilde m_{d_R}$ &  $600$ GeV & 
 $A_b$ & $-600$ GeV \\ \hline
$\mu$ & $800$ GeV & $m_{A^0}$ & 350 GeV\\ \hline
$M_1$ & $300$ GeV & $M_2$ & $400$ GeV \\ \hline
$M_3$ & $500$ GeV & $\varphi_{A_t}$ & $3\pi/2$ \\ \hline
 $\tan\beta$ & $50$ & & \\ \hline
\end{tabular}
\caption{Parameter point used for the numerical analyses of $C_8(m_b)$
  in \fig{c8overat} and $S_{\phi K_S}$ in \fig{Soverat}.}
\label{fixedpara}
\end{ntable}
This feature is illustrated in \fig{c8overat} where we plot
$|C_8(m_b)|$ over $|A_t|$ while fixing the other MSSM parameters to the
values given in Tab.~\ref{fixedpara} and applying the same constraints
as above. We see that a wide range of values is allowed for $|A_t|$
(this range corresponds to the plot range) and that the importance of
the gluino-squark contributions to $|C_8(m_b)|$ grows with $|A_t|$.

\begin{nfigure}{t}
\centering
\includegraphics[width=13cm]{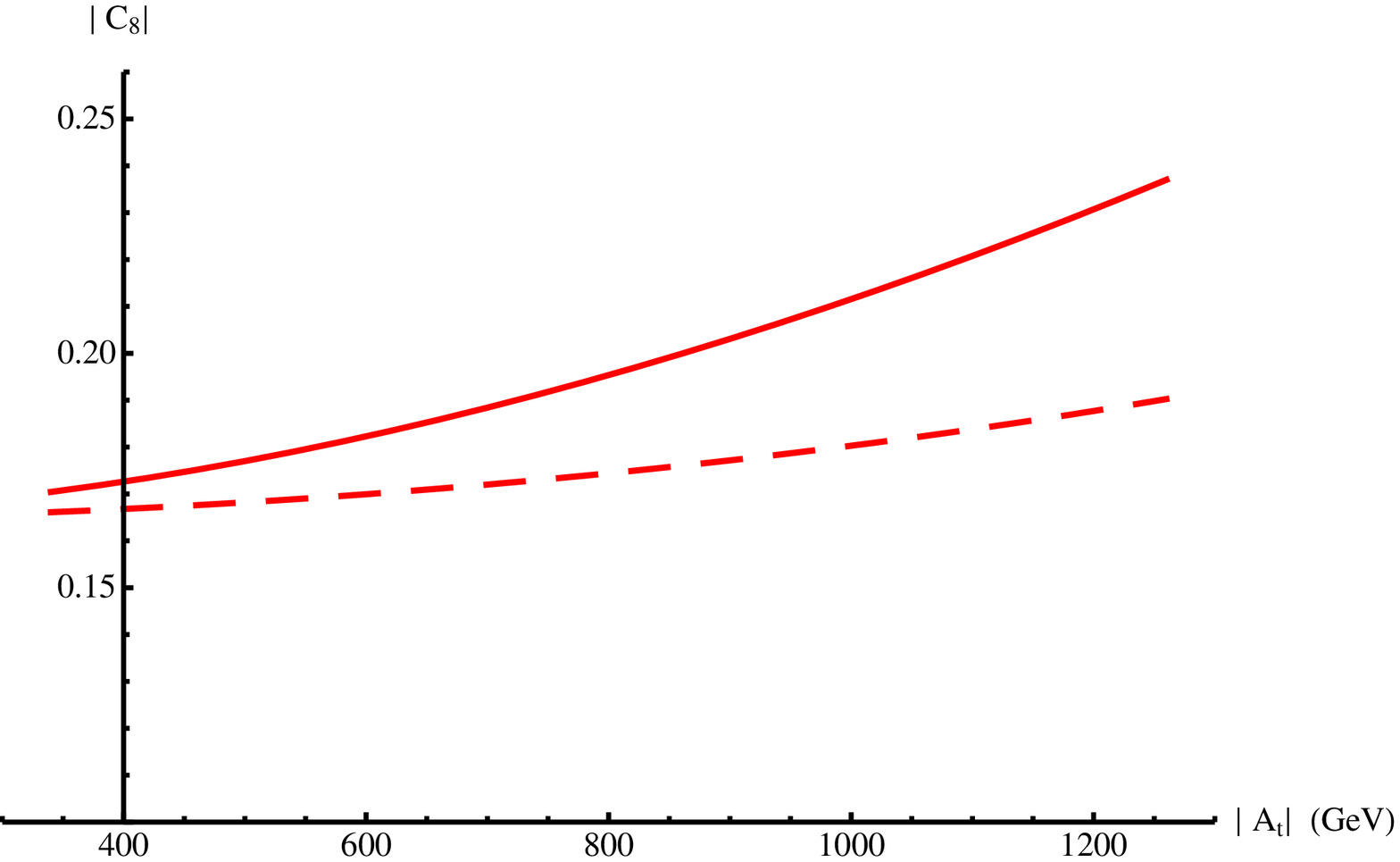}
\caption{$|C_8(m_b)|$ as a function of $|A_t|$ for the parameter point
  of Tab.~\ref{fixedpara}: full result (solid) and result without the
  gluino contribution (dashed).}
\label{c8overat}
\end{nfigure}
Our finding affects some important low-energy observables which depend
on $C_8(m_b)$. As an example, we have estimated the mixing-induced CP
asymmetry $S_{\phi K_S}$ of the FCNC decay $\bar B^0 \rightarrow \phi
K_S$. This decay is generated by $b\rightarrow s\, \bar s s$ QCD
penguins and can thus arise from the operator $\mathcal O_8$ with the
gluon coupling to $\bar s s$. Here we only want to give a qualitative
picture of the importance of the new contribution to the coefficient of
$\mathcal O_8$. Therefore we have calculated the matrix element only in
the leading-order of QCD factorisation \cite{Beneke:2000ry,Beneke:2003zv}, i.e.  dropping $\mathcal
O(\Lambda_\text{QCD}/m_b)$ and $\mathcal O(\alpha_s)$ corrections. Only
the $\tan\beta$-enhanced chargino and gluino contributions to
$C_8(m_b)$ are taken into account and their sum is denoted by
$C_8^{NP}$. The result presented here is therefore not to be seen as a
precise quantitative prediction. A more detailed study including
next-to-leading order effects will be performed in an upcoming
publication.

For the moment, we will follow the analyses of Refs.~\cite{Buchalla:2005us} and \cite{Altmannshofer:2008hc} and write
\begin{equation} 
  \mathcal A_{\phi K_S} \equiv 
  \langle \phi K_S| \mathcal H_\text{eff} | B^0\rangle 
   = A^c_{\phi K_S} \left[ 1 + a^u_{\phi K_S}e^{i\gamma} + 
 (b^c_{\phi K_S} + b^u_{\phi K_S}e^{i\gamma})C_{8}^{NP*}(m_W) \right]\label{aphiks}
\end{equation}
for the $B^0\to \phi K_S $ decay amplitude and $\bar{\mathcal A}_{\phi
  K_S}$ as the CP-conjugate $\bar B^0$ decay amplitude.  We remark that
the complex conjugation of $C_{8}^{NP}$ is missing in Ref.~\cite{Altmannshofer:2008hc}. With the standard definition
\begin{equation}
  \lambda_{\phi K_S} = 
 -e^{-i\phi_B} \frac{ \bar {\mathcal A}_{\phi K_S} }{\mathcal A_{\phi K_S}}
\end{equation}
the mixing-induced CP asymmetry reads 
\begin{equation}
  S_{\phi K_S} =  \frac{ 2 \,\text{Im}(\lambda_{\phi K_S})}{ 
    1 + |\lambda_{\phi K_S}|^2} .
\end{equation}
In this section we have not considered possible new-physics contributions
to the phase $\phi_B$ of \bbm, which are necessarily small in our naive MFV 
scenario. We have found agreement with
the numerical values of $a^u_{\phi K_S}$ and $b^c_{\phi K_S}$ in
Ref.\cite{Buchalla:2005us} and have used 
$b^u_{\phi K_S} \approx |V_{us}^*V_{ub}|/|V_{cs}^*V_{cb}|\,b^c_{\phi
  K_S}$.  In \fig{Soverat} we plot $S_{\phi K_S}$ versus $|A_t|$ for the
parameter point of Tab.~\ref{fixedpara}.  We can see a large impact of
the gluino-squark contribution on $C_8^{NP}(m_b)$, especially for large
$|A_t|$. 
\begin{nfigure}{t}
\centering
\includegraphics[width=12cm]{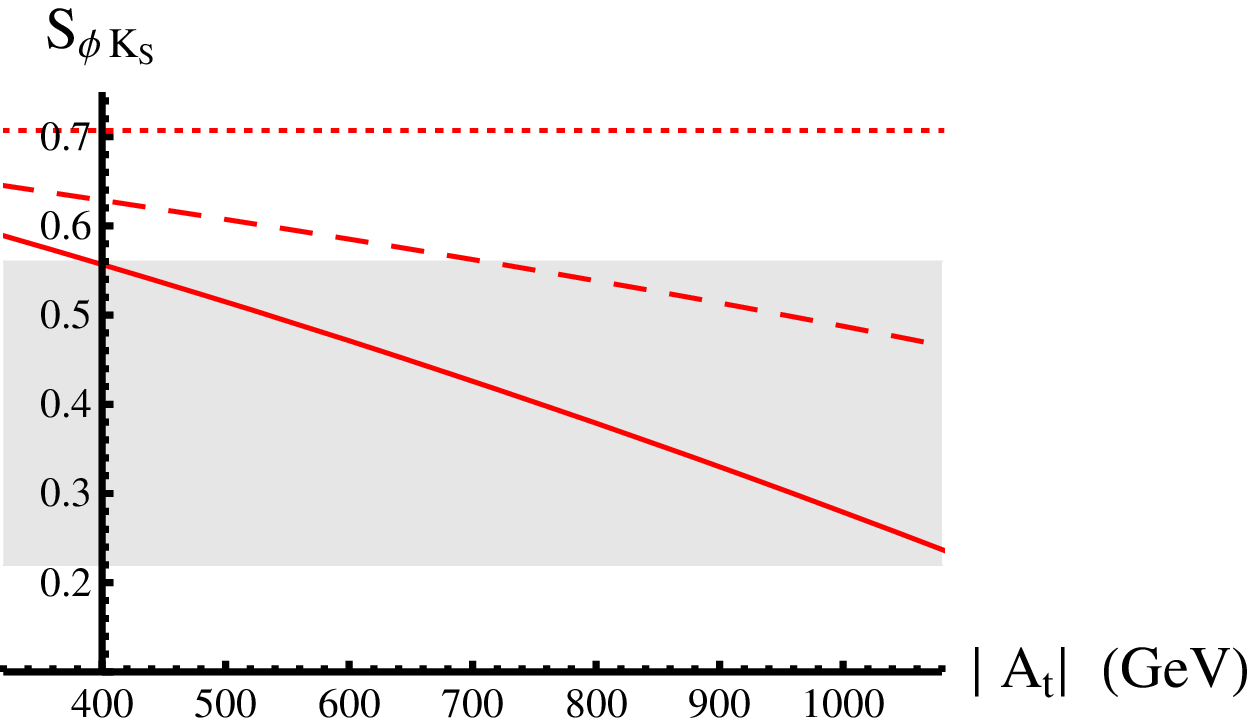}
\caption{$S_{\phi K_S}$ as a function of $|A_t|$ at the parameter point
  of Tab.~\ref{fixedpara}: full result (solid) and result without the
  gluino contribution (dashed). The shaded area represents the
  experimental $1\sigma$ range and the dotted line is the Standard-Model
  value.}
\label{Soverat}
\end{nfigure}

\section{Conclusions}

This paper addresses the MSSM for  large values of $\tan
\beta$.  We have considered a version of Minimal Flavour
Violation (MFV) in which all elementary couplings of neutral bosons to
(s)quarks are flavour-diagonal and the flavour structures of $W$,
charged-Higgs and chargino couplings are governed by the CKM matrix.
Complex phases of flavour-conserving parameters like the trilinear
SUSY-breaking term $A_t$ are consistently included in our results.  It
is well-known that loop suppression factors can be compensated by a
factor of $\tan\beta$, so that $\tan\beta$-enhanced loop diagrams must
be resummed to all orders in perturbation theory
\cite{Hall:1993gn,Carena:1994bv,
  Carena:2000uj,cgnw,Blazek:1995nv,Hempfling:1993kv,mmns}. Further
$\tan\beta$-enhanced loop-induced FCNC couplings of neutral Higgs bosons
lead to a plethora of interesting effects in $B$ physics, which can be
probed with current data from $B$ factories and the Tevatron
\cite{Hamzaoui:1998nu,Babu:1999hn,bcrs,Buras:2002wq,D'Ambrosio:2002ex,
  Isidori:2001fv,Dedes:2002er,Buras:2001mb,ddn,gjnt}. The subject is
usually treated with the help of an effective field theory, a general
two-Higgs-doublet model. This model is found by integrating out the
genuine supersymmetric particles and is therefore valid for $M_{\rm
  SUSY} \gg v, M_{A^0,H^0,H^\pm}$. In this paper we derive resummation
formulae which do not assume any hierarchy between $M_{\rm SUSY}$, the
electroweak scale $v$ and the Higgs masses. We use the diagrammatic
resummation developed in Ref.~\cite{cgnw} and extend the method to the
case of flavour-changing interactions.

As a first result we derive the dependence of the resummation formula on
the renormalisation scheme of the MSSM parameters. In particular we find
that the familiar expression of \eq{eq:ResForm1} is modified if the
sbottom mixing angle $\tilde{\theta}_b$ is used as input. This result is
useful if high-$p_T$ collider physics is studied in conjunction with
low-energy data from $B$ physics. While the focus of large-$\tan\beta$
collider physics has been on Higgs physics so far
\cite{cgnw,Carena:1998gk,Ellis:2007ss,Heinemeyer:2004xw}, our result
permits the correct treatment of $\tan\beta$-enhanced effects in
production and decay of bottom squarks.  We then resum
$\tan\beta$-enhanced loop corrections to flavour-changing processes for
arbitrary values of the supersymmetric particle masses. We find that the
renormalisation of CKM elements and the loop-induced neutral-Higgs
couplings to quarks have the same form as in the decoupling limit
$M_{\rm SUSY} \gg v, M_{A^0,H^0,H^\pm}$, but the loop-induced couplings
depend on the supersymmetric parameters in a different way. As novel
results we find $\tan\beta$-enhanced loop-induced couplings of gluinos
and neutralinos and determine the analogous corrections
to chargino couplings. These results permit the study of
$\tan\beta$-enhanced corrections to processes involving a decoupling
supersymmetric loop.  Since these processes vanish for $M_{\rm SUSY}\to
\infty$, they cannot be studied with the effective-field-theory method
employed in
Refs.~\cite{Hamzaoui:1998nu,Babu:1999hn,bcrs,Buras:2002wq,D'Ambrosio:2002ex,
  Isidori:2001fv,Buras:2001mb,gjnt}.  Other applications are
flavour-changing processes with squark final states, which may be a
topic for the ILC. All new FCNC couplings share a feature which was
found for the flavour-conserving Higgs couplings to quarks in
Ref.~\cite{cgnw}: The resummed $\tan\beta$-enhanced effects can be
absorbed into judiciously chosen counterterms. Therefore they can be
viewed as effective \emph{local}\ couplings, irrespective of the
hierarchy between $M_{\rm SUSY}$ and $v$.  We exploit this feature to
derive effective Feynman rules (collected in Appendix~\ref{feynman}) for
all affected MSSM couplings.  However, $\tan\beta$-enhanced corrections
to suppressed tree-level couplings of order $\cot \beta$ are non-local
and involve process-dependent form factors.

We have further performed an exhaustive phenomenological analysis of
FCNC processes in $B$ physics. The new gluino-squark loop contributions
are negligible for \bbm\ and are small in $b\to s \gamma$, where they
are of similar size as the non-enhanced two-loop contributions
\cite{Degrassi:2006eh}. The latter feature stems from an accidental
numerical suppression factor in the Wilson coefficient $C_7$. This
suppression is absent in $C_8$: Here the gluino-squark loop can
contribute as much as the known chargino-squark diagram. We have studied
the impact on the mixing-induced CP asymmetry $S_{\phi K_S}$ in the decay
$B_d\to \phi K_S$. The result in \fig{Soverat} complies with ${\cal
  B}(\ov B \to X_s \gamma) $ and the experimental lower bounds on the
masses of sparticles and the lightest Higgs boson. Since no MSSM Higgs
bosons are involved, the size of $S_{\phi K_S}$ is uncorrelated with
${\cal B} (B_s \to \mu^+\mu^-)$. Therefore tighter future bounds on the
latter quantity can be evaded by increasing $M_{A^0}$ without
suppressing $S_{\phi K_S}$. We have further generalised the known
neutral-Higgs mediated contributions to $B_s \to \mu^+\mu^-$ and \bbms\
to the case of arbitrary $M_{\rm SUSY}$. Our more accurate expression
for ${\cal B} (B_s \to \mu^+\mu^-)$ is especially useful once LHCb
measures this branching fraction in excess of the SM prediction.
Finally we have identified a new contribution to \bbms:
The parameter $r$ in \eq{defr} can alter the phase of the  \bbms\
amplitude and may affect the mixing-induced CP asymmetry in 
$B_s \to J/\psi \phi$.  

\section*{Acknowledgements}
The authors thank Andreas Crivellin for stimulating discussions and
Wolfgang Altmannshofer for communications concerning \eq{aphiks}.  We
further thank Leonardo Vernazza for helpful discussions on QCD
factorisation and Thomas Hahn for checking eqs. (\ref{eq:Self2},\ref{eq:Self3}). This work is supported by the DFG grant No.~NI 1105/1--1,
by project C6 of the DFG Research Unit SFB--TR 9
\emph{Computergest\"utzte Theoretische Teilchenphysik}, by the EU
Contract No.~MRTN-CT-2006-035482, \lq\lq FLAVIAnet'', and by BMBF grant
no.~05H09VKF. L.H.\ and D.S.\ acknowledge the support by Ev.~Studienwerk
Villigst and by Cusanuswerk, respectively, and by the DFG Graduate
College \emph{High Energy Physics and Particle Astrophysics}.

\begin{appendix}

\section{Conventions}\label{conventions}
Throughout this paper, our notation for SUSY parameters, sparticle masses and
mixing matrices follows the conventions of the SLHA \cite{Skands:2003cj}. In
Sect.~\ref{squarks} we extend the SLHA to accommodate complex phases in the
squark mass matrices. In Sect.~\ref{charginos} we give explicit expressions for certain combinations of elements of the chargino mixing matrices. Sect.~\ref{loopfunctions} lists the loop functions 
entering our results.

\subsection{Squark mixing}\label{squarks}

In the naive MFV scenario the squark mass-matrices are hermitian $2\times 2$-matrices. For top- and bottom-squarks they can be expressed in the basis $(\tilde q_L, \tilde q_R)$ with $q = t,b$ as 
\begin{equation}\label{squark_mm}
\mathcal M_{\tilde q}^2 = \begin{pmatrix} m_{\tilde q_{L}}^2  & X_{\tilde q}  \\  X^*_{\tilde q}  &  m_{\tilde q_{R}}^2 \end{pmatrix}.
\end{equation}
The diagonal elements can be chosen real and are given by
\begin{align}
   m_{\tilde q_{L}}^2 =& \widetilde{m}_{Q_L}^2 + m_q^2 + (T^3_q -Q_q  \sin^2\theta_W) m_Z^2 \cos 2\beta, \\
   m_{\tilde q_{R}}^2 =& \widetilde{m}_{q_{R}}^2 + m_q^2 + Q_q \sin^2 \theta_W  m_Z^2 \cos 2\beta.
\end{align}
Neglecting terms proportional to the small $v_d$ in the off-diagonal elements we obtain
\begin{align}
X_{\tilde t} &= m_t A^*_t, \\
X_{\tilde b} &= -y_b^{(0)*} v_u \mu .
\label{eq:xb}
\end{align}

The mass eigenstates $\tilde q_{1,2}$ are related to the weak eigenstates via 
\begin{equation}
\left(  \tilde q_1 , \tilde q_2 \right)^T = \widetilde{R}^q \left(  \tilde q_L , \tilde q_R \right)^T 
\end{equation}
with a unitary matrix $\widetilde{R}^q$ which diagonalises the mass matrix:
\begin{equation}\label{squark_diag}
\widetilde{R}^q \mathcal M_{\tilde q}^2 \widetilde{R}^{q\dagger} = \text{diag}(m_{\tilde q_1}^2, m_{\tilde q_2}^2), 
\end{equation}
\begin{align}\label{eq:SquarkMasses}
m^2_{\tilde q_{1,2}} & = \frac{1}{2}\left( m_{\tilde q_{L}}^2 + m_{\tilde q_{R}}^2 \pm \sqrt{ (m_{\tilde q_{L}}^2 - m_{\tilde q_{R}}^2)^2 + 4 |X_q|^2} \right).
\end{align}

If the diagonal elements of the mass matrix are chosen real, the mixing matrix contains only one physical phase and can thus be parameterised as
\begin{equation}
\widetilde{R}^q  = \begin{pmatrix} \cos \tilde{\theta}_{q} & \sin \tilde{\theta}_{q} e^{i\tilde{\phi}_{q}} \\  -\sin \tilde{\theta}_{q} e^{-i\tilde{\phi}_{q}} & \cos\tilde{\theta}_{q} \end{pmatrix},
\end{equation}
i.e. by two real parameters, the mixing-angle $\tilde{\theta}_q$ and the
phase $\tilde{\phi}_q$. In practical calculations where squarks are
involved, elements of the mixing matrices appear in the Feynman rules.
One then has the choice either to consider $\tilde{\theta}_{q}$ and
$\tilde{\phi}_{q}$ as input parameters or to express them by means of
the relation
\begin{align}\label{eq:MixingAngle}
e^{i\tilde{\phi}_{q}}\, \sin 2\tilde{\theta}_{q}  =&   
   \frac{2 X_{\tilde q}}{m_{\tilde q_1}^2 - m_{\tilde q_2}^2},
\end{align}
that can be derived from eq. (\ref{squark_diag}). To give separate
relations for $\tilde{\theta}_q$ and $\tilde{\phi}_q$ one has to specify
the allowed range for both parameters. Choosing $\tilde{\theta}_{q} \in
[0,\pi/4]$ and $\tilde{\phi}_{q} \in [0,2\pi)$ for example results in
\begin{equation}
  \sin 2\tilde{\theta}_q = \left|\frac{2 X_{\tilde q}}{m_{\tilde q_1}^2
      - m_{\tilde q_2}^2}\right|,\qquad
  \tilde{\phi}_q= \arg\left(\frac{2 X_{\tilde q}}{m_{\tilde q_1}^2 - 
      m_{\tilde q_2}^2}\right).
 \end{equation}
 Constraining $\tilde{\theta}_{q}$ to this interval amounts to defining
 $\tilde q_1$ ($\tilde q_2$) as the eigenstate which is predominantly
 left-handed
 (right-handed). 

 We emphasize that in the sbottom mass-matrix we have defined the
 off-diagonal element $X_{\tilde b}$ in terms of the Yukawa coupling
 $y_b^{(0)}$ instead of the bottom mass. This parameterisation is valid
 irrespective of the renormalisation scheme used for the
 $\tan\beta$-enhanced corrections to $m_b$. In practical calculations,
 one can use one of the resummation formulae given in section
 \ref{flavour_cons} to relate $y_b^{(0)}$ to the measured bottom mass.
 The corresponding corrections to $m_b^2$ in the diagonal elements of
 the sbottom mass-matrix are negligible since $m_b^2\ll
 \widetilde{m}_{Q_L}^2,\widetilde{m}_{b_{R}}^2$.

\subsection{Chargino mixing}\label{charginos}
In our conventions the chargino mass-matrix is given by
\begin{equation}
   \mathcal{M}_{\widetilde{\chi}^{\pm}}=\begin{pmatrix} M_2 & \sqrt{2}M_W\sin\beta \\ \sqrt{2}M_W\cos\beta & \mu \end{pmatrix}.
   \label{eq:CharMass}
\end{equation}
We define the biunitary transformation which brings it into diagonal form as
\begin{equation}
    \widetilde{U}^*\mathcal{M}_{\widetilde{\chi}^{\pm}}\widetilde{V}^{\dagger}=
        \textrm{diag}(m_{\widetilde{\chi}^{\pm}_1},m_{\widetilde{\chi}^{\pm}_2})
    \label{eq:CharMix}.
\end{equation}
The matrices $\widetilde{U}$ and $\widetilde{V}$ can be determined by
diagonalising the matrices
$\mathcal{M}_{\widetilde{\chi}^{\pm}}^{\dagger}\mathcal{M}_{\widetilde{\chi}^{\pm}}$
and
$\mathcal{M}_{\widetilde{\chi}^{\pm}}\mathcal{M}_{\widetilde{\chi}^{\pm}}^{\dagger}$.
In Feynman amplitudes for diagrams with chirality-flipping propagators
only certain combinations of matrix-elements of $\widetilde{U}$ and
$\widetilde{V}$ appear. These combinations can be expressed as
\begin{align}
   \widetilde{U}_{11}\widetilde{V}_{11}&=\frac{m_{\widetilde{\chi}_1^{\pm}}M_2-m_{\widetilde{\chi}_2^{\pm}}\mu^*
      \,e^{i\psi}}{m_{\widetilde{\chi}_1^{\pm}}^2-m_{\widetilde{\chi}_2^{\pm}}^2},
   & \widetilde{U}_{11}\widetilde{V}_{12}&=\sqrt{2}M_W\frac{m_{\widetilde{\chi}_1^{\pm}}\sin\beta+
      m_{\widetilde{\chi}_2^{\pm}}\cos\beta\,e^{i\psi}}{m_{\widetilde{\chi}_1^{\pm}}^2
       -m_{\widetilde{\chi}_2^{\pm}}^2}, \\
  \widetilde{U}_{12}\widetilde{V}_{12}&=\frac{m_{\widetilde{\chi}_1^{\pm}}\mu-m_{\widetilde{\chi}_2^{\pm}}M_2^*\,e^{i\psi}}
  {m_{\widetilde{\chi}_1^{\pm}}^2-m_{\widetilde{\chi}_2^{\pm}}^2},
   & \widetilde{U}_{12}\widetilde{V}_{11}&=\sqrt{2}M_W\frac{m_{\widetilde{\chi}_1^{\pm}}\cos\beta+
  m_{\widetilde{\chi}_2^{\pm}}\sin\beta\,e^{i\psi}}{m_{\widetilde{\chi}_1^{\pm}}^2-
  m_{\widetilde{\chi}_2^{\pm}}^2}, \\
   \widetilde{U}_{21}\widetilde{V}_{21}&=\frac{m_{\widetilde{\chi}_1^{\pm}}\mu^*\,e^{i\psi}-
    m_{\widetilde{\chi}_2^{\pm}}M_2}{m_{\widetilde{\chi}_1^{\pm}}^2-m_{\widetilde{\chi}_2^{\pm}}^2},
   &\widetilde{U}_{21}\widetilde{V}_{22}&=-\sqrt{2}M_W\frac{m_{\widetilde{\chi}_1^{\pm}}\cos\beta\,e^{i\psi}+
      m_{\widetilde{\chi}_2^{\pm}}\sin\beta}{m_{\widetilde{\chi}_1^{\pm}}^2-m_{\widetilde{\chi}_2^{\pm}}^2}, \\
  \widetilde{U}_{22}\widetilde{V}_{22}&=\frac{m_{\widetilde{\chi}_1^{\pm}}M_2^*\,e^{i\psi}-m_{\widetilde{\chi}_2^{\pm}}
    \mu}{m_{\widetilde{\chi}_1^{\pm}}^2-m_{\widetilde{\chi}_2^{\pm}}^2},
   &\widetilde{U}_{22}\widetilde{V}_{21}&=-\sqrt{2}M_W\frac{m_{\widetilde{\chi}_1^{\pm}}\sin\beta\,e^{i\psi}+
  m_{\widetilde{\chi}_2^{\pm}}\cos\beta}{m_{\widetilde{\chi}_1^{\pm}}^2-m_{\widetilde{\chi}_2^{\pm}}^2}
\end{align}
with
\begin{equation}
   e^{i\psi}=(M_2\mu-M_W^2\sin 2\beta)/(m_{\widetilde{\chi}_1^{\pm}}m_{\widetilde{\chi}_2^{\pm}}).
\end{equation}
For large $\tan\beta$ the $\cos\beta$-terms can be neglected and the above expressions reduce to
\begin{align} \label{eq:ChMixMat1}
   \widetilde{U}_{11}\widetilde{V}_{11}&=\frac{M_2}{m_{\widetilde{\chi}_1^{\pm}}}\cdot
     \frac{m_{\widetilde{\chi}_1^{\pm}}^2-|\mu|^2}
      {m_{\widetilde{\chi}_1^{\pm}}^2-m_{\widetilde{\chi}_2^{\pm}}^2},
   & \widetilde{U}_{11}\widetilde{V}_{12}&=\frac{\sqrt{2}M_W m_{\widetilde{\chi}_1^{\pm}}\sin\beta}
      {m_{\widetilde{\chi}_1^{\pm}}^2-m_{\widetilde{\chi}_2^{\pm}}^2}, \\
  \widetilde{U}_{12}\widetilde{V}_{12}&=\frac{\mu}{m_{\widetilde{\chi}_1^{\pm}}}\cdot
    \frac{m_{\widetilde{\chi}_1}^2-|M_2|^2}{m_{\widetilde{\chi}_1^{\pm}}^2-m_{\widetilde{\chi}_2^{\pm}}^2},
   & \widetilde{U}_{12}\widetilde{V}_{11}&=\frac{M_2}{m_{\widetilde{\chi}_1^{\pm}}}\cdot
     \frac{\sqrt{2}M_W \mu\sin\beta}{m_{\widetilde{\chi}_1^{\pm}}^2-m_{\widetilde{\chi}_2^{\pm}}^2}, \\
   \widetilde{U}_{21}\widetilde{V}_{21}&=\frac{M_2}{m_{\widetilde{\chi}_2^{\pm}}}\cdot
     \frac{|\mu|^2-m_{\widetilde{\chi}_2^{\pm}}^2}{m_{\widetilde{\chi}_1^{\pm}}^2-m_{\widetilde{\chi}_2^{\pm}}^2},
   &\widetilde{U}_{21}\widetilde{V}_{22}&=-\frac{\sqrt{2}M_W m_{\widetilde{\chi}_2^{\pm}}\sin\beta}
     {m_{\widetilde{\chi}_1^{\pm}}^2-m_{\widetilde{\chi}_2^{\pm}}^2}, \\\label{eq:ChMixMat2}
  \widetilde{U}_{22}\widetilde{V}_{22}&=\frac{\mu}{m_{\widetilde{\chi}_2^{\pm}}}\cdot
   \frac{|M_2|^2-m_{\widetilde{\chi}_2^{\pm}}^2}{m_{\widetilde{\chi}_1^{\pm}}^2-m_{\widetilde{\chi}_2^{\pm}}^2},
   &\widetilde{U}_{22}\widetilde{V}_{21}&=-\frac{\mu}{m_{\widetilde{\chi}_2^{\pm}}}\cdot
     \frac{\sqrt{2}M_W M_2\sin\beta}
    {m_{\widetilde{\chi}_1^{\pm}}^2-m_{\widetilde{\chi}_2^{\pm}}^2}.
\end{align}

\subsection{Loop functions}\label{loopfunctions}

In the calculation of quark self-energies with internal SUSY particles, we use the scalar integrals
\begin{align}
B_0(m_1,m_2) &= \frac{(2\pi\mu)^{4-d}}{i\pi^2} \int{ \frac{d^dq}{(q^2-m_1^2)(q^2-m_2^2)}},\\
C_0(m_1,m_2,m_3) &= \frac{(2\pi\mu)^{4-d}}{i\pi^2} \int \frac{d^dq}{(q^2-m_1^2)(q^2-m_2^2)(q^2-m_3^2)},\\
D_0(m_1,m_2,m_3,m_4) &= \frac{(2\pi\mu)^{4-d}}{i\pi^2} \int \frac{d^dq}{(q^2-m_1^2)(q^2-m_2^2)(q^2-m_3^2)(q^2-m_4^2)},
\end{align}
where $\mu$ is the renormalisation scale. This corresponds to the well-known Passarino-Veltman notation with vanishing external momenta. Besides, we use the function 
\begin{equation}
D_2(m_1,m_2,m_3,m_4)=\frac{(2\pi\mu)^{4-d}}{i\pi^2} \int \frac{q^2\, d^dq}{(q^2-m_1^2)(q^2-m_2^2)(q^2-m_3^2)(q^2-m_4^2)}.
\end{equation}
Explicit expressions for these integrals read \small
\begin{align}
& B_0(m_1,m_2) = \frac{2}{4-d} - \gamma_E + \log 4\pi  + 1 -\log\frac{m_1^2}{\mu^2} + \frac{m_2^2}{m_2^2-m_1^2} \log\frac{m_1^2}{m_2^2}, \\
& C_0(m_1,m_2,m_3) = \frac{m_2^2}{(m_1^2-m_2^2)(m_3^2-m_2^2)}\log\frac{m_1^2}{m_2^2} + \frac{m_3^2}{(m_1^2-m_3^2)(m_2^2-m_3^2)} \log\frac{m_1^2}{m_3^2} ,\\
& D_0(m_1,m_2,m_3,m_4) = \frac{m_2^2}{(m_2^2-m_1^2)(m_2^2-m_3^2)(m_2^2-m_4^2)} \log\frac{m_1^2}{m_2^2} + \notag\\ &\qquad \frac{m_3^2}{(m_3^2-m_1^2)(m_3^2-m_2^2)(m_3^2-m_4^2)} \log\frac{m_1^2}{m_3^2}
 + \frac{m_4^2}{(m_4^2-m_1^2)(m_4^2-m_2^2)(m_4^2-m_3^2)} \log\frac{m_1^2}{m_4^2}, \label{d0def}\\
& D_2(m_1,m_2,m_3,m_4) = \frac{m_2^4}{(m_2^2-m_1^2)(m_2^2-m_3^2)(m_2^2-m_4^2)} \log\frac{m_1^2}{m_2^2} +\notag\\  & \qquad \frac{m_3^4}{(m_3^2-m_1^2)(m_3^2-m_2^2)(m_3^2-m_4^2)} \log\frac{m_1^2}{m_3^2} 
 + \frac{m_4^4}{(m_4^2-m_1^2)(m_4^2-m_2^2)(m_4^2-m_3^2)} \log\frac{m_1^2}{m_4^2} .
\end{align}\normalsize
The divergence in $B_0$ always drops out when we sum over the internal squarks and gauginos. 

In our expressions for the Wilson coefficients $C_{7,8}$, we use the loop functions 
\begin{align}
   f_1(x)=&\frac{5-7x}{6(x-1)^2}+\frac{x(3x-2)}{3(x-1)^3}\log x,\\
   f_2(x)=&\frac{x+1}{2(x-1)^2}-\frac{x}{(x-1)^3}\log x,\\
   f_3(x)=&\frac{1}{2(x-1)}-\frac{x}{2(x-1)^2}\log x.
\end{align}

\section{QCD corrections to flavour-changing self-energies}\label{sec:QCDcorrections}

Here we want to discuss the issue of the bottom mass appearing in
calculations following the approach of Sect.~\ref{self_en_approach}.
In that section, we have introduced $\tan\beta$-enhanced
flavour-mixing via flavour-changing self-energies $\Sigma^{RL}_{bs}$
in external legs.  As a consequence the quark pole-mass
$m_b^{\textrm{pole}}$ enters the resulting expression through the
Dirac equation $\slashed{p}b=m_b^{\textrm{pole}}b$. However, as we
will show in this section, QCD corrections add in such a way that the
final result does not depend on $m_b^{\textrm{pole}}$ but only on the
$\overline{\rm MS}$-mass $m_b$.

To see this we consider an effective theory at $\mu\sim\mathcal{O}(m_b)$ where
the SUSY-particles are integrated out. The self-energy $\Sigma^{RL}_{bs}$ then
appears as Wilson coefficient of the (on-shell vanishing) operator
$\bar{b}P_Ls$. Comparing QCD corrections to this operator to QCD corrections
to the bottom mass $m_b$ (see \fig{fig:QCD1})
\begin{nfigure}{t}
  \begin{picture}(200,50)(-200,-10)
    \SetWidth{0.5}
    \SetColor{Black}
    \ArrowLine(-180,0)(-150,0)
    \ArrowLine(-150,0)(-120,0)
    \ArrowLine(-110,0)(-80,0)
    \ArrowLine(-80,0)(-50,0)
    \GlueArc(-115,0)(35,0,180){5}{8}
    \GOval(-115,0)(5,5)(0){0.882}
    \Text(-170,-15)[lb]{\Black{$s_L$}}
    \Text(-70,-15)[lb]{\Black{$b_R$}}
    \Text(-140,-15)[lb]{\Black{$s_L$}}
    \Text(-100,-15)[lb]{\Black{$b_R$}}
    \Text(-125,5)[lb]{\small{\Black{$-i\Sigma^{RL}_{bs}$}}}
    \Text(-45,5)[lb]{\Black{$\equiv$}}
    \Text(-30,2)[lb]{\Black{$-i{\Sigma_{bs}^{RL(1)}}$}}
    \ArrowLine(60,0)(90,0)
    \ArrowLine(90,0)(120,0)
    \ArrowLine(120,0)(150,0)
    \ArrowLine(150,0)(180,0)
    \GlueArc(120,0)(30,0,180){5}{8}
    \Vertex(120,0){2}
    \Text(70,-15)[lb]{\Black{$b_L$}}
    \Text(100,-15)[lb]{\Black{$b_L$}}
    \Text(130,-15)[lb]{\Black{$b_R$}}
    \Text(160,-15)[lb]{\Black{$b_R$}}
    \Text(110,5)[lb]{\small{\Black{$-im_b$}}}
    \Text(185,5)[lb]{\Black{$\equiv$}}
    \Text(200,0)[lb]{\Black{$-i{\Sigma^{QCD}_{b}}$}}
  \end{picture}

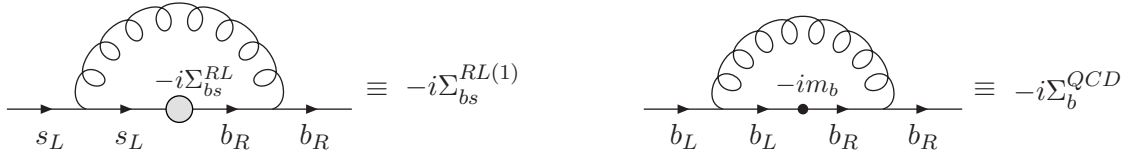
\captionof{figure}{QCD corrections to the self-energy 
   $\Sigma^{RL}_{bs}$ (left) and the bottom mass $m_b$ (right).}
\label{fig:QCD1}
\end{nfigure}
we find
\begin{equation}
   \frac{\Sigma^{RL(1)}_{bs}(p)}{\Sigma^{RL}_{bs}}=\frac{\Sigma^{QCD}_{b}(p)}{m_b} \label{eq:QCDRel},
\end{equation}
where $p$ denotes the external momentum. Therefore the Wilson coefficient $\Sigma^{RL}_{bs}$ and the $\overline{\rm MS}$-mass $m_b$ renormalise the same way. To make the behaviour under renormalisation explicit we write 
\begin{equation}
   \Sigma^{RL}_{bs}=m_bA
   \label{eq:ParamSelf} 
\end{equation}
where now $A$ is renormalisation-scale-independent (note the analogy
to the definitions of $\epsilon_b$ and $\epsfc$ in
Eqs.~(\ref{eq:Self0}), (\ref{eq:dbk}) and (\ref{eq:FCSelf}) which are
thus renormalisation-scale independent).

Now we calculate QCD corrections to the diagrams in 
\fig{fig:ExtLegSelf}. Using the parameterisation (\ref{eq:ParamSelf}) for
$\Sigma^{RL}_{bs}$ and neglecting the s-quark mass the Feynman
amplitudes for the diagrams in figure \ref{fig:ExtLegSelf} read
\begin{eqnarray}
  \mathcal{M}^{(1)}_1&=&\mathcal{M}^{\textrm{rest}}_1\cdot{\left.\frac{i(\slashed{p}+m_b)}{p^2-m_b^2}\right|}_{\slashed{p}=0}(-i\Sigma^{RL}_{bs})=-\mathcal{M}^{\textrm{rest}}_1\cdot A, \label{eq:ExtLegQCD1}\\
  \mathcal{M}^{(2)}_2&=&\mathcal{M}^{\textrm{rest}}_2\cdot{\left.\frac{i(\slashed{p}+m_s)}{p^2-m_s^2}\right|}_{\slashed{p}=m_b^{\textrm{pole}}}(-i\Sigma^{RL*}_{bs})=+\mathcal{M}^{\textrm{rest}}_2\cdot A^*\frac{m_b}{m_b^{\textrm{pole}}}\label{eq:ExtLegQCD2}.
\end{eqnarray}
Since we want to perform a calculation up to order $\alpha_s$ in the
effective theory we have to determine $A$ from two-loop matching at the
SUSY scale and we make this explicit by writing
\begin{equation}
    A=A^{(0)}+A^{(1)}
\end{equation}
where $A^{(1)}$ contains $\mathcal{O}(\alpha_s)$ QCD-corrections. The
one-loop corrections to $\mathcal{M}_1$ and $\mathcal{M}_2$ in the
effective theory are given in Figs.~\ref{fig:QCD2} and \ref{fig:QCD3},
respectively, with diagrams (1b) and (2b) taking into account the
counterterm to the Wilson coefficient $\Sigma^{RL}_{bs}=m_b\,A$. As a
consequence of (\ref{eq:QCDRel}), the contributions of (1a) and (1c) and
of (1b) and (1d) cancel pairwise so that the expression for
$\mathcal{M}_1$ in (\ref{eq:ExtLegQCD1}) still holds at one loop with
$A=A^{(0)}+A^{(1)}$ instead of $A=A^{(0)}$. For the contributions of
(2a) and (2b) we find with the help of (\ref{eq:QCDRel})
\begin{nfigure}{t}
  \begin{picture}(200,190) (-220,-140)
    \SetWidth{0.5}
    \SetColor{Black}
    \ArrowLine(-200,0)(-170,0)
    \ArrowLine(-170,0)(-140,0)
    \GOval(-135,0)(5,5)(0){0.882}
    \ArrowLine(-130,0)(-100,0)
    \ArrowLine(-100,0)(-70,0)
    \ArrowLine(-70,0)(-40,0)
    \Vertex(-70,0){2}
    \GlueArc(-135,0)(35,0,180){5}{8}
    \Text(-190,-15)[lb]{\Black{$s_L$}}
    \Text(-160,-15)[lb]{\Black{$s_L$}}
    \Text(-120,-15)[lb]{\Black{$b_R$}}
    \Text(-90,-15)[lb]{\Black{$b_R$}}
    \Text(-60,-15)[lb]{\Black{$b_L$}}
    \Line(-40,-20)(-40,20)
    \Line(-40,-20)(-30,-10)
    \Line(-40,-10)(-30,0)
    \Line(-40,0)(-30,10)
    \Line(-40,10)(-30,20)
    \Text(-93,-34)[lb]{\Black{(1a)}} 
    \Text(-150,8)[lb]{\small{\Black{$-im_bA$}}}
    \ArrowLine(50,0)(80,0)
    \ArrowLine(90,0)(120,0)
    \GOval(85,0)(5,5)(0){0.882}
    \SetWidth{1}
    \Line(80,5)(90,-5)
    \Line(80,-5)(90,5)
    \SetWidth{0.5}
    \ArrowLine(120,0)(150,0)
    \Vertex(120,0){2}
    \Text(60,-15)[lb]{\Black{$s_L$}}
    \Text(100,-15)[lb]{\Black{$b_R$}}
    \Text(130,-15)[lb]{\Black{$b_L$}}
    \Line(150,-20)(150,20)
    \Line(150,-20)(160,-10)
    \Line(150,-10)(160,0)
    \Line(150,0)(160,10)
    \Line(150,10)(160,20) 
    \Text(107,-34)[lb]{\Black{(1b)}}
    \Text(70,8)[lb]{\small{\Black{$-i\delta m_bA$}}}
    \ArrowLine(-200,-100)(-170,-100)
    \ArrowLine(-160,-100)(-140,-100)
    \GOval(-165,-100)(5,5)(0){0.882}
    \ArrowLine(-140,-100)(-120,-100)
    \ArrowLine(-120,-100)(-100,-100)
    \ArrowLine(-100,-100)(-80,-100)
    \ArrowLine(-80,-100)(-60,-100)
    \ArrowLine(-60,-100)(-40,-100)
    \Vertex(-140,-100){2}
    \Vertex(-100,-100){2}
    \Vertex(-60,-100){2}
    \GlueArc(-100,-100)(20,0,180){3}{5}
    \Text(-190,-115)[lb]{\Black{$s_L$}}
    \Text(-155,-115)[lb]{\Black{$b_R$}}
    \Text(-135,-115)[lb]{\Black{$b_L$}}
    \Text(-115,-115)[lb]{\Black{$b_L$}}
    \Text(-95,-115)[lb]{\Black{$b_R$}}
    \Text(-75,-115)[lb]{\Black{$b_R$}}
    \Text(-55,-115)[lb]{\Black{$b_L$}}
    \Line(-40,-120)(-40,-80)
    \Line(-40,-120)(-30,-110)
    \Line(-40,-110)(-30,-100)
    \Line(-40,-100)(-30,-90)
    \Line(-40,-90)(-30,-80)
    \Text(-93,-134)[lb]{\Black{(1c)}} 
    \Text(-180,-92)[lb]{\small{\Black{$-im_bA$}}}
    \ArrowLine(50,-100)(80,-100)
    \ArrowLine(90,-100)(110,-100)
    \GOval(85,-100)(5,5)(0){0.882}
    \SetWidth{1}
    \Line(125,-95)(135,-105)
    \Line(125,-105)(135,-95)
    \SetWidth{0.5}
    \ArrowLine(110,-100)(130,-100)
    \ArrowLine(130,-100)(150,-100)
    \ArrowLine(150,-100)(170,-100)
    \Vertex(110,-100){2}
    \Vertex(150,-100){2}
    \Text(60,-115)[lb]{\Black{$s_L$}}
    \Text(95,-115)[lb]{\Black{$b_R$}}
    \Text(115,-115)[lb]{\Black{$b_L$}}
    \Text(135,-115)[lb]{\Black{$b_R$}}
    \Text(155,-115)[lb]{\Black{$b_L$}}
    \Line(170,-120)(170,-80)
    \Line(170,-120)(180,-110)
    \Line(170,-110)(180,-100)
    \Line(170,-100)(180,-90)
    \Line(170,-90)(180,-80) 
    \Text(107,-134)[lb]{\Black{(1d)}}
    \Text(125,-92)[lb]{\small{\Black{$-i\delta m_b$}}}
    \Text(70,-92)[lb]{\small{\Black{$-im_bA$}}} 
  \end{picture}
  
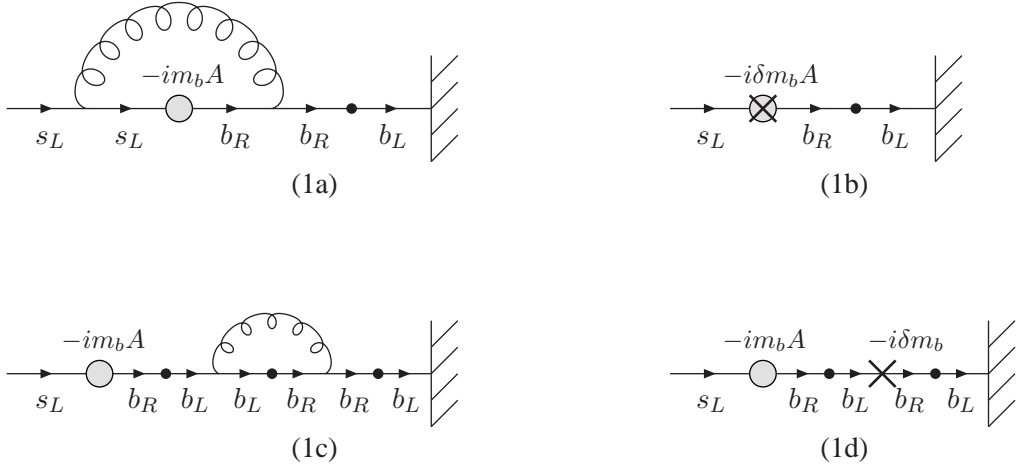
\captionof{figure}{QCD corrections to diagram (1) in 
    \fig{fig:ExtLegSelf}.}
  \label{fig:QCD2}
\end{nfigure}
\begin{eqnarray}
  \mathcal{M}^{(2a)}_2&=&
      \mathcal{M}^{\textrm{rest}}_2\cdot{\left.\frac{i(\slashed{p}+m_s)}{p^2-m_s^2}
      \left(-i\Sigma^{RL(1)*}_{bs}(p)\right)\right|}_{\slashed{p}=m_b^{\textrm{pole}}}=
      \mathcal{M}^{\textrm{rest}}_2\cdot A^{(0)*}
      \left.\frac{\Sigma^{QCD}_b(p)}{m_b^{\textrm{pole}}}\right|_{\slashed{p}=m_b^{\textrm{pole}}}\\
      \mathcal{M}^{(2b)}_2&=&\mathcal{M}^{\textrm{rest}}_2\cdot{\left.\frac{i(\slashed{p}+m_s)}{p^2-m_s^2}\right|}_{\slashed{p}=m_b^{\textrm{pole}}}(-i\delta
      m_b A^{(0)*})=\mathcal{M}^{\textrm{rest}}_2\cdot
      A^{(0)*}\frac{\delta m_b}{m_b^{\textrm{pole}}}.
\end{eqnarray}
Adding these to \eq{eq:ExtLegQCD2} one gets
\begin{equation}
   \mathcal{M}_2=\mathcal{M}^{(2)}_2+\mathcal{M}^{(2a)}_2+\mathcal{M}^{(2b)}_2=
                \mathcal{M}_2^{\textrm{rest}}\cdot \frac{A^{(0)*}}{m_b^{\textrm{pole}}}\left(m_b+m_b\frac{A^{(1)*}}{A^{(0)*}}+\left.\Sigma^{QCD}_b(p)\right|_{\slashed{p}=m_b^{\textrm{pole}}}+\delta m_b\right).
   \label{eq:M2}
\end{equation}
Plugging in
\begin{equation}
  m_b^{\textrm{pole}}=m_b+\left.\Sigma^{QCD}_b(p)\right|_{\slashed{p}=m_b^{\textrm{pole}}}+\delta m_b
\end{equation}
and dropping terms of order $\mathcal{O}(\alpha_s^2)$ we get the final
result
\begin{equation}
   \mathcal{M}_2=\mathcal{M}^{\textrm{rest}}_2\cdot (A^{(0)*}+A^{(1)*})= \mathcal{M}^{\textrm{rest}}_2\cdot A^*\label{eq:FinRes}
\end{equation}
which now does not depend on $m_b^{\textrm{pole}}$ anymore.

Applying this result to our case by expressing $A$ in \eq{eq:FinRes}
through $\Sigma^{RL}_{bs}$ via \eqsand{eq:ParamSelf}{eq:FCSelf} 
we find \eq{eq:ExtLegSelf2}. Since
\eq{eq:FCSelf} is linear in $m_b$, the parameterisation
of \eq{eq:ParamSelf} is quite natural.  When one considers
a more general $\Sigma^{RL}_{bs}$ which is no longer linear in $m_b$
(for example in the generic MSSM), the parameter $A$ depends on $m_b$ via
(\ref{eq:ParamSelf}) but in any case it does not involve
$m_b^{\textrm{pole}}$.

\section{Feynman rules}\label{feynman}
In this appendix, we explain how $\tan\beta$-enhanced loop corrections
can be incorporated into calculations in the MSSM with naive MFV by simple
modifications of the Feynman rules. The resulting modified rules are
valid beyond the decoupling limit and refer to input scheme (i) for the
sbottom parameters specified in section \ref{flavour_cons}. They can
also be used for processes with external SUSY particles. The
modifications, which can easily be implemented into computer programs
like FeynArts, are given as follows:

\begin{nfigure}{tb}
  \begin{picture}(200,90) (-220,-30)
    \SetWidth{0.5}
    \SetColor{Black}
    \ArrowLine(-200,0)(-170,0)
    \ArrowLine(-170,0)(-140,0)
    \GOval(-105,0)(5,5)(0){0.882}
    \ArrowLine(-140,0)(-110,0)
    \ArrowLine(-100,0)(-70,0)
    \ArrowLine(-70,0)(-40,0)
    \Vertex(-170,0){2}
    \GlueArc(-105,0)(35,0,180){5}{8}
    \Text(-190,-15)[lb]{\Black{$b_L$}}
    \Text(-160,-15)[lb]{\Black{$b_R$}}
    \Text(-130,-15)[lb]{\Black{$b_R$}}
    \Text(-90,-15)[lb]{\Black{$s_L$}}
    \Text(-60,-15)[lb]{\Black{$s_L$}}
    \Line(-40,-20)(-40,20)
    \Line(-40,-20)(-30,-10)
    \Line(-40,-10)(-30,0)
    \Line(-40,0)(-30,10)
    \Line(-40,10)(-30,20)
    \Text(-93,-34)[lb]{\Black{$(2a)$}} 
    \Text(-120,8)[lb]{\small{\Black{$-im_bA^*$}}}
    \ArrowLine(50,0)(80,0)
    \ArrowLine(80,0)(110,0)
    \GOval(115,0)(5,5)(0){0.882}
    \SetWidth{1}
    \Line(110,5)(120,-5)
    \Line(110,-5)(120,5)
    \SetWidth{0.5}
    \ArrowLine(120,0)(150,0)
    \Vertex(80,0){2}
    \Text(60,-15)[lb]{\Black{$b_L$}}
    \Text(90,-15)[lb]{\Black{$b_R$}}
    \Text(130,-15)[lb]{\Black{$s_L$}}
    \Line(150,-20)(150,20)
    \Line(150,-20)(160,-10)
    \Line(150,-10)(160,0)
    \Line(150,0)(160,10)
    \Line(150,10)(160,20) 
    \Text(107,-34)[lb]{\Black{$(2b)$}}
    \Text(100,8)[lb]{\small{\Black{$-i\delta m_bA^*$}}}
  \end{picture}
  
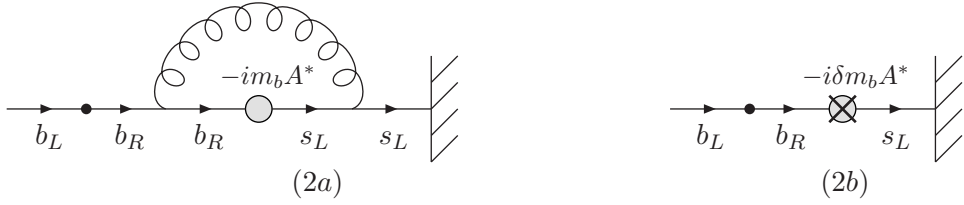
\captionof{figure}{QCD corrections to diagram (2) in \fig{fig:ExtLegSelf}.}
  \label{fig:QCD3}
\end{nfigure}
\begin{itemize}
 \item[(i)] Express the Feynman rules in terms of the down-type Yukawa couplings $y_{d_i}$ and replace them according to relation (\ref{eq:ResForm1}) by
 \begin{equation}
     y_{d_i}\to y_{d_i}^{(0)}=\frac{m_{d_i}}{v_d(1+\epsilon_i\tan\beta)}.
 \end{equation}
 It should be stressed that the same replacement has to be performed for
 the Yukawa coupling appearing in the sbottom mass matrix
 $\mathcal{M}_{\tilde{b}}$ in (\ref{squark_mm}) before determining the
 mixing angle via (\ref{eq:MixingAngle}). In case one wants to rely on
 input scheme (iii) the sbottom mixing matrix has to be calculated
 iteratively as described in section \ref{sect:smr}.

 \item[(ii)] Replace CKM-elements involving the third quark generation according to
   \begin{align}
     V_{ti} &\longrightarrow V_{ti}^{(0)}=\frac{1+\epsilon_b\tan\beta}{1+(\epsilon_b-\epsfc)\tan\beta}V_{ti} \qquad (i=d,s)\\
     V_{ib} &\longrightarrow V_{ib}^{(0)}=
     \frac{1+\epsilon^*_b\tan\beta}{1 +
       (\epsilon^*_b-\epsilon^*_{FC})\tan\beta}V_{ib} \qquad (i=u,c).
   \end{align}
   All other CKM-elements remain unchanged. The $V_{ij}$ appearing after
   these replacements correspond to the physical ones which can be
   measured from the $W^+ u_i d_j$-vertex.

 \item[(iii)] This last rule concerns vertices involving down-type
   quarks. Into these one has to include the flavour-changing
   wave-function counterterms
   \begin{align}
     \frac{ \delta Z^L_{bi}}{2} &=
     -\frac{\epsfc \tan\beta}{1+\epsilon_b\tan\beta} V_{tb}^{*} V_{ti}^{(0)}\\
     \frac{ \delta Z^R_{bi}}{2} &= -\frac{m_i}{m_b}\left[ \frac{ \epsfc
         \tan\beta}{1+\epsilon_b\tan\beta} +
       \frac{\epsilon^*_{FC}\tan\beta } {1+\epsilon^*_i \tan\beta}
     \right] V_{tb}^{*} V_{ti}^{(0)}
     \end{align}
     for $i=d,s$. This leads to additional flavour-changing vertices and
     occasionally cancels the corrections from rule (ii).
\end{itemize}

If one uses our Feynman rules, $\tan\beta$-enhanced loop corrections of
the form $(\epsilon\tan\beta)^n$ are automatically resumed to all
orders. There is one exception: Proper vertex-corrections to the
$\tan\beta$-suppressed $h^0 d^i d^j$- and $H^+ d^i_L u^j_R$-vertices and
to the corresponding Goldstone-boson vertices can not be accounted for
by this method.

As mentioned above, additional flavour-changing vertices are generated
by replacement rule (iii) in the case of external down-quarks. In the
following we give explicit Feynman rules for these vertices, suppressing
therein colour indices of (s)quarks. Repeated indices are not summed
over.
\bigskip\\
\begin{minipage}{0.25\textwidth}
\includegraphics[width=\textwidth]{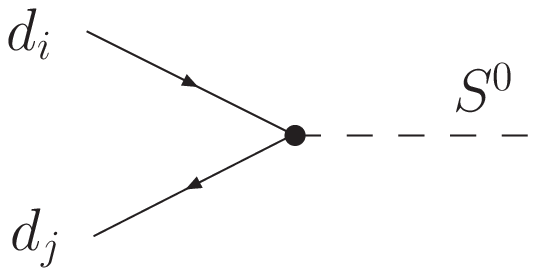}
\end{minipage}
\begin{minipage}{0.74\textwidth}
   \begin{align}\nonumber
     -\frac{i}{\sqrt{2}}&\left[ x_d^S\left(\delta_{ji}\,y_{d_j}^{(0)}
         +\frac{\delta Z^L_{ji}}{2}\,y_{d_j}^{(0)} -
         \frac{\delta Z^R_{ji}}{2}\,y_{d_i}^{(0)}\right)P_L\right.\\
     & \left. + (x_d^S)^{*}\left(\delta_{ji}\,y_{d_j}^{(0)*} +
         \frac{\delta Z^R_{ji}}{2}\,y_{d_j}^{(0)*} - \frac{\delta
           Z^L_{ji}}{2}\,y_{d_i}^{(0)*}\right)P_R\right]
   \end{align}
 \end{minipage}
 \begin{equation*}
   \textrm{with}\qquad x_d^S=
   (\cos\alpha,-\sin\alpha,i\sin\beta,-i\cos\beta)\quad \textrm{for} 
 \quad S^0=(H^0,h^0,A^0,G^0)
\end{equation*}
\bigskip\\ 
\begin{minipage}{0.25\textwidth}
  \includegraphics[width=\textwidth]{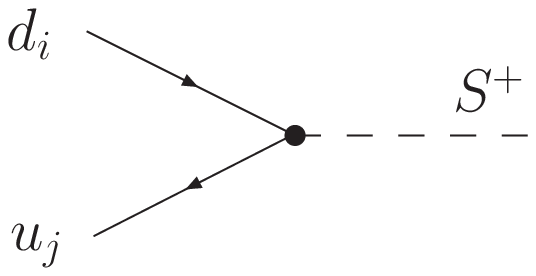}
\end{minipage}
\begin{minipage}{0.74\textwidth}
\begin{equation}
  i\xi_L^S\,  y_{u_j}\,V_{ji}\,P_L + i\xi_R^S
  \left( y^{(0)*}_{d_i}\,V^{(0)}_{ji} + 
    \frac{\delta Z^R_{ji}}{2}\,y^{(0)*}_{d_j}\,V_{jj}\right)P_R 
\label{feynsp}
\end{equation}
\end{minipage}
\begin{equation}
  \textrm{with}\qquad \xi^S_L = (\cos\beta,\sin\beta) \quad\textrm{and} 
  \quad\xi^S_R=(\sin\beta,-\cos\beta)
  \qquad \textrm{for} \quad S^+=(H^+,G^+) 
\end{equation}

\bigskip
\begin{minipage}{0.25\textwidth}
\includegraphics[width=\textwidth]{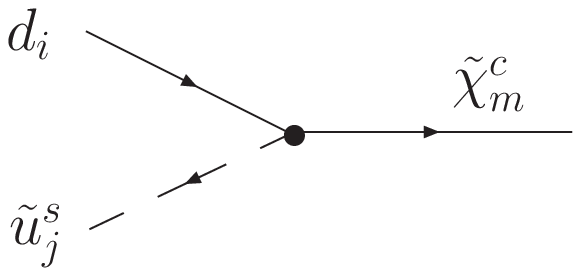} 
\end{minipage}
\begin{minipage}{0.74\textwidth}
\begin{align}\nonumber
  & i V_{ji} \left(y_{u_j} \widetilde{R}^{u_j}_{s2} \widetilde{V}_{m2}^*
    - g \widetilde{R}^{u_j}_{s1} \widetilde{V}_{m1}^*\right)P_L \\ & + i
  \widetilde{R}^{u_j}_{s1}\, \widetilde{U}_{m2} \left( y_{d_i}^{(0)*}
    V_{ji}^{(0)} + \frac{\delta Z^R_{ji}}{2} \, y_{d_j}^{(0)*} V_{jj}
  \right) P_R
\end{align}
\end{minipage}
\bigskip\\
\begin{minipage}{0.25\textwidth}
\includegraphics[width=\textwidth]{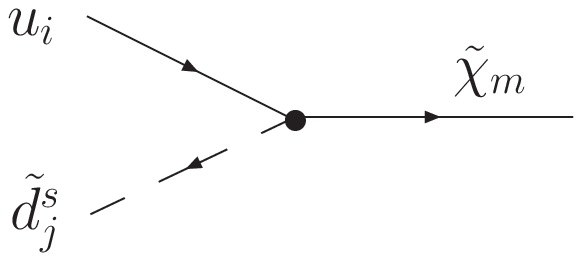} 
\end{minipage}
\begin{minipage}{0.74\textwidth}
\begin{equation}
  iV_{ij}^{(0)*} \left[ \left(y_{d_j}^{(0)} 
      \widetilde{R}^{d_j}_{s2} \widetilde{U}_{m2}^* - 
      g \widetilde{R}^{d_j}_{s1} \widetilde{U}_{m1}^*\right)P_L  + 
    y_{u_i} \widetilde{R}^{d_j}_{s1} \widetilde{V}_{m2} P_R      \right]
\end{equation}
\end{minipage}
\bigskip\\
\begin{minipage}{0.25\textwidth}
  \includegraphics[width=\textwidth]{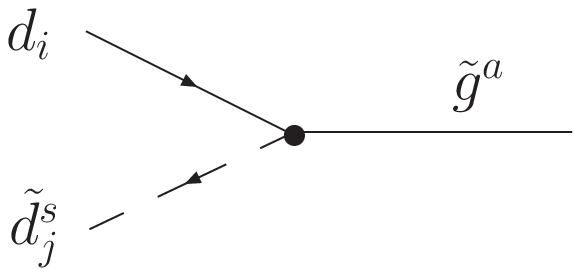}
\end{minipage}
\begin{minipage}{0.74\textwidth}
\begin{equation}
  -i \sqrt{2} g_s T^a \left[\left(\delta_{ji}+\frac{\delta Z^L_{ji}}{2} 
    \right) \widetilde{R}^{d_j}_{s1} P_L - 
    \left(\delta_{ji} + \frac{\delta Z^R_{ji}}{2}\right) 
    \widetilde{R}^{d_j}_{s2} P_R  \right]
\end{equation}
\end{minipage}
\bigskip\\
\begin{minipage}{0.25\textwidth}
\includegraphics[width=\textwidth]{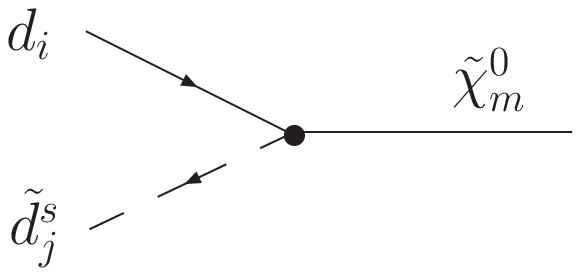} 
\end{minipage}
\begin{minipage}{0.74\textwidth}
\begin{align}\nonumber
  & i \left(\delta_{ji} + \frac{\delta Z^{L}_{ji}}{2}\right) \left[
    \sqrt{2} \widetilde{R}^{d_j}_{s1}\left( \frac{g}{2}
      \widetilde{N}^*_{m2} - \frac{g'}{6} \widetilde{N}^*_{m1} \right) -
    y_{d_j}^{(0)} \widetilde{R}^{d_j}_{s2} \widetilde{N}^*_{m3}\right]  P_L \\
  & - i\left(\delta_{ji}+\frac{\delta Z^{R}_{ji}}{2}\right)\left[ \frac{
      \sqrt{2}}{3}g' \widetilde{R}^{d_j}_{s2} \widetilde{N}_{m1} +
    y_{d_j}^{(0)*} \widetilde{R}^{d_j}_{s1} \widetilde{N}_{m3} \right]
  P_R
\end{align}
\end{minipage}
\bigskip\\
Occasionally, the flavour-changing counterterms have to be explicitly
inserted into external or internal quark lines.  In these cases, they
cancel insertions of $\tan\beta$-enhanced flavour-changing self-energies
up to corrections which are
suppressed by at least one power of $m_b/\msusy$. The Feynman rule reads\\\\
\begin{minipage}{0.25\textwidth}
\includegraphics[width=\textwidth]{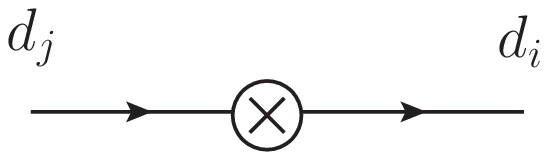} 
\end{minipage}
\begin{minipage}{0.74\textwidth}
\begin{align}\nonumber
  &-i\left( \frac{m_i}{1+\epsilon_i\tan\beta} \frac{\delta Z^L_{ij}}{2} - \frac{m_j}{1+\epsilon_j\tan\beta} \frac{\delta Z^R_{ij}}{2} \right) P_L\\
  & -i \left( \frac{m_i}{1+\epsilon_i^*\tan\beta} \frac{\delta
      Z^R_{ij}}{2} - \frac{m_j}{1+\epsilon_j^*\tan\beta} \frac{\delta
      Z^L_{ij}}{2} \right) P_R .
\end{align}
\end{minipage}

\end{appendix}

\bibliographystyle{JHEP}

\bibliography{paper}

\end{document}